\documentclass[11pt,a4paper]{article}
\pdfoutput=1
\usepackage{jheppub}

\usepackage{amsfonts,amssymb}
\usepackage[mathscr]{eucal}
\usepackage{amsmath}
\usepackage{graphicx}

\newcommand{\be}{\begin{equation}}
\newcommand{\ee}{\end{equation}}
\newcommand{\bea}{\begin{eqnarray}}
\newcommand{\eea}{\end{eqnarray}}
\newcommand{\ba}{\begin{eqnarray}}
\newcommand{\ea}{\end{eqnarray}}

\newcommand{\beq}{\begin{equation}}
\newcommand{\eeq}{\end{equation}}
\newcommand{\beqa}{\begin{eqnarray}}
\newcommand{\eeqa}{\end{eqnarray}}
\newcommand{\beqar}{\begin{eqnarray*}}
\newcommand{\eeqar}{\end{eqnarray*}}

\newcommand{\ratio}{\mathbf{r}} 
\newcommand{\Cblk}{C_\text{blk}}
\newcommand{\Cbdy}{C_{{\cal O}}} 
\newcommand{\tCbdy}{\widetilde C_{{\cal O}}} 
\newcommand{\RAdS}{R_{\text{\tiny AdS}}}
\newcommand{\mAdS}{m_{\text{\tiny AdS}}}

\newcommand{\Mdiamonds}{\mathcal{M}_\lozenge^{(d)}}
\newcommand{\Casimir}{\mathcal{C}_2}
\newcommand{\BoxM}{\nabla_\lozenge^2}

\def\half{\frac{1}{2}}

\newcommand{\eg}{{\it e.g.,}\ } 
\newcommand{\ie}{{\it i.e.,}\ }
\newcommand{\reef}[1]{(\ref{#1})}
\newcommand{\ssc}{\scriptscriptstyle}

\newcommand{\see}{S_\mt{EE}}
\newcommand{\hm}{H_{\ssc m}}
\newcommand{\rhoo}{\rho_{\ssc 0}}
\newcommand{\mt}[1]{\textrm{\tiny #1}}
\newcommand{\coset}{\lozenge} 

\newcommand{\labell}[1]{\label{#1}} %
\newcommand{\dS}[2]{Q({\cal O};#1,#2)} 
\newcommand{\dSh}[2]{Q_\mt{holo}({\cal O};#1,#2)} 
\newcommand{\dSO}{Q({\cal O})} 
\newcommand{\dSOh}{Q_\mt{holo}({\cal O})} 
\newcommand{\dSt}{Q({\cal O};u,\bar{u};v,\bar{v})}

\newcommand{\DSR}{\Delta S_{\ssc R}}
\newcommand{\DSL}{\Delta S_{\ssc L}}
\newcommand{\tDSR}{\widetilde\Delta S_{\ssc R}}

\newcommand{\Rs}{R_\mt{sph}}
\newcommand{\Rads}{R_\mt{AdS}}
\newcommand{\lp}{\ell_\mt{P}}
\newcommand{\cO}{{\cal O}}
\newcommand{\cW}{{\cal W}}

\title{Entanglement, Holography and Causal Diamonds}

\author[a]{Jan de Boer,}
\author[b]{Felix M.\ Haehl,}
\author[c,*]{Michal P.\ Heller\note[*]{On leave from: \emph{National Centre for Nuclear Research, Ho{\.z}a 69, 00-681 Warsaw, Poland}.}}
\author[c]{and Robert C.\ Myers}

 \affiliation[a]{Institute of Physics, Universiteit van Amsterdam,\\
 Science Park 904, 1090 GL Amsterdam, The Netherlands}

 \affiliation[b]{Centre for Particle Theory \& Department of Mathematical Sciences,\\
Durham University, South Road, Durham DH1 3LE, UK.}

 \affiliation[c]{Perimeter Institute for Theoretical Physics,\\
 31 Caroline Street North, Waterloo, Ontario N2L 2Y5, Canada}

\emailAdd{j.deboer@uva.nl}
\emailAdd{f.m.haehl@gmail.com}
\emailAdd{mheller@perimeterinstitute.ca}
\emailAdd{rmyers@perimeterinstitute.ca}

\abstract{We argue that the degrees of freedom in a $d$-dimensional CFT can be re-organized in an insightful way by studying observables on the moduli space of causal diamonds (or equivalently, the space of pairs of timelike separated points).  This $2d$-dimensional space naturally captures some of the fundamental nonlocality and causal structure inherent in the entanglement of CFT states. For any primary CFT operator, we construct an observable on this space, which is defined by smearing the associated one-point function over causal diamonds. Known examples of such quantities are the entanglement entropy of vacuum excitations and its higher spin generalizations. We show that in holographic CFTs, these observables are given by suitably defined integrals of dual bulk fields over the corresponding Ryu-Takayanagi minimal surfaces. Furthermore, we explain connections to the operator product expansion and the first law of entanglement entropy from this unifying point of view. We demonstrate that for small perturbations of the vacuum, our observables obey linear two-derivative equations of motion on the space of causal diamonds. In two dimensions, the latter is given by a product of two copies of a two-dimensional de Sitter space. For a class of universal states, we show that the entanglement entropy and its spin-three generalization obey nonlinear equations of motion with local interactions on this moduli space, which can be identified with Liouville and Toda equations, respectively. This suggests the possibility of extending the definition of our new observables beyond the linear level more generally and in such a way that they give rise to new dynamically interacting theories on the moduli space of causal diamonds. Various challenges one has to face in order to implement this idea are discussed.
}

\begin{document}  

\maketitle

\section{Introduction}
\label{sec:intro}

It has now been a decade since Ryu and Takayanagi \cite{rt1,rt2} discovered an elegant geometric prescription to evaluate entanglement entropy in gauge/gravity duality.  In particular, the entanglement entropy between a (spatial) region
$V$ and its complement $\bar V$ in the boundary theory is computed as
 \be
\see(V) = \mathrel{\mathop {\rm
ext}_{\scriptscriptstyle{v\sim V}} {}\!\!} \left[\frac{A(v)}{4G_\mt{N}}\right]\,.
 \label{define}
 \ee
That is, one determines the extremal value of the Bekenstein-Hawking formula evaluated on bulk surfaces $v$ which are homologous to the boundary region $V$. 
In the subsequent years, holographic entanglement entropy has proven to be a remarkably fruitful topic of study. In particular, it provides a useful diagnostic with which to examine the boundary theory. For example, it was shown to be an effective probe to study thermalization in quantum quenches, \eg \cite{Lopez1,Clifford1,vijay1,vijay2} or to distinguish different phases of the boundary theory, \eg \cite{Igor1,Clifford2,Cai44}. In fact, such holographic studies have even revealed new universal properties that extend beyond holography and hold for generic CFTs, \eg \cite{fthem1,fthem2,pablo1,pablo2}. 

However, holographic entanglement entropy has also begun to provide new insights into the nature of quantum gravity in the bulk. As first elucidated in \cite{mvr0,mvr}, the Ryu-Takayanagi prescription indicates the essential role which entanglement plays in creating the connectivity of the bulk geometry or more generally in the emergence of the holographic geometry. In fact, this has lead to a new prescription to reconstruct the bulk geometry in terms of a new boundary observable known as  `differential entropy', which provides a novel prescription for sampling the entanglement throughout the boundary state \cite{Balasubramanian:2013lsa,diff1,diff2,diff3}. 

The distinguished role of extremal surfaces in describing entanglement entropy has led to several other important insights. 
There is by now significant evidence that the bulk region which can be described by a particular boundary causal domain is not determined by causality alone, as one might have naively thought, but rather it corresponds to the so-called `entanglement wedge,' which in general extends deeper into the bulk, \eg \cite{Czech77,Czech66,matt88,Dong:2016eik}. That is, the bulk region comprised of points which are spacelike-separated from extremal surfaces attached to the boundary region and connected to the corresponding boundary causal domain \cite{matt88}. This entanglement wedge reconstruction in turn led to the insight that local bulk operators must have simultaneous but different approximate descriptions in 
various spatial subregions of the boundary theory, which resulted in intriguing connections to quantum error correction 
\cite{Almheiri:2014lwa,Mintun:2015qda,Pastawski:2015qua}.
We also notice that while it is not at all clear that a suitable factorization of the full quantum gravity Hilbert space 
corresponding to the inside and outside of an arbitrary spatial domain exists (there certainly is no obvious choice of tensor subfactors
on the CFT Hilbert space), the RT prescription does provide a natural choice for such a factorization for extremal
surfaces, and entanglement wedge reconstruction supports this point of view. It is therefore conceivable that a reorganization
of the degrees of freedom which crucially relies on extremal surfaces will shed some light on the (non)locality of the
degrees of freedom of quantum gravity, and this was in fact one of the original motivations for this work.

One interesting result that was brought to light by holographic studies of the relative entropy \cite{Blanco:2013joa} was the `first law of entanglement'. The relative entropy is again a general diagnostic that allows one to compare different states reduced to the same entangling geometry \cite{wehrl1,vedral2}. For `nearby' states, the leading variation of the relative entropy yields a result reminiscent of the first law of thermodynamics, \ie
\beq
\delta \see =\delta\langle \hm \rangle\, ,
\label{rambo}
\eeq
where $\hm$ is the modular or entanglement Hamiltonian for the given reference state $\rhoo$, \ie $\hm=-\log \rhoo$. While the latter is a useful device at a formal level \cite{haag2}, in generic situations, the modular Hamiltonian is a nonlocal operator, \ie $\hm$ cannot be expressed as a local expression constructed from fields within the region of interest. However, a notable exception to this general rule arises in considering a spherical region in the vacuum state of a CFT and in this case, the first law \reef{rambo} becomes  
\be
\label{deltaS}
\delta \see
=\delta\langle \hm \rangle= 2 \pi \int_{B}\!\! d^{d-1} x' \ \frac{R^{2} - |\vec{x} - \vec{x}'|^{2}}{2 R} \ \langle T_{t t}(\vec{x}')\rangle \,.
\ee
Here $B$ denotes a ball of radius $R$ centred at $\vec{x}$ on a fixed time slice, while $\langle T_{t t}\rangle$ is the energy density in the excited state being compared to the vacuum. Examining this expression holographically, the energy density is determined by the asymptotic behaviour of the metric near the AdS boundary, \eg \cite{haro}. In contrast, through Eq.~\reef{define}, the variation of the entanglement entropy is determined by variations of the geometry deep in the bulk spacetime. Hence Eq.~\reef{deltaS} imposes a nonlocal constraint on perturbations of the AdS geometry which are dual to excitations of the boundary CFT. However, if one examines this constraint for all balls of all sizes and all positions, as well as on all time slices, this can be re-expressed in terms of a local constraint on the bulk geometry \cite{eom1,eom2,eom3}, namely, {\it perturbations of the AdS vacuum geometry must satisfy the linearized Einstein equations!}

In terms of the boundary theory, the holographic results above point towards the utility of considering the entanglement entropy as a functional on the space of all entangling surfaces (or at least a broad class of such geometries) to characterize various excited states of a given quantum field theory. In this regard,  one intriguing observation \cite{deBoer:2015kda} is that the perturbations of the entanglement entropy of any CFT naturally live on an auxiliary de Sitter geometry. In particular, the functional $\delta\see(R,\vec{x})$, defined by Eq.~\reef{deltaS}, satisfies the Klein-Gordon equation
\be
\label{eq.wave}
\left(\nabla^2_{\mathrm{dS}} 
- m^{2} \right) \delta \see = 0\,,
\ee in the following de-Sitter (dS) geometry:
\be
\label{eq.deSitter}
ds_{\mathrm{dS}_{d}}^{2} =  \frac{L^{2}}{R^{2}} \left( - dR^{2} + d\vec{x}^{2} \right)\,.
\ee
Note that the radius of the spheres $R$  plays the role of time in dS space. The mass above is given by
\be
\label{eq.mass}
m^{2} L^{2} = -d\,,
\ee
where $d$ is the spacetime dimension of the CFT.\footnote{We should note that for $d=2$ essentially the same dS geometry appeared in \cite{Czech:2015qta}, which used integral geometry to describe the relation of MERA tensor networks \cite{MERA} to the AdS$_\mt{3}$/CFT$_\mt{2}$ correspondence.} In CFTs with higher spin symmetries, one can extend this construction using the corresponding conserved currents to produce additional scalars, which also propagate on the dS geometry according to a Klein-Gordon equation with an appropriate mass \cite{deBoer:2015kda} --- see section \ref{sec:spin} below.  

The proposal of \cite{deBoer:2015kda} was that this new dS geometry may provide the  foundation on which to construct an alternative `holographic' description of any CFT. That is, it may be possible to reorganize any CFT in terms a local theory of interacting fields propagating in the auxiliary spacetime. We stress that here the CFT  under consideration need not be holographic in the conventional sense of the AdS/CFT correspondence, and hence there is no requirement of a large central charge or strong coupling. Of course, the discussion in \cite{deBoer:2015kda} only provided some preliminary steps towards establishing this new holographic dictionary and such a program faces a number of serious challenges. For example, the dS scale only appears as an overall factor of $L^2$ in Eq.~\reef{eq.wave} and so remains an undetermined constant. Of course, our experience from the AdS/CFT correspondence suggests that $L$ would be determined in terms of CFT data through the gravitational dynamics of the holographic geometry and so here one faces the question of understanding whether the new auxiliary geometry is actually dynamical.

Another challenge would be to produce a holographic description of the time dependence of quantities in the CFT, since the above construction was firmly rooted on a fixed time slice. A natural extension is to consider all spherical regions throughout the $d$-dimensional spacetime of the CFT, \ie all of the ball-shaped regions of all sizes and at all positions on all time slices. As described in \cite{deBoer:2015kda}, this extended perspective yields an auxiliary geometry which is $SO(2,d)/[SO(1,d-1)\times SO(1,1)]$ and  the perturbations $\delta\see$ can be seen to obey a wave equation on this coset. Further it was noted that this auxiliary space is 2$d$-dimensional and has {\it multiple} time-like directions. 

This new expanded auxiliary geometry is the starting point for the present paper. As we will describe, in the context where we are considering all spheres throughout the spacetime, it is more natural to think in terms of the causal diamonds, where each causal diamond is the domain of dependence of a spherical region. Following \cite{Czech:2015qta}, our nomenclature will be to refer to the moduli space of all causal diamonds as generalized kinematic space, since it is a natural generalization of the kinematic space introduced there, \ie the space of ordered intervals on a time slice in $d=2$. Our focus will be to construct interesting nonlocal CFT observables on causal diamonds, similar to the perturbation $\delta\see$ in Eq.~\reef{deltaS}.\footnote{As we review in appendix \ref{dead}, conservation and tracelessness of the stress tensor allows the modular Hamiltonian to be evaluated on any time slice in spanning the corresponding causal diamond.} Our objective will be two-fold: The first is to examine if these new observables and the generalized kinematic space provide a natural forum to construct a complete description of the underlying CFT. The second is to investigate how the new perspective of the nonlocal observables interfaces with the standard holographic description given by the AdS/CFT correspondence. 

The remainder of the paper is organized as follows: Section \ref{sec.2} contains a detailed discussion of the geometry of the moduli space of causal diamonds. In section \ref{sec:linearized} we define linearized observables associated with arbitrary CFT primaries. These observables are local fields obeying two-derivative equations of motion on the space of causal diamonds and they explain and generalize various known statements about the first law of entanglement entropy, the OPE expansion of twist operators, and the holographic Ryu-Takayanagi prescription. From section \ref{kintwo} onwards, we focus on $d=2$ and the question of extending the previous framework to nonlinearly interacting fields on the space of causal diamonds. Section \ref{kintwo} is concerned with a certain universal class of states, for which the entanglement entropy satisfies a nonlinear equation with local interactions on the moduli space. Section \ref{sec.hs} generalizes this discussion to higher spin theories. In particular, we construct a framework where the entanglement and its spin-three generalization are described by two nonlinearly interacting fields on the space of causal diamonds. Some challenges for the definition of more general nonlinearly interacting fields are discussed in section \ref{loco9}. In section \ref{discuss}, we conclude with a discussion of open questions and future directions for this program of describing general CFTs in terms of nonlocal observables on the moduli space of causal diamonds, and also formulating the AdS/CFT correspondence within this framework for holographic CFTs. Appendix \ref{app:geometric} discusses various geometric details and generalizations. Some of our conventions are fixed in appendix \ref{app:conventions}. Appendix \ref{app:normalization} contains explicit computations to verify the AdS/CFT version of our generalized first law.  

\paragraph{Note:} While this work was in progress, the preprint \cite{bartek66} by Czech, Lamprou, McCandlish, Mosk and Sully appeared on the arXiv, which explores ideas very similar to the ones presented here.\\

\section{The geometry of causal diamonds in Minkowski space \label{sec.2}}

In this section, we examine the geometry of the generalized kinematic space introduced in \cite{deBoer:2015kda}. We begin by deriving the natural metric on this
moduli space of all causal diamonds in a $d$-dimensional CFT. As noted above, this $2d$-dimensional metric will turn out have multiple time directions, and in particular, has signature $(d,d)$. We will also discuss how to intuit this signature geometrically in terms of containment relations between causal diamonds.

\subsection{Metric on the space of causal diamonds}
\label{sec:geometry}

\begin{figure}[t]
\centerline{\includegraphics[width=.4\textwidth]{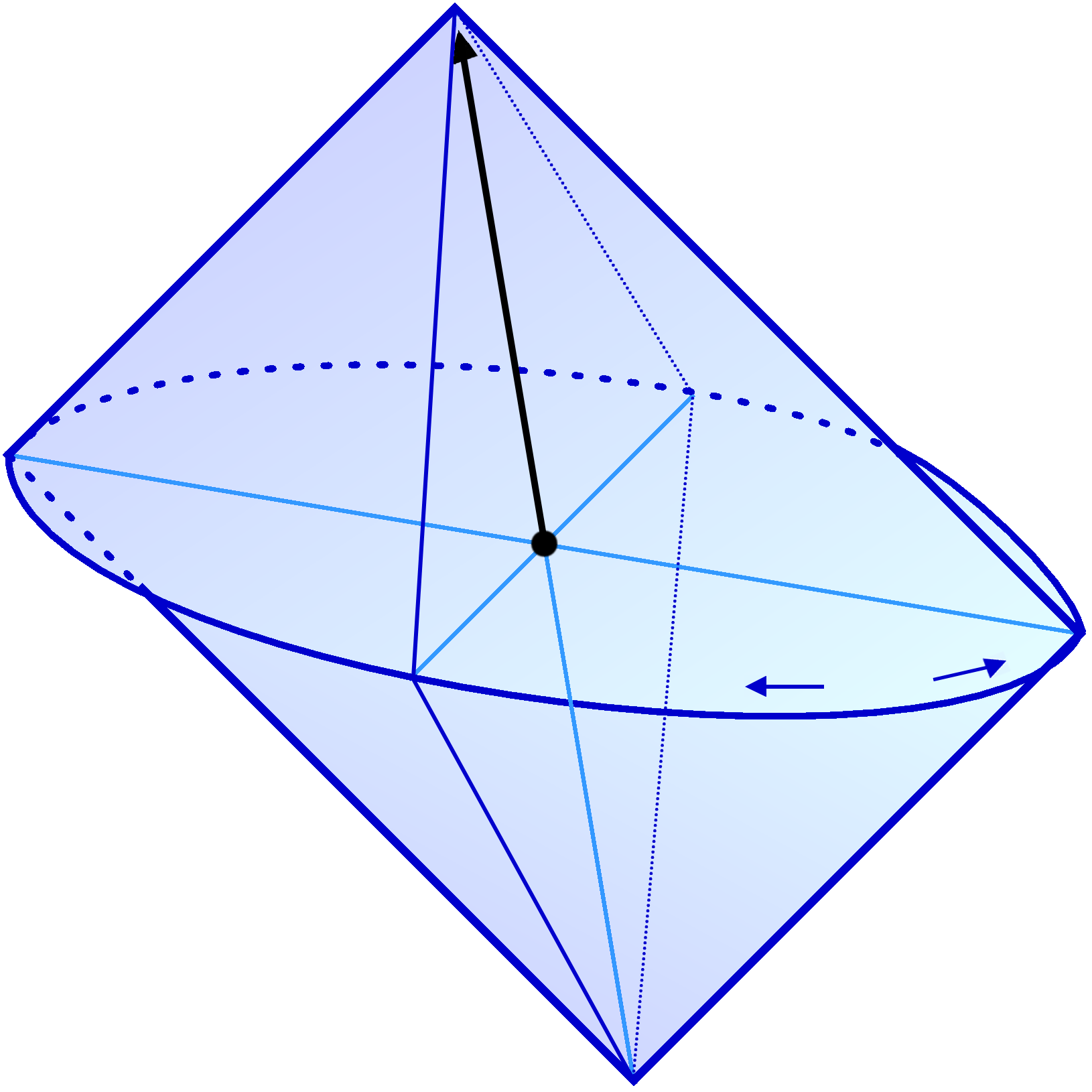}}
\setlength{\unitlength}{0.1\columnwidth}
\begin{picture}(0.3,0.4)(0,0)
\Large{
\put(4.7,4.6){\makebox(0,0){${\color{blue}y^\mu}$}}
\put(5.4,0.3){\makebox(0,0){${\color{blue}x^\mu}$}}
\put(6.25,1.97){\makebox(0,0){${\color{blue}w^\mu}$}}
}
\normalsize{
\put(5.11,2.67){\makebox(0,0){$c^\mu$}}
\put(4.95,3.67){\makebox(0,0){$\ell^\mu$}}
}
\end{picture}
\caption{A causal diamond (in $d=3$ dimensions) and our basic coordinates. Specifying the timelike separated pair of points $(x^\mu,y^\mu)$ is equivalent to specifying a spacelike $(d-2)$-sphere which consists of all points $w^\mu$ null separated from both $x^\mu$ and $y^\mu$, \ie satisfying Eq.~\reef{eq.spherecond}. The alternative parametrization in terms of $c^\mu = \frac{1}{2}(y^\mu+x^\mu)$ and $\ell^\mu = \frac{1}{2} (y^\mu-x^\mu)$ will prove convenient in section \ref{sec:CausalDi}. 
}
\label{fig.sphere}
\end{figure}
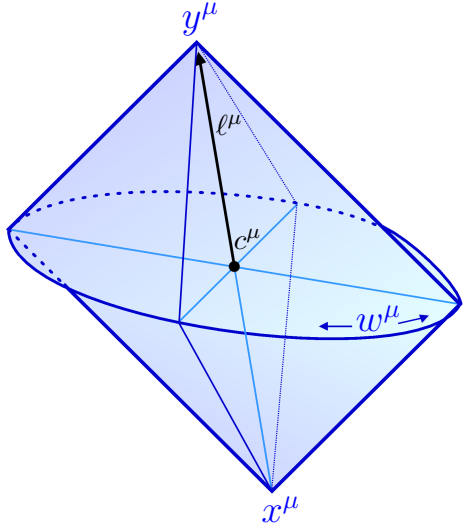

Spheres are destined to play a special role in CFTs, as the conformal group $SO(2,d)$ in $d$ dimensions maps them into each other. The past and future development of the region enclosed by a $(d-2)$-sphere form a causal diamond and hence the space of all $(d-2)$-spheres is the same as the space of all causal diamonds.\footnote{Implicitly, then we are assigning an orientation to the spheres, \ie the interior is distinguished from the exterior. One could also consider unoriented spheres, which would amount to an additional $Z_2$ identification in the coset given in Eq.~\reef{eq:Mdiamonds}. See \cite{deBoer:2015kda} for further discussion. \label{footy}} Therefore a generic $(d-2)$-sphere can be parametrized in terms of the positions of the tips of the corresponding causal diamond. That is, given these positions, $x^{\mu}$ and $y^{\mu}$, the $(d-2)$-sphere is the intersection of the past light-cone of the future tip and the future light-cone of the past tip, as shown in Figure \ref{fig.sphere}. Of course, these points are necessarily timelike separated,\footnote{Our notation here and throughout the following is that for $d$-dimensional vectors, \\[-1.2em]
$$(y-x)^2=\eta_{\mu\nu}(y-x)^\mu(y-x)^\nu\,.$$} \ie
\be
(x-y)^{2} < 0\,.\label{timsep}
\ee
The corresponding sphere comprising the intersection of the light-cones illustrated in the figure can be defined as the set of points $w^{\mu}$ which are null-separated from both $x^{\mu}$ and~$y^{\mu}$:
\be
\label{eq.spherecond}
( w - x )^{2} = 0 \quad \mathrm{and} \quad ( w - y )^{2} = 0\,.
\ee
Due to these considerations, in what follows we will interchangeably use the notions of spheres, causal diamonds, and pairs of timelike separated points.

The generalized kinematic space is the moduli space of all causal diamonds. The easiest way to construct the metric on this space is to start with an $(d+2)$-dimensional embedding space parametrized by coordinates 
\be
X^b = (X^{-}, X^{\mu}, X^{d})\,, \label{points}
\ee
with $\mu=0,\cdots,d-1$. Further this embedding space has a flat metric with signature $(2,d)$:
\be
\label{eq.metricaux}
ds_{\ssc (2,d)}^{2} = - (dX^{-})^{2} + \eta_{\mu \nu} \, dX^{\mu} dX^{\nu} +(dX^{d})^{2}\,,
\ee
where $\eta_{\mu \nu}={\rm diag}(-1,+1,\ldots,+1)$ is the usual $d$-dimensional Minkowski metric. Of course, this geometry is invariant under Lorentz group $SO(2,d)$ --- which, of course, matches the conformal group acting on a $d$-dimensional CFT.

As a warm-up, let us discuss the familiar example of anti-de Sitter space in this language. 
The $(d+1)$-dimensional anti-de Sitter (AdS) space with curvature radius $\RAdS$ corresponds to a hyperboloid defined as
\be
\label{eq.hyperboloid}
\langle X, X\rangle \, = - \RAdS^2\,,
\ee
where $\langle\,\cdot\,,\,\cdot\,\rangle$ denotes the inner product with respect to the metric \eqref{eq.metricaux}.
It can be thought of as a set of all the points in the embedding space that can be reached by acting with $SO(2,d)$ transformations on a unit timelike vector, \eg on the vector $(1, 0, \ldots, 0)$. Since any timelike vector in~\eqref{eq.metricaux} is preserved by an $SO(1,d)$ subgroup of the conformal group, $(d+1)$-dimensional anti-de Sitter space is a coset space $SO(2,d)/SO(1,d)$. The metric on this coset is induced by the embedding space metric \eqref{eq.metricaux}. For example, the Poincar{\'e} patch AdS metric
\be
\label{eq.AdSmetric}
ds^{2}_\mt{AdS} = \frac{\RAdS^2}{z^{2}} \left(dz^{2} + \eta_{\mu \nu} \, dw^{\mu} dw^{\nu}\right)
\ee
is obtained from the metric \eqref{eq.metricaux} upon using the following parametrization of the AdS hyperboloid~\eqref{eq.hyperboloid}:
\bea
X^{-} &=&  \frac{z}{2} + \frac{1}{2 z}\, (\RAdS^2 + \eta_{\mu\nu}w^\mu w^\nu)\,,\nonumber \\
X^{\mu} &=& \frac{\RAdS}{z} \,w^{\mu}\,, \labell{poinc}\\
X^{d} &=& \frac{z}{2} - \frac{1}{2z}\, (\RAdS^2-\eta_{\mu\nu}w^\mu w^\nu)\,.\nonumber
\eea
Of course, the asymptotic boundary of AdS space is reached by taking the limit $z \to 0$. In the context of the AdS/CFT correspondence, $SO(2,d)$ transformations leaving the embedding geometry \reef{eq.metricaux} invariant become the conformal transformations acting on the boundary theory. Of course, this highlights the advantage of the embedding space approach. Namely, the $SO(2,d)$ transformations act linearly on the points \reef{points} in the embedding space. 
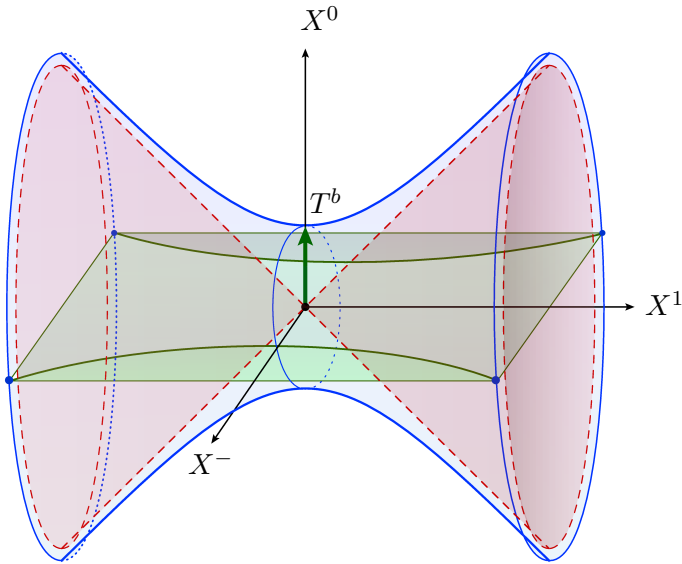
\begin{figure}[t]
\centerline{\includegraphics[width=.55\textwidth]{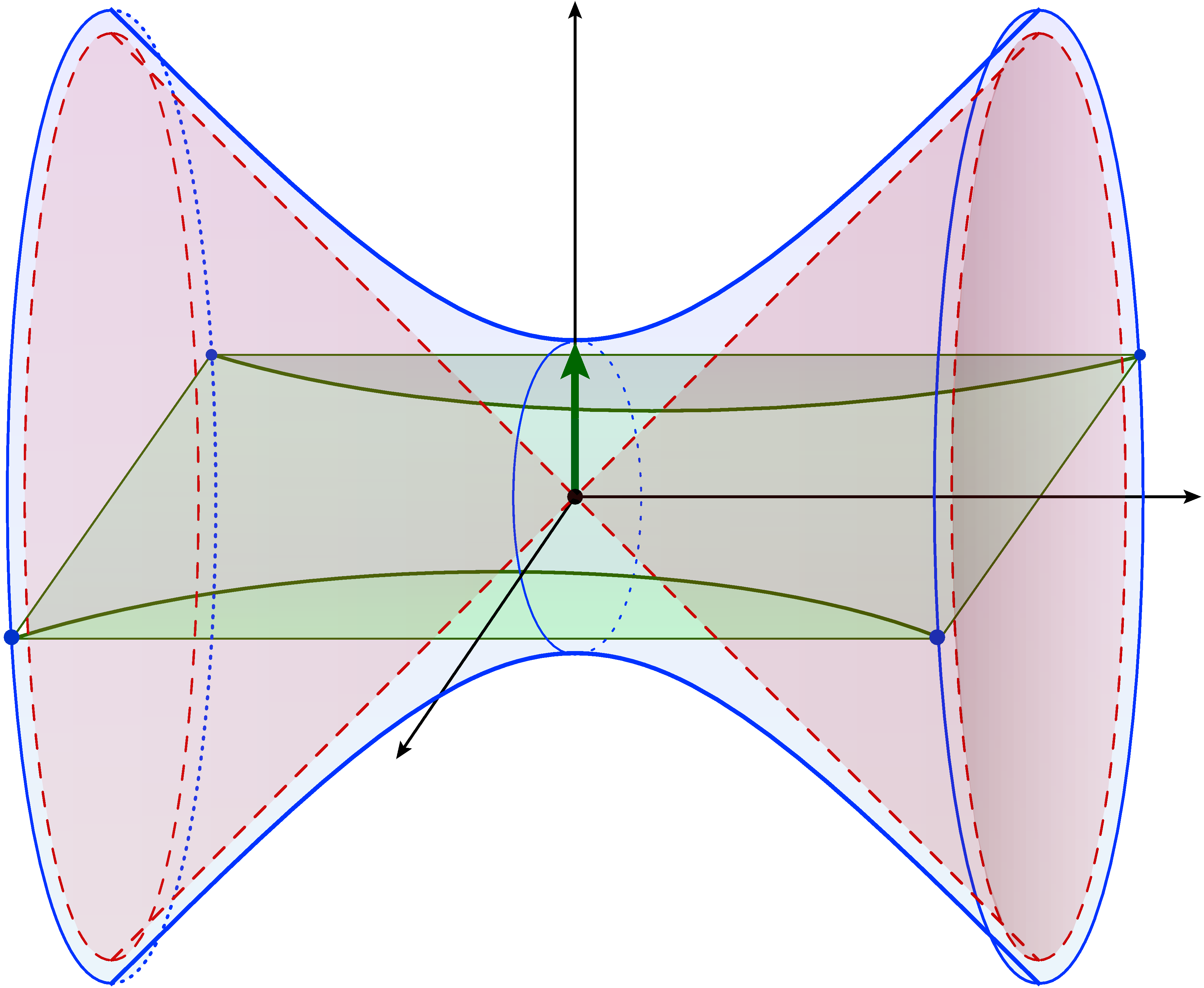}}
\setlength{\unitlength}{0.1\columnwidth}
\begin{picture}(0.3,0.4)(0,0)
\normalsize{
\put(5,5.2){\makebox(0,0){$X^{0}$}}
\put(4.05,1.3){\makebox(0,0){$X^{-}$}}
\put(8,2.7){\makebox(0,0){$X^1$}}
\put(5.05,3.6){\makebox(0,0){$T^b$}}
}
\end{picture}
\caption{Anti-de Sitter hyperboloid in flat embedding space $\mathbb{R}^{2,d}$ is indicated in blue. The timelike embedding coordinates are $X^{-}$ and $X^0$. The remaining directions (including the $d-1$ suppressed dimensions $X^{2,\,\cdots,d}$ at each point) are spacelike. The green $d$-plane is orthogonal to the timelike vector $T^b$ and to  the spacelike vector $S^b$ (the latter being hidden in the suppressed dimensions). The intersection of the $d$-plane with AdS$_{d+1}$ yields the green minimal surface. Its boundary as the hyperboloid approaches the red lightcone defines a $(d-1)$-sphere in the CFT.
}
\end{figure}

In the following, we will phrase our discussion in terms of the geometry of the CFT background being defined by the boundary of the AdS hyperboloid \reef{eq.hyperboloid} because we feel that it is an intuitive picture familiar to most readers. However, with only minor changes, the entire discussion can be phrased in terms of the embedding space formalism, \eg \cite{embed1,embed2,embed3}, which can be used to consider any CFT and makes no reference to the AdS/CFT correspondence. Hence we stress that the geometry of the generalized kinematic space that emerges below applies for general $d$-dimensional CFTs.

We now turn to the moduli space of causal diamonds in a CFT, which we construct using the language of cosets, in similar manner to that introduced above in discussing the AdS geometry \reef{eq.hyperboloid}. In order to describe a sphere in a CFT, we choose a unit timelike vector $T^b$ and an orthogonal unit spacelike vector $S^b$, both of which are anchored at the origin of the $(d+2)$-dimensional embedding space. That is, we choose two vectors satisfying
\be
\label{eq.condTSnorm}
\langle T, T \rangle  =-1\,, \qquad
\langle S, S\rangle  = 1\,, \qquad
\langle T, S\rangle  = 0\,.
\ee
The sphere is now specified by considering asymptotic points in the AdS boundary that are orthogonal to both of these unit vectors, \ie 
\be 
\label{eq.condTSorth}
\langle T ,X\rangle \big|_{z \rightarrow 0} = 0 \quad{\rm and}\quad
\langle S, X\rangle  \big|_{z \rightarrow 0} =0\,.
\ee 
To explicitly illustrate this construction of a sphere in the CFT, let us consider the coordinates \reef{poinc} yielding the Poincar\'e patch metric \reef{eq.AdSmetric}.
A convenient choice of the unit vectors is then 
\bea
T^b&=&(0,1,0,\ldots,0)\quad\longrightarrow\ \  w^0=0\,,\nonumber\\
S^b&=&(0,0,0,\ldots,1)\quad\longrightarrow\ \ \eta_{\mu\nu}\,w^\mu w^\nu=1\,.
\labell{example}
\eea
The expressions on the right denote the surfaces in the asymptotic geometry that are picked out by the orthogonality constraints \reef{eq.condTSorth}, \ie $T^b$ selects a particular time slice in the boundary metric while $S^b$ selects a timelike hyperboloid. Of course, the intersection of these two surfaces then yields the unit $(d-2)$-sphere $\delta_{ij}w^i w^j=1$ (on the time slice $w^0=0$). 
   
Now a particular choice of the unit vectors, $T^b$ and $S^b$, picks out a particular sphere in the boundary geometry. Acting with $SO(2,d)$ transformations, we can then reach all of the other spheres throughout the $d$-dimensional spacetime where the CFT lives. To determine the coset describing the space of all spheres, we must first find the symmetries preserved by any particular choice of the unit vectors. Given two unit vectors satisfying Eq.~\reef{eq.condTSnorm}, we have defined a timelike two-plane in the embedding space. Hence the $SO(2,d)$ symmetry broken to $SO(1,d-1)$ transformations acting in the $d$-dimensional hyperplane orthogonal to this $(T,S)$-plane, as well as the $SO(1,1)$ transformations acting in the two-plane.
Thus, in analogy with AdS coset construction above, the natural coset describing the moduli space of spheres, or alternatively of causal diamonds, in $d$-dimensional CFTs is  
\be \label{eq:Mdiamonds}
 \Mdiamonds \equiv \frac{SO(2,d)}{SO(1,d-1)\times SO(1,1)} \,.
\ee
Of course, this is precisely auxiliary geometry already described in \cite{deBoer:2015kda}.

The interpretation of the stabilizer group, which preserves a given sphere in the CFT, is as follows: The $SO(1,d-1)$ factor of the stabilizer group is the subgroup of $SO(2,d)$ comprising of $(d-1)(d-2)/2$ rotations and $d-1$ spatial special conformal transformations leaving a given sphere invariant. While it is obvious that the former transformations preserve spheres centred at the origin, it can also be verified that the latter do so as well. Further, let us note that these transformation also leave invariant the time slice in which the sphere is defined. The additional $SO(1,1)$ represents a combination of special conformal transformations and translations, which both involve the timelike direction and leads to a modular flow generated by the conformal Killing vector $K^{\mu}$ --- see appendix \ref{dead}. The latter was constructed precisely in such a way to preserve a given spherical surface.

We can also perform a simple cross-check at the level of counting dimensions. The moduli space of causal diamonds can parametrized by a set of $2d$ coordinates: $x^{\mu}$ and $y^{\mu}$, \ie the positions of the tips of the causal diamonds. Now, the number of generators of the isometry group $SO(2,d)$ is $(d+2)(d+1)/2$, whereas for the stabilizer group $SO(1,d-1)\times SO(1,1)$ we have $d(d-1)/2 + 1 = d(d+1)/2$ generators. The difference between the two numbers matches the dimensionality of the space of causal diamonds, \ie $2d$, as it must.

In the context of the AdS/CFT correspondence, we can remove the asymptotic limit from the orthogonality constraints \reef{eq.condTSorth}, \ie consider $\langle T ,X\rangle = 0$  and $\langle S, X\rangle  =0$. These constraints now specify not only the sphere on a constant time slice of the AdS boundary (at $z = 0$), but the entire minimal surface anchored to this sphere. With the simple example of $T^b$ and $S^b$ given in Eq.~\reef{example}, these constraints yield the unit hemisphere
$z^2+\delta_{ij}w^i w^j=1$ on the time slice $w^0=0$. Of course, using the Ryu-Takayanagi prescription \reef{define}, the area of this surface computes the entanglement entropy of the region enclosed by the (asymptotic) sphere in the vacuum of the boundary CFT.

Let us now move to the object of prime interest for us, which is the metric on the coset $\Mdiamonds$ induced by the flat geometry of the $(d+2)$-dimensional embedding space. Towards this end, we parameterize motions in this generalized kinematic space by variations of the unit vectors  $T^{b}$ and $S^{b}$. Of course, these are naturally contracted with the embedding space metric \reef{eq.metricaux} and so the most general $SO(1,d-1)$-invariant metric can be written as:
\be
\label{eq.cosetansatz}
ds^{2} = \alpha_{TT}\,\langle  d T, d T \rangle  + \alpha_{SS}\,\langle  d S, dS \rangle  + \alpha_{TS}\, \langle d T, d S\rangle \,,
\ee
where $\alpha_{TT}$, $\alpha_{SS}$ and $\alpha_{TS}$ are constant coefficients. Also requiring invariance under $SO(1,1)$ transformations, \ie under boosts in the  $(T,S)$-plane,
requires that we set $\alpha_{TS} = 0$ and $ \alpha_{SS}=-\alpha_{TT}  \equiv L^2$ --- only the relative sign of $\alpha_{SS}$ and $\alpha_{TT}$ is determined by boost invariance but we choose $\alpha_{SS}>0$ here for later convenience. This then yields
\be
\label{eq.metricpartialcoset}
ds^{2}_\coset = L^2 \left( - \langle  d T, d T \rangle + \langle  d S, dS \rangle\right)\,.
\ee
Next, we must impose the conditions \eqref{eq.condTSnorm} and \eqref{eq.condTSorth} in the above expression to fix the metric (up to an overall prefactor) in terms of geometric data in the CFT. This calculation is straightforward but somewhat tedious, and we refer the interested reader to Appendix \ref{app:metric} for the details. Our final result for the metric on the coset $\Mdiamonds$ given in \eqref{eq:Mdiamonds} becomes:
\be
\label{eq.metriccosetd}
ds^{2}_\coset
= h_{\mu \nu}\, dx^{\mu} dy^{\nu} =  \frac{4 L^2}{(x-y)^{2}} \left( -\eta_{\mu \nu} + \frac{2(x_{\mu}-y_{\mu})(x_{\nu}-y_{\nu})}{(x-y)^{2}}\right) dx^{\mu} dy^{\nu}\,,
\ee
where $x^{\mu}$ and $y^{\mu}$ denote the past and future tips of the corresponding causal diamond, as illustrated in Figure \ref{fig.sphere}.
This metric is the main result of the present section and the starting point for our investigations of the generalized kinematic space in the subsequent sections. 

Some comments are now in order: First, it is straightforward to verify that this metric~\eqref{eq.metriccosetd} is invariant under the full conformal group. Second, the pairs $(x^{\mu},y^{\mu})$ appear as pairs of null coordinates in the metric~\eqref{eq.metriccosetd}. As a result, this metric on the coset~\eqref{eq.metriccosetd} has the highly unusual signature $(d,d)$. Third, it is amusing
to notice that while AdS geometrizes scale transformations, the coset geometrizes yet another $d-1$ additional conformal
transformations. \\

\noindent Let us now discuss two special cases for which the general result \eqref{eq.metriccosetd} simplifies:

\paragraph{Example 1: Fixed time slice.}
The first example concerns the moduli space of spheres lying on a given constant time slice, which we can always take to be $t=0$. That is, we choose $y^{0} = -x^{0} = R$ and $\vec{x} = \vec{y}$ and then we are considering spheres on the $t=0$ slice with radius $R$ and with $\vec{x}$ giving the spatial position of their centres. Constraining the coordinates $x^{\mu}$ and $y^{\mu}$ in this way, the coset metric \eqref{eq.metriccosetd} reduces to 
\be
\label{eq.metricdSd}
ds_\coset^{2}\Big|_{\vec{x} = \vec{y};\,y^{0} = -x^{0} = R}= \frac{L^2}{R^{2}} \left(- dR^2 + d\vec{x}^{2} \right)
\equiv ds_{\mathrm{dS}_{d}}^{2} \,.
\ee
That is, we have recovered precisely the $d$-dimensional de Sitter space appearing in Eq.~\reef{eq.deSitter} as a submanifold of the full coset $\Mdiamonds$.

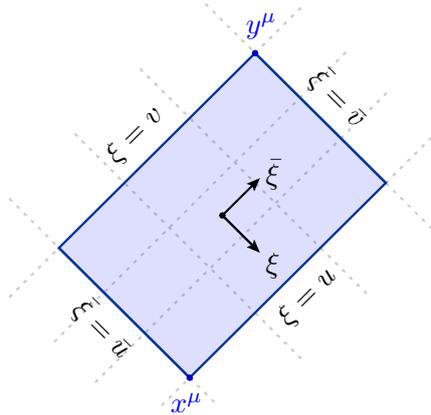
\begin{figure}[t]
\centerline{\includegraphics[width=.37\textwidth]{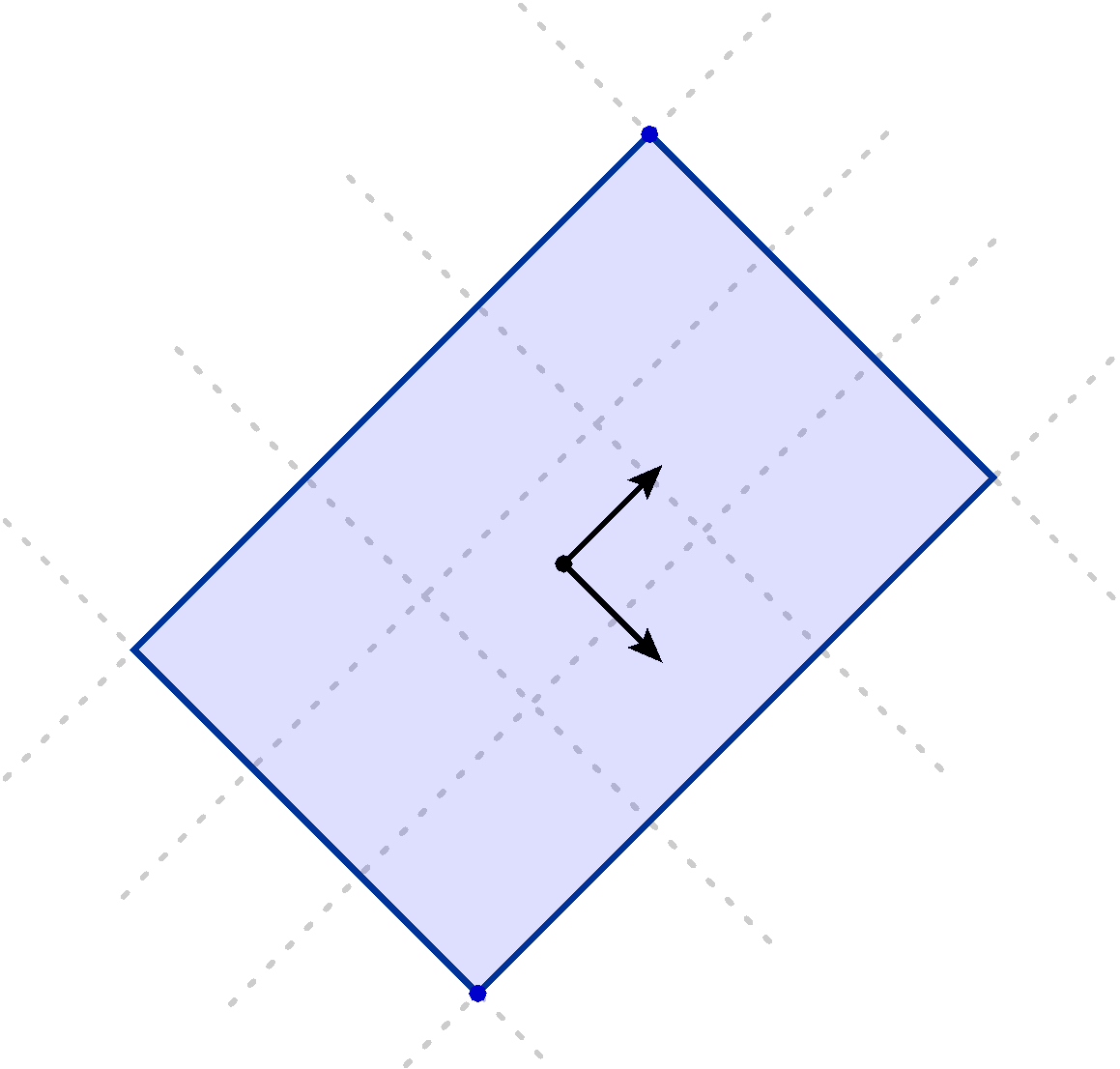}}
\setlength{\unitlength}{0.1\columnwidth}
\begin{picture}(0.3,0.4)(0,0)
\normalsize{
\put(4.7,.45){\makebox(0,0){${\color{blue}x^\mu}$}}
\put(5.35,3.75){\makebox(0,0){${\color{blue}y^\mu}$}}
\put(5.45,2.45){\makebox(0,0){$\bar\xi$}}
\put(5.45,1.65){\makebox(0,0){$\xi$}}
\put(5.75,1.38){\rotatebox{45}{\makebox(0,0){$\xi=u$}}}
\put(4.25,2.8){\rotatebox{45}{\makebox(0,0){$\xi=v$}}}
\put(6.02,3.12){\rotatebox{-45}{\makebox(0,0){$\bar\xi=\bar v$}}}
\put(3.97,1.1){\rotatebox{-45}{\makebox(0,0){$\bar\xi=\bar u$}}}
}
\end{picture}
\caption{Lightcone coordinates for two-dimensional causal diamonds. The coordinates $(\xi,\bar \xi)$ will provide a useful parametrization of the given diamond in section \ref{sec.genO}. Changing the endpoints corresponds to moving in the moduli space of causal diamonds parametrized by $(u,\bar u,v,\bar v)$; thereby $u$ is constant if $x^\mu$ moves along the line $\xi=u$, and so forth.
}
\label{fig:2dcoords}
\end{figure}

\paragraph{Example 2: CFT in two dimensions.}
A second special case of interest is the restriction to $d=2$. The metric on the coset in two dimensions has a structure of a direct product of two copies of two-dimensional de Sitter space. One can see this explicitly by introducing right- and left-moving light-cone coordinates, \eg we replace the Minkowski coordinates $(\xi^0,\xi^1)$ with
\be
\label{eq.LCdef}
\xi = \xi^{1} - \xi^{0} \quad \mathrm{and} \quad \bar{\xi} = \xi^{1} + \xi^{0}\,.
\ee
Then we may specify the two-dimensional causal diamonds, defined by $(x^\mu,y^\mu)$ above, in terms of the positions of their four null boundaries 
--- see Figure \ref{fig:2dcoords}, 
\be 
 (u,\bar u) \equiv (x^1-x^0,x^1+x^0) \,,\qquad (v,\bar v) \equiv (y^1-y^0,y^1+y^0) \,. \label{nulle}
\ee 
Finally re-expressing the coset metric \eqref{eq.metriccosetd} in terms of these coordinates yields
\be
\label{eq.dS2dS2}
ds_{\coset}^2\Big|_{d=2}= 2L^2 \left\{ \frac{du\, dv}{(u-v)^{2}} + \frac{d\bar{u}\, d\bar{v}}{(\bar{u}-\bar{v})^{2}} \right\}
\equiv \frac{1}{2} \left\{ ds_{\mathrm{dS}_{2}}^{2} + ds_{\overline{\mathrm{dS}}_{2}}^{2} \right\}\,.
\ee
Notice that the first copy of de Sitter metric is only a function of the right-moving coordinates, whereas the second copy depends only on the left-moving coordinates. We chose the normalization on the right hand side of Eq.~\eqref{eq.dS2dS2} in such a way that $L$ is the curvature scale in each de Sitter component and upon restricting to a timeslice (\ie $\bar u=v \equiv x-R$ and $\bar v = u\equiv x+R$),  Eq.~\eqref{eq.metricdSd} obviously emerges. This way we can heuristically think of each of the two copies of dS$_2$ in \eqref{eq.dS2dS2} as a copy of the geometry in Eq.~\eqref{eq.metricdSd}.

Of course, the product structure found in the moduli space metric here has its origins in the fact that for two dimensions, the conformal group itself decomposes into a direct product, \ie $SO(2,2) \simeq SO(2,1)\times SO(2,1)$, where the two factors act separately on the right- and left-moving coordinates. Hence the moduli space \reef{eq:Mdiamonds} of intervals in $d=2$ CFTs becomes 
\be
\label{twodM}
\mathcal{M}_\lozenge^{(2)}=\frac{SO(2,1)}{SO(1,1)}\times \frac{SO(2,1)}{SO(1,1)}\,,
\ee
where we recognize that each of factors corresponds to a two-dimensional de Sitter space.

\subsection{The causal structure on the space of causal diamonds}
\label{sec:CausalDi}

Given the metric \eqref{eq.metriccosetd} on the moduli space of causal diamonds, we are in the position to study the causal structure of this space. The essential feature of this causal structure comes from the fact that the space possesses $d$ spacelike and $d$ timelike directions. 

We start by writing the metric \eqref{eq.metriccosetd} in terms of the coordinates
\be\label{eq:coordsDdiamond}
 c^\mu \equiv \frac{y^\mu+x^\mu}{2} \quad \text{and} \quad \ell^\mu \equiv \frac{y^\mu-x^\mu}{2} \,.
\ee
Here, $c^\mu$ denotes the position of the centre of the causal diamond or, equivalently, the centre of the corresponding sphere. Similarly, $\ell^\mu$ denotes the vector from the centre to the future tip of the causal diamond ---
see Figure \ref{fig.sphere}. The metric \eqref{eq.metriccosetd} then becomes 
\be 
\label{eq:dsspheres2}
 ds_\coset^2 = -\frac{L^2}{\ell^2} \left( \eta_{\mu\nu} - \frac{2}{\ell^2}\, \ell_\mu \ell_\nu \right) \left(dc^\mu\, dc^\nu - d\ell^\mu \, d\ell^\nu \right) \,.
\ee
First, we note that $\ell^2<0$ from Eq.~\reef{timsep}, \ie the tips of the causal diamond are timelike separated. Further, we observe that the tensor $\left( \eta_{\mu\nu} - \frac{2}{\ell^2} \, \ell_\mu \ell_\nu \right)$ is positive definite again because $\ell^\mu$ is a timelike vector. This is easily verified by picking a frame where, say, $\ell^\mu \propto \delta^\mu_0$. In such a frame, the metric \reef{eq:dsspheres2} reduces to
\be
 ds_\coset^2\Big|_{\ell^\mu = \delta^\mu_0} = -\frac{L^2}{\ell^2} \, \delta_{\mu\nu} \left( dc^\mu \, dc^\nu - d\ell^\mu\, d\ell^\nu \right) \,.
 \label{goshwoggle}
\ee

\begin{figure}[t]
\centerline{\includegraphics[width=\textwidth]{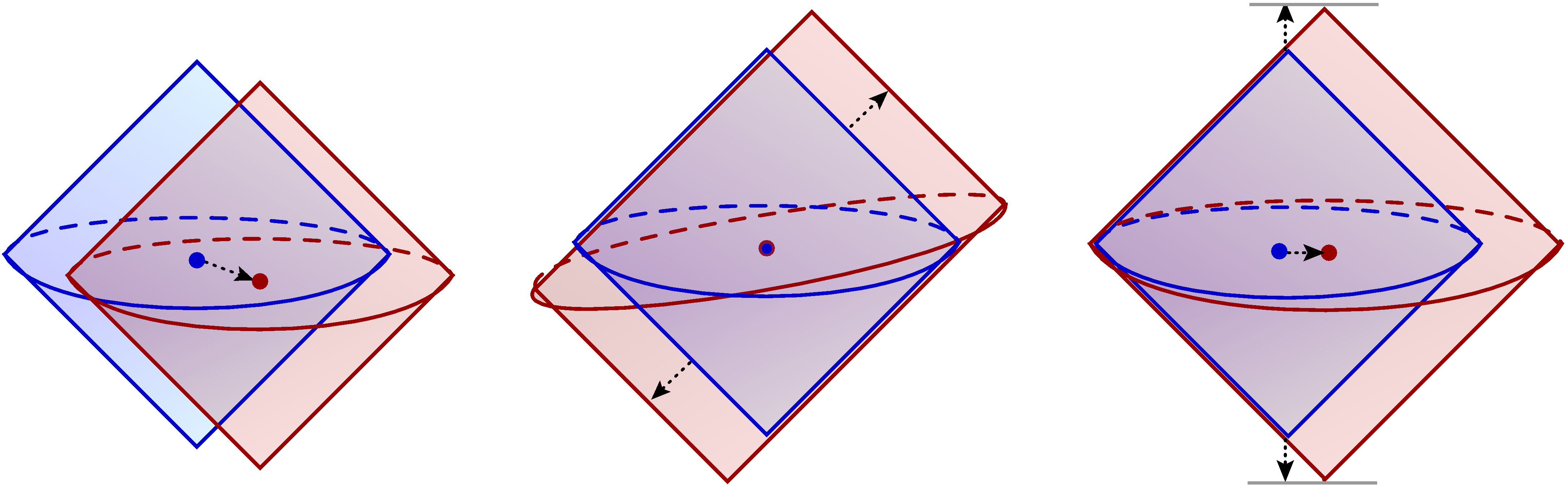}}
\setlength{\unitlength}{0.1\columnwidth}
\begin{picture}(0.3,0.4)(0,0)
\normalsize{
\put(1.3,.2){\makebox(0,0){(a) $ds_\coset^2>0$}}
\put(5,.2){\makebox(0,0){(b) $ds_\coset^2<0$}}
\put(8.4,.2){\makebox(0,0){(c) $ds_\coset^2=0$}}
}
\end{picture}
\caption{Three basic types of infinitesimal moves in the moduli space of causal diamonds (in $d=3$ dimensions): (a) spacelike moves correspond to translations of the diamond, (b) timelike moves correspond to deformations of the diamond which leaves its centre fixed, (c) null moves correspond to a combination of the previous two by the `same' amounts.
}
\label{fig:stretch1}
\end{figure}
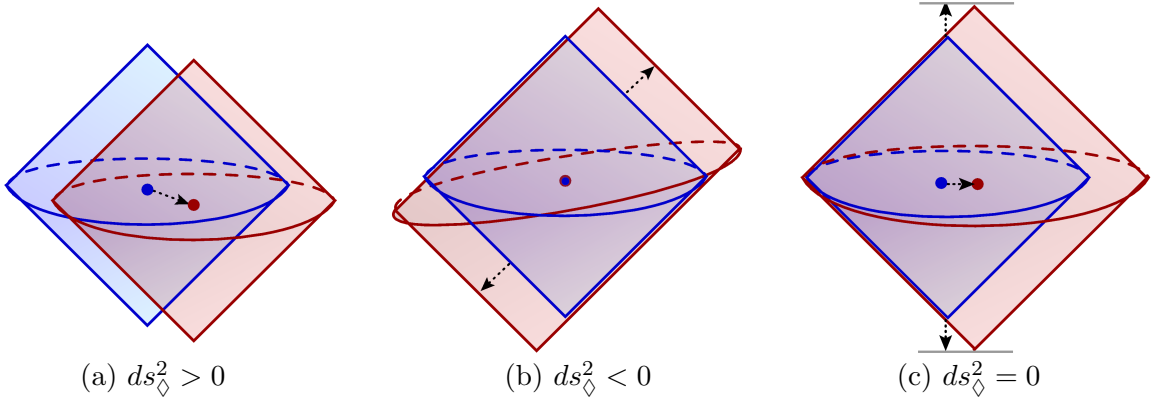

Therefore, the sign of $ds^2_\coset$ is determined solely by the last factor in Eq.~\eqref{eq:dsspheres2} containing the differentials. In particular, we can now see that $c^\mu$ are the $d$ spacelike directions in the space of causal diamonds, while $\ell^\mu$ are the $d$ timelike directions.  To make this precise, consider two infinitesimally close causal diamonds specified by their coordinates $\lozenge_1=(c^\mu,\ell^\mu)$ and $\lozenge_2=(c^\mu+dc^\mu,\ell^\mu+d\ell^\mu)$, we say that their separation is spacelike, timelike or null if $ds_\coset^2(c^\mu,\ell^\mu)$ is positive, negative or zero, respectively. 
From this, it is now easy to intuit the timelike, spacelike and null directions in the moduli space of causal diamonds as follows: 
\begin{itemize}
 \item[(a)] Moving the centre $c^\mu$ of a causal diamond by an infinitesimal amount $dc^\mu$ in any of the $d$ directions of the background Minkowski spacetime of the CFT corresponds to moving in a {\it spacelike} direction in the coset space. Geometrically, this corresponds to translating the diamond without deforming it. 
 \item[(b)] Moving any of the `relative' coordinates $\ell^\mu$ by some $d\ell^\mu$ corresponds to a {\it timelike} displacement in the coset space. In the diamond picture, this corresponds to stretching the diamond in one of $d$ independent ways while holding the centre of the diamond fixed. 
 \item[(c)] {\it Null} movements correspond heuristically to deforming the diamond by the `same' amount as it is translated in spacetime, as quantified by the condition $ds^2_\coset=0$. 
\end{itemize} 
These cases are illustrated in Figure \ref{fig:stretch1} for infinitesimal displacements. It is noteworthy that moving the centre of causal diamond in the time direction, \ie with $dc^0$, produces a spacelike displacement in the kinematic space. We return to discuss this point in section \ref{discuss}. \\

Let us now give a slightly different perspective on the measure of distances on this moduli space. Consider two causal diamonds, specified by the coordinates of their tips, $\lozenge_1=(x_1^\mu,y_1^\mu)$ and $\lozenge_2=(x_2^\mu,y_2^\mu)$. The conformal symmetry ensures that there exists a natural conformally invariant measure of distance, namely, the cross ratio 
\be\label{eq:crossratio}
\ratio(x_1,y_1;x_2,y_2) \equiv \frac{(y_1-x_2)^2\,(y_2-x_1)^2}{(y_1-x_1)^2\,(y_2-x_2)^2}
\,.
\ee
As we will show the cross ratio paves the way to understanding the global causal structure of the moduli space of diamonds, however, first we relate this expression to the previous discussion. Hence we translate it to the
`centre of mass' coordinates and consider the two causal diamonds with $\lozenge_1=(c^\mu,\ell^\mu)$ and $\lozenge_2=(c^\mu+\Delta c^\mu,\ell^\mu + \Delta \ell^\mu)$. Then the invariant cross ratio reads
\bea
 \ratio(\lozenge_1,\lozenge_2) &=& \frac{(2\ell+\Delta\ell +\Delta c)^2 (2\ell+\Delta\ell -\Delta c)^2}{16\,\ell^2 \, (\ell+\Delta \ell)^2}
 \labell{eq:DistFinal}\\
&=&1+ \frac{1}{2\ell^2} \left( \eta_{\mu\nu} - \frac{2}{\ell^2}\, \ell_\mu \ell_\nu \right) \left(\Delta c^\mu\, \Delta c^\nu - \Delta\ell^\mu \, \Delta\ell^\nu \right)+
\cdots \,. \nonumber
\eea
In the second line, we are expanding the cross ratio for infinitesimal displacements and the ellipsis indicates terms of cubic order in $\Delta c^\mu$ and $\Delta\ell^\mu$. Comparing to Eq.~\reef{eq:dsspheres2}, we see that causal diamonds that are very nearby
\be
\ratio(\lozenge_1,\lozenge_2) \simeq 1 - \frac{1}{2L^2}\,ds^2_\coset + \cdots\,.
\label{nearby}
\ee
That is, for infinitesimal displacements, the cross ratio encodes the invariant line element \reef{eq:dsspheres2} of the generalized kinematic space. Further, we  observe that Eq.~\reef{nearby} shows that timelike, spacelike and null displacements in this moduli space correspond, respectively, to $\ratio>1$, $\ratio<1$ and $\ratio=1$. 

Two other observations about the cross ratio in Eq.~\reef{eq:DistFinal}: We note that  the centre of mass coordinates $c^\mu$ are Killing coordinates of the metric  \reef{eq:dsspheres2}, \ie the metric is independent of these coordinates. However, this feature also extends to finite separations, as is apparent from the first line of Eq.~\reef{eq:DistFinal}. That is, the position $c^\mu$ of the reference diamond $\lozenge_1$ is irrelevant for the distance to $\lozenge_2$ and only the relative $\Delta c^\mu$ appears in this expression.
Similarly, $d c^\mu = d \ell^\mu$ yields a null displacement in Eq.~\reef{eq:dsspheres2} but two diamonds separated by finite displacements with $\Delta c^\mu = \Delta \ell^\mu$ are also null separated, \ie it is straightforward to show that the first line of Eq.~\reef{eq:DistFinal} yields $\ratio =1$ in this situation.  Geometrically, $\Delta c^\mu = \Delta \ell^\mu$ corresponds to two diamonds whose past tips coincide (and similarly, $\Delta c^\mu = -\Delta \ell^\mu$ corresponds to diamonds whose future tips coincide).

We can go further and define an invariant `geodesic distance' function between two diamonds $\lozenge_1=(x_1^\mu,y_1^\mu)$ and $\lozenge_2=(x_2^\mu,y_2^\mu)$ in terms of the cross ratio as
\be\label{eq:GeodDist}
 \mathbf{d}(\lozenge_1,\lozenge_2) = 
 \left\{
 \begin{aligned}
  &\quad\ L\,\cos^{-1}\! \left(2\sqrt{\ratio(x_1,y_1;x_2,y_2)}-1\right)  \qquad &\text{if } 0\leq \ratio \leq 1\,, \\
  &-L\,\cosh^{-1}\! \left(2\sqrt{\ratio(x_1,y_1;x_2,y_2)}-1\right)  \qquad &\text{if } \ratio > 1\,. \quad\ \ \, 
 \end{aligned} 
 \right.
\ee 
As we will show in examples, this distance function computes geodesic distance between finitely separated diamonds, within the range of validity specified above. Note then that the corresponding cross ratio is greater than, less than or equal to $1$ if two diamonds may be connected by a timelike, spacelike or null geodesic. However, the converse need not be true, \ie, even if the cross ratio is positive, there may not be a geodesic connecting the corresponding diamonds --- see further discussion below. Further, note that as $\ratio \rightarrow \infty$, the corresponding causal diamonds become infinitely timelike separated. However, there is a maximal spacelike separation that can achieved by following geodesics through the coset, \ie at $\ratio=0$, we find $\mathbf{d}_\mt{max}=\pi L$.

Equipped with the distance function \reef{eq:GeodDist}, let us briefly comment on the structure of the cross ratio  \eqref{eq:crossratio}. We have the following interesting cases in general:
\begin{itemize}
\item $(x_1-y_1)^2\rightarrow 0$ or $(x_2-y_2)^2 \rightarrow 0$: if one of the diamonds' volumes shrinks to zero,\footnote{The tips may not coincide in this limit rather they only need to be null separated. \label{call8}} the cross ratio and the distance function both diverge, in particular, $\mathbf{d}(\lozenge_1,\lozenge_2)\rightarrow -\infty$. This is just the statement that zero-volume diamonds lie at the timelike infinity of the coset space $\Mdiamonds$.
\item $y_1\rightarrow y_2$ or $x_1 \rightarrow x_2$: if either the past or future tips of two diamonds coincide, the cross ratio becomes one and the invariant distance $\mathbf{d}(\lozenge_1,\lozenge_2)$ vanishes, \ie the diamonds become null separated. 
\item $(x_1-y_1)^2\rightarrow \infty$ or $(x_2-y_2)^2 \rightarrow \infty$: if either of the diamonds' volumes grows to infinity, the cross ratio vanishes and the distance function reaches its maximal value, $\mathbf{d}_\mt{max}=\pi L$. 
\item $(y_1-x_2)^2 \rightarrow 0$ or $(y_2-x_1)^2\rightarrow 0$: if the future (past) tip of one causal diamond approaches the lightcone of the past (future) tip of the other diamond (as illustrated in Figure \ref{fig:singularities}), the cross ratio vanishes and the corresponding separation again reaches the maximal value $\mathbf{d}_\mt{max}=\pi L$. 
\end{itemize}
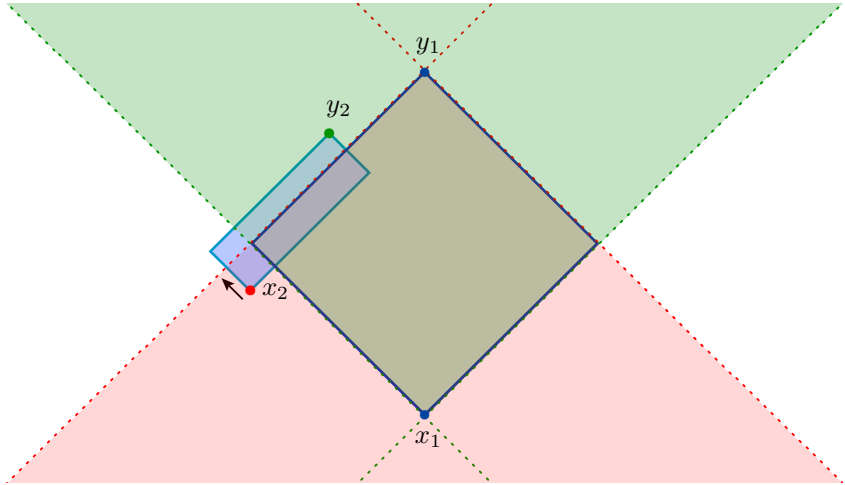
\begin{figure}[t]
\centerline{\includegraphics[width=.73\textwidth]{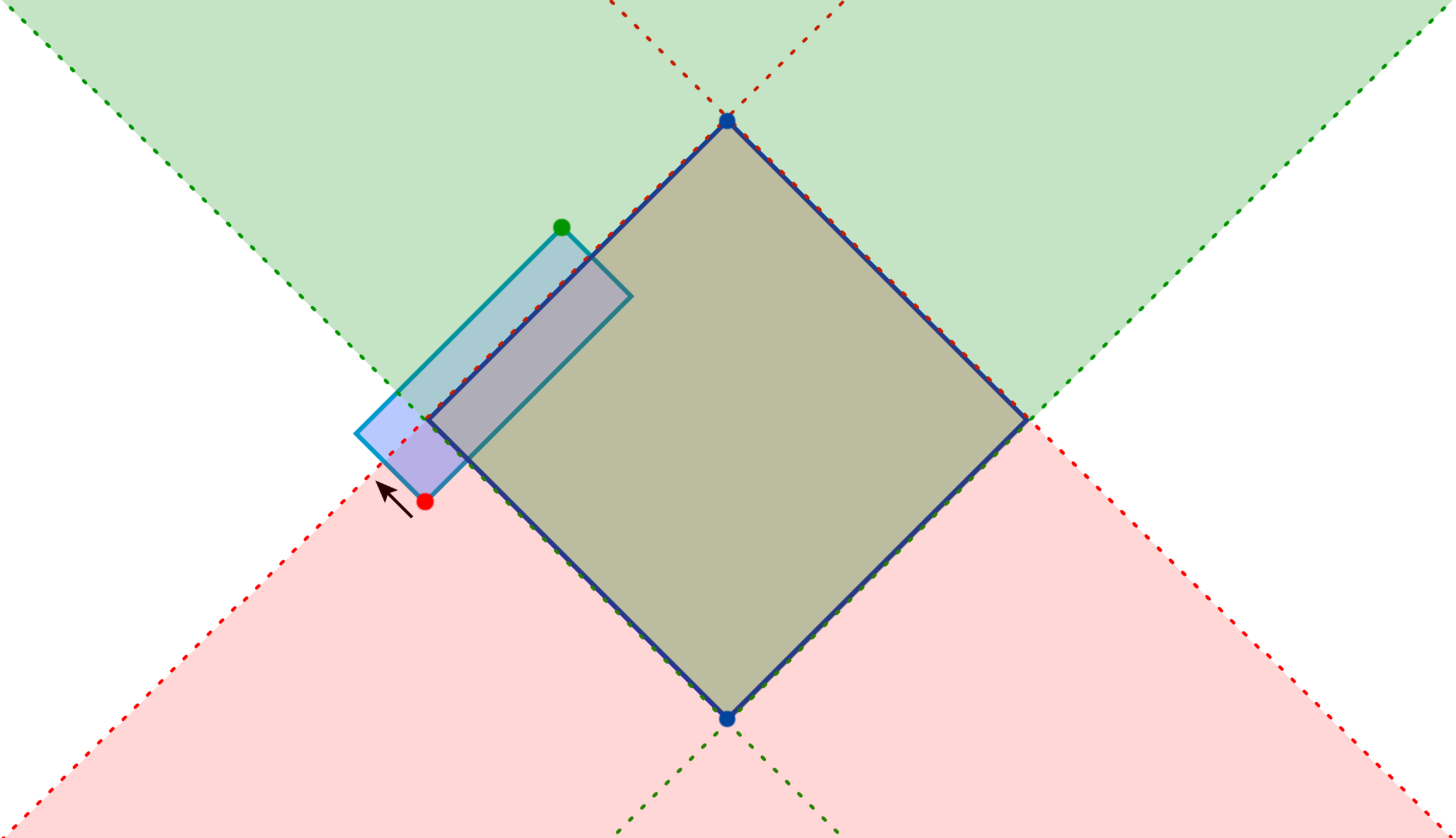}}
\setlength{\unitlength}{0.1\columnwidth}
\begin{picture}(0.3,0.4)(0,0)
\small{
\put(5.03,.8){\makebox(0,0){$x_1$}}
\put(5.03,4.25){\makebox(0,0){$y_1$}}
\put(3.7,2.1){\makebox(0,0){$x_2$}}
\put(4.25,3.7){\makebox(0,0){$y_2$}}
}
\end{picture}
\caption{Illustration of lightcone singularities in the moduli space of causal diamonds. We compare the big blue reference diamond $\lozenge_1$ with the small blue diamond $\lozenge_2$. If the red (green) tip of $\lozenge_2$ leaves the red (green) shaded lightcone region, the geodesic distance $\mathbf{d}(\lozenge_1,\lozenge_2)$ becomes infinite, i.e., the diamonds are no longer geodesically connected. An example of this happening would be by moving the tip $x_2$ along the arrow towards the lightcone of $y_1$. 
}
\label{fig:singularities}
\end{figure}

Let us comment further on the domain of validity of our geodesic distance function.
As defined in Eq.~(\ref{eq:GeodDist}), this function is well-defined for $\ratio \geq 0$. However, as commented above, merely having $\ratio \geq 0$ does not ensure that the corresponding causal diamonds are connected by a geodesic. Further, certain pairs of causal diamonds will also yield $\ratio<0$. Examining Eq.~\reef{eq:crossratio}, we see that both factors in the denominator are negative by construction, \ie the tips of each casual diamond must be timelike separated, and hence the sign of $\ratio$ is determined by the numerator. 

Let us consider beginning with two nearby diamonds, $\lozenge_1$ and $\lozenge_2$. Both $(y_1-x_2)^2<0$ and $(y_2-x_1)^2<0$ so that the cross ration is positive. As indicated by Eq.~\reef{eq:DistFinal}, we will have $\ratio \approx 1$ in this situation. If we deform the second diamond away from $\lozenge_1$ in a spacelike direction, (not necessarily along a geodesic), the cross ratio will decrease.
As described above,  if the future (past) tip of $\lozenge_2$ reaches the lightcone of the past (future) tip of $\lozenge_1$, the cross ratio and the corresponding distance vanishes --- see Figure \ref{fig:singularities}. If we continue deforming in the same direction, one of the factors in the numerator is now positive and $\ratio$ becomes negative, \eg pushing the future tip of $\lozenge_2$ out of causal contact with the past  tip of $\lozenge_1$ gives $(y_2-x_1)^2>0$. Now in this range of $\ratio$, the distance function \reef{eq:GeodDist} is not defined and there is no geodesic connecting the corresponding causal diamonds. Hence submanifold of configurations where $\ratio$ (first) vanishes defines the `maximum' range which the geodesics originating at $\lozenge_1$ can reach  in the kinematic space.

Note that generically if $\lozenge_2$ lies on this boundary where $\ratio=0$, then
the two diamonds will not be connected by a geodesic. However there are exceptional configurations with a vanishing cross ratio, which are connected. These are `antipodal' points in the kinematic space, which are in fact connected by multiple geodesics --- see further discussion below. As noted above, this configuration yields to the maximal spacelike separation that  can be reached along a geodesic, \ie $\mathbf{d}_\mt{max}=\pi L$. 

One can further deform $\lozenge_1$ and $\lozenge_2$ so that the two diamonds become completely out of causal contact with each other, \ie both $(y_1-x_2)^2>0$ and $(y_2-x_1)^2>0$. In this case, the cross ratio passes through zero again to reach positive values. However, even though Eq.~\reef{eq:GeodDist} is well defined for these diamonds, there will still be no geodesic connecting them.

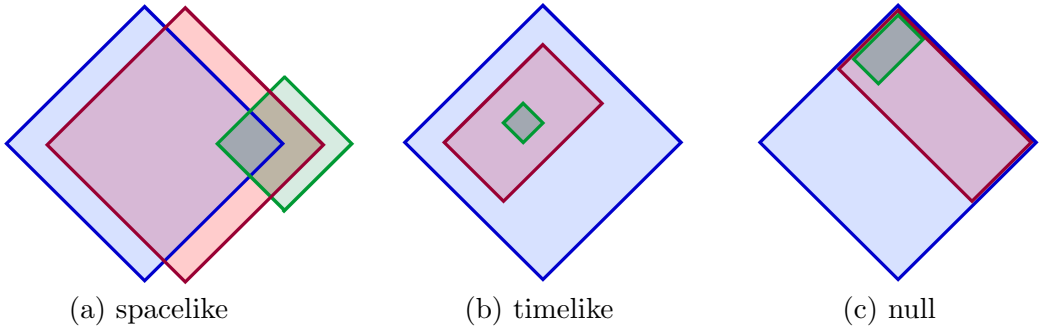
\begin{figure}[t]
\centerline{\includegraphics[width=.9\textwidth]{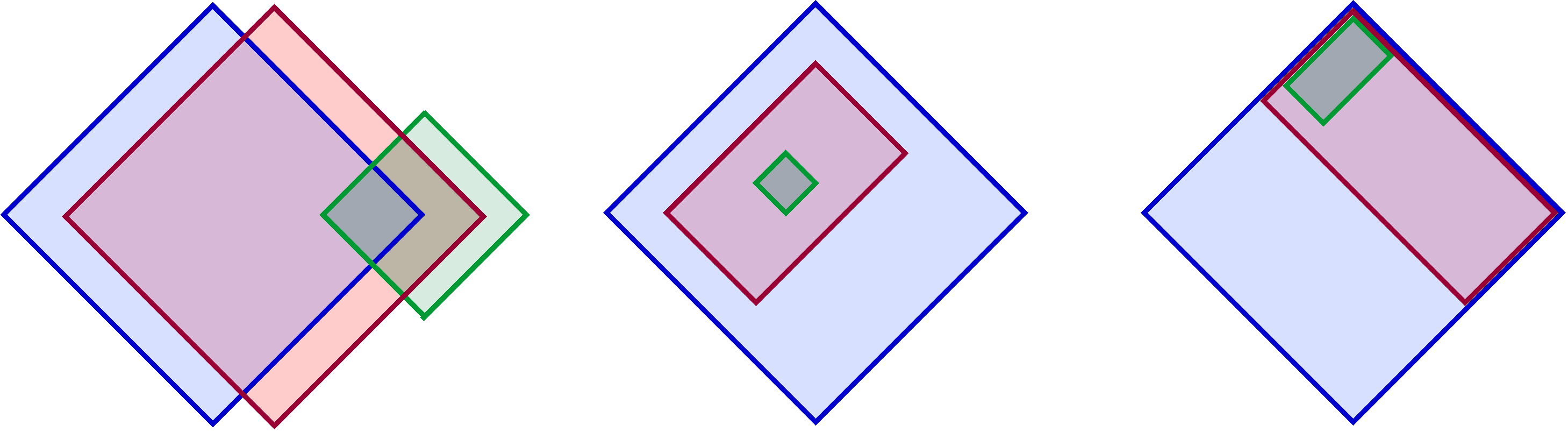}}
\setlength{\unitlength}{0.1\columnwidth}
\begin{picture}(0.3,0.4)(0,0)
\normalsize{
\put(1.75,.2){\makebox(0,0){(a) spacelike}}
\put(5.15,.2){\makebox(0,0){(b) timelike}}
\put(8.2,.2){\makebox(0,0){(c) null}}
}
\end{picture}
\caption{Causal structure on the moduli space of causal diamonds in two-dimensional case. In (a) all three diamonds are spacelike separated from each other. Case (b) shows three timelike separated causal diamonds. Finally, all diamonds in (c) are null separated. 
}
\label{fig.shifts3}
\end{figure}

Figure \ref{fig.shifts3} shows some more examples of the causal structure on the moduli space of (two-dimensional) causal diamonds. In particular, note the cases (a) and (b) of that figure, which illustrate two statements that are generally true (in any number of dimensions):
\begin{enumerate}
 \item[$(i)$] If two causal diamonds are contained within one another, then they are timelike separated.
 \item[$(ii)$] If two causal diamonds touch in at least one corner, then they are null separated.
\end{enumerate}

\noindent Let us now return to the two examples which we identified as being of particular interest in section \ref{sec:geometry}:

\paragraph{Example 1: Fixed time slice.} If we compare diamonds $\lozenge_{1,2}$ on a given time slice, we know from our previous discussion that we are restricting to a submanifold with the geometry of $d$-dimensional de Sitter space. Taking the time slice to be $t=0$, we have $c_1^0 = c_2^0 = 0$ and $\ell^i_1 = \ell^i_2 = 0$. Using the same coordinates as before, $x^i \equiv c^i$ and $R \equiv \ell^0 >0$, the cross ratio simplifies as
\be \label{beast}
 \ratio_{_{\text{dS}_d}}(R_1,\vec{x}_1;R_2,\vec{x}_2) = \frac{ \left[ - (R_1+R_2)^2 + (\vec{x}_1 - \vec{x}_2)^2 \right]^2}{16\, R_1^2 \, R_2^2}\,\geq0 \,.
\ee
We observe the following causal relations between spatial spheres lying on a common time slice:\footnote{We assume here that $(\vec{x}_1-\vec{x}_2)^2 \leq (R_1+R_2)^2$, for otherwise the spheres would not be geodesically connected --- see further discussion in the following.}
\begin{itemize}
 \item $\ratio_{_{\text{dS}_d}} \geq 1$ if $(\vec{x}_1-\vec{x}_2)^2 \leq (R_1-R_2)^2$, \ie one sphere is contained within the other.
 \item $\ratio_{_{\text{dS}_d}}\leq 1$ if $(\vec{x}_1-\vec{x}_2)^2 \geq (R_1-R_2)^2$, \ie the spheres overlap but neither is fully contained within the other.
 \item $\ratio_{_{\text{dS}_d}}=1$ if and only if $(\vec{x}_1-\vec{x}_2)^2 = (R_1-R_2)^2$, \ie the spheres tangentially touch in at least one point. 
 \item Note that $\ratio_{_{\text{dS}_d}} \rightarrow 0$ as $(\vec{x}_1-\vec{x}_2)^2 \rightarrow (R_1+R_2)^2$, which corresponds to the point where the two spheres become disjoint.  
\end{itemize}

It is straightforward to show that this de Sitter geometry is a `completely geodesic' submanifold of the full kinematic space \reef{eq:Mdiamonds}. That is, all of the geodesics within dS$_d$ are also geodesics of $\Mdiamonds$. Hence upon substituting Eq.~\reef{beast}, it is sensible to compare Eq.~\reef{eq:GeodDist} to the geodesic distances in de Sitter space with the metric \reef{eq.metricdSd} and
one can easily verify that $\mathbf{d}(\lozenge_1,\lozenge_2)$ reduces to the expected geodesic distances. 

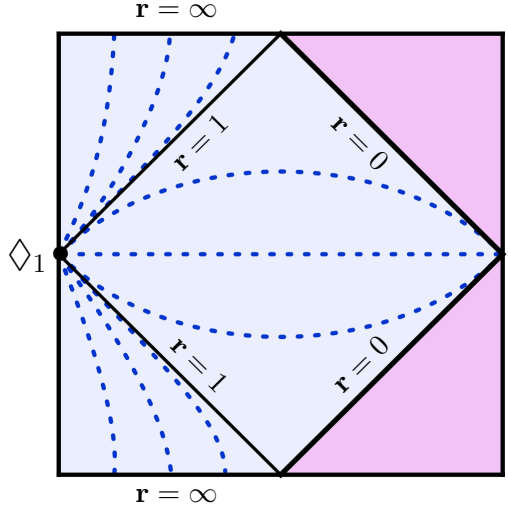
\begin{figure}[t]
\centerline{\includegraphics[width=.4\textwidth]{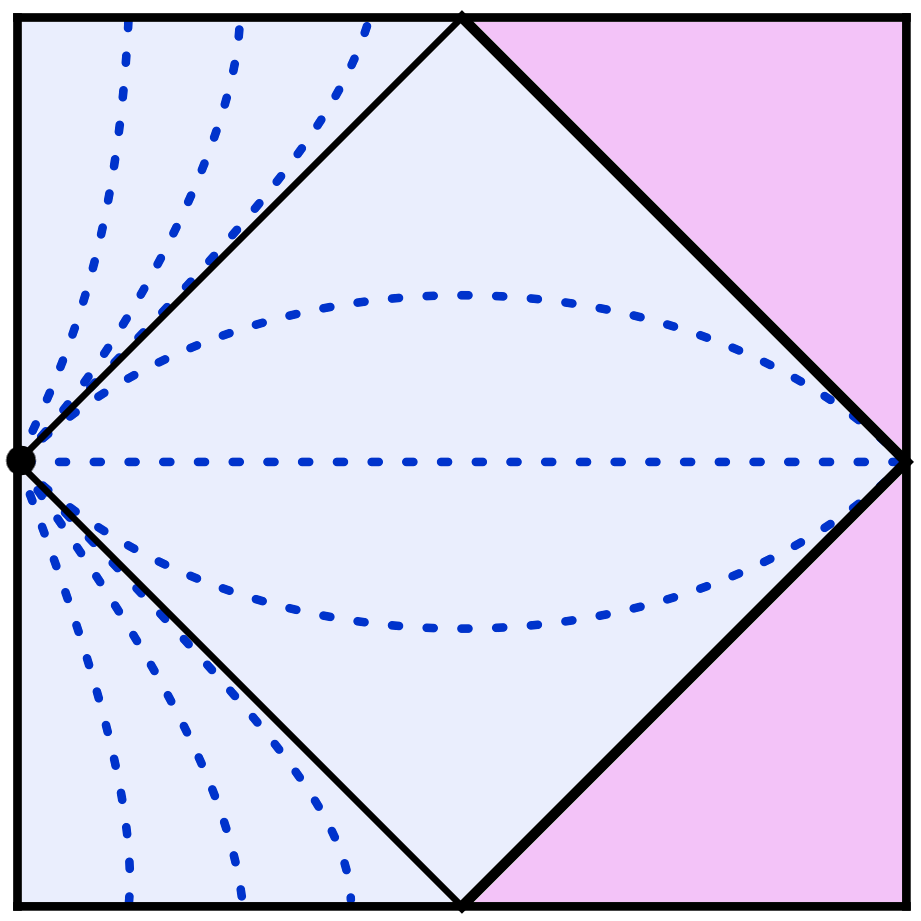}}
\setlength{\unitlength}{0.1\columnwidth}
\begin{picture}(0.3,0.4)(0,0)
\Large{
\put(2.8,2.4){\makebox(0,0){$\lozenge_1$}}
}
\normalsize{
\put(4.1,4.55){\makebox(0,0){$\ratio=\infty$}}
\put(4.1,.3){\makebox(0,0){$\ratio=\infty$}}
\put(5.7,1.45){\rotatebox{45}{\makebox(0,0){$\ratio=0$}}}
\put(5.7,3.4){\rotatebox{-45}{\makebox(0,0){$\ratio=0$}}}
\put(4.3,3.4){\rotatebox{45}{\makebox(0,0){$\ratio=1$}}}
\put(4.3,1.45){\rotatebox{-45}{\makebox(0,0){$\ratio=1$}}}
}
\end{picture}
\caption{Penrose diagram illustrating some reference geodesics (dashed blue lines) and lines of constant cross ratio $\ratio$ in dS$_d$. The pink shaded "shadow" region is not connected to the diamond $\lozenge_1$ by any geodesic. It can naturally be reached through geodesics starting at the antipodal of $\lozenge_1$. 
}
\label{fig:penrose}
\end{figure}

To provide some intuition for our previous discussion, Figure \ref{fig:penrose} illustrates representative geodesics emanating from a particular point in the dS geometry.\footnote{The planar coordinates used in Eq.~\reef{eq.metricdSd} and above actually only cover half of the de Sitter geometry. The surface $R=\infty$ would correspond to a diagonal running across the Penrose diagram in Figure \ref{fig:penrose}. The figure and our discussion here assume a suitable continuation of the cross ratio to the entire geometry. Let us add here that the additional $Z_2$ identification discussed in footnote \ref{footy} would here identify points by an inversion in the square in figure \ref{fig:penrose}, as well as an inversion on the corresponding $S^{d-2}$ at each point on the diagram, to produce elliptic de Sitter space. With regards to the minimal geodesic distances, this identification would essentially remove the right half of the square, \eg there would no longer be any shadow regions.} We observe here that the cross ratio \reef{beast} never becomes negative for spheres restricted to a fixed time slice, however, it does reach zero as noted above just as the spheres become disjoint. As illustrated in the figure, the boundary where $\ratio=0$ corresponds to the past and future null cone emerging from the antipodal point to $\lozenge_1$. Hence there are `shadow regions' in the dS space which cannot be reached along a single geodesic originating from this reference point. Note, however, that there are an infinite family of spacelike goedesics that extend from  $\lozenge_1$ to this antipodal point.

\paragraph{Example 2: CFT in two dimensions.}
In our previous discusion, we showed that for  $d=2$, the coset factorizes into dS$_2\times$dS$_2$, with the metric as in Eq.~\eqref{eq.dS2dS2}. The cross ratio $\ratio$ also factorizes when written in the $\{u,v,\bar u,\bar v\}$ coordinates:
\be
 \ratio_{_{\text{dS}_2\times\text{dS}_2}}((u,v)_1,(\bar u,\bar v)_1;(u,v)_2,(\bar u,\bar v)_2) = \ratio_{_{\text{dS}_2}}(u_1,v_1;u_2,v_2) \, \ratio_{_{\text{dS}_2}}(\bar{u}_2,\bar{v}_2;\bar u_2,\bar v_2) \,,
\ee
where the conformally invariant cross ratio for two points on the dS$_2$ factor is given by
\be 
\ratio_{_{\text{dS}_2}}(u_1,v_1;u_2,v_2) \equiv \frac{(u_2-v_1)(u_1-v_2)}{(u_1-v_1)(u_2-v_2)}
\ee
and similarly with bars. Using this factorization of the cross ratio, one can then compute the geodesic distance on dS$_2\times$dS$_2$ using Eq.~\eqref{eq:GeodDist}.\\

We close this section with two explicit examples of simple geodesics on the full kinematic space $\Mdiamonds$. First, consider some diamond $\lozenge_1 = (c^\mu_1,\ell^\mu_1)$. We wish to compare it with the family of diamonds $\lozenge^{(\lambda)} = (c^\mu_1,\sqrt{\lambda}\,\ell^\mu_1)$ for $0<\lambda<\infty$. One can verify that $\lambda$ parameterizes a timelike geodesic in the space of causal diamonds. As $\lambda \rightarrow 0$, the diamond shrinks to zero size and approaches a locus in the asymptotic past. Similarly, $\lambda \rightarrow \infty$ follows a geodesic to future asymptotia. 
The geodesic distance in this case can be computed explicitly:
\begin{equation}
\begin{split}
 \mathbf{d}(\lozenge_1,\lozenge^{(\lambda_0)}) &= - \bigg{|}\int_{\lambda=1}^{\lambda=\lambda_0} \sqrt{ds_\lozenge^2(\lozenge^{(\lambda)})} \; \bigg{|}
 = - \cosh^{-1} \left( \frac{1+\lambda_0}{2\sqrt{\lambda_0}} \right) \,.
\end{split}
\end{equation}

A second simple example corresponds to a class of null geodesics $\lozenge(\lambda)=(c^\mu(\lambda),\ell^\mu(\lambda))$ with $\partial_\lambda c^\mu= \pm \partial_\lambda \ell^\mu$, where $\lambda$ denotes the affine parameter along the geodesic. Here we begin by noting that because the center of mass coordinates are Killing coordinates for the metric \reef{eq:dsspheres2}, the following are conserved quantities along any geodesics in the kinematic space:
\be
\label{oplk}
P_\mu=\frac{L^2}{\ell^2} \left( \eta_{\mu\nu} - \frac{2}{\ell^2}\, \ell_\mu \ell_\nu \right)\, \partial_\lambda c^\nu\,.
\ee
Further, the full geodesic equations for $\ell^\mu(\lambda)$ simplify greatly upon substituting $\partial_\lambda c^\mu=\pm\partial_\lambda \ell^\mu$ and one finds that 
they are solved by
\be
\label{oplk2}
P_\mu=\pm\frac{L^2}{\ell^2} \left( \eta_{\mu\nu} - \frac{2}{\ell^2}\, \ell_\mu \ell_\nu \right)\, \partial_\lambda \ell^\nu\,,
\ee
which consistently maintains the desired equality between $\partial_\lambda c^\mu$ and $\pm\partial_\lambda \ell^\mu$. As noted above,  $\Delta c^\mu = \pm\Delta \ell^\mu$ corresponds to two diamonds whose past/future tips coincide and so these geodesics correspond to a simple monotonic trajectory through a family of causal diamonds where one tip remains fixed. A simple example is given by choosing $\ell^\mu=\delta^\mu_0\,R(\lambda)$ and $c^0=\delta^\mu_0\,t(\lambda)$, which yields
\be
\label{simp44}
R= R_1/\lambda = \pm t\,,
\ee
where $R_1$ is a constant determining the radius of the corresponding sphere at $\lambda=1$.

\section{Observables in a linearized approximation}
\label{sec:linearized}

As discussed in the introductions, we are interested in trying to construct new nonlocal observables $S_{\cal O}(x,y)$ with a (local) primary operator ${\cal O}$ in the CFT and associated to a causal diamond with past and future tips, $x$ and $y$. 
Our motivation in the present section is to construct extensions of the first law of entanglement for spherical regions in the CFT vacuum. Again, as shown in Eq.~\reef{deltaS}, the perturbations in the entanglement entropy is given by the expectation value of a local operator, the energy density, integrated over the region enclosed by the sphere. This result was used in \cite{deBoer:2015kda} to show that such first order perturbations obey a free wave equation on the corresponding kinematic space, \ie $d$-dimensional de Sitter space. Moreover,
a generalization of the first law was constructed for a conserved higher spin current, which yields an analogous charge $Q^{(s)}$ defined on the spherical region which also obeys a free wave equation on de Sitter space. Here, we would like to extend these results characterizing small excitations of the vacuum to arbitrary scalar primaries.\footnote{We will briefly comment on non-scalar primaries later in this section; for two-dimensional conformal field theories we will present results for
general primaries in section~\ref{sec.genO}.} 

We propose that a natural generalization of the first law to arbitrary primaries takes the following form\footnote{We are using the standard notation here that $(y-x)^2=\eta_{\mu\nu}(y-x)^\mu(y-x)^\nu$ and hence each of the three inner products in the kernel is negative.}:
\be \label{gen:firstlaw}
\delta S_{\cal O}(x,y) 
\equiv \dS x y= \Cbdy \int_{D(x,y)} \!\!\!\!\!d^d \xi\, \left( \frac{(y-\xi)^2(\xi-x)^2 }{-(y-x)^2} \right)^{\half(\Delta_{\cal O}-d)}\ 
\langle {\cal O}(\xi)\rangle\,,
\ee
where the integral is over the causal diamond $D(x,y)$ with past and future endpoints $x,y$, and $\Delta_{\cal O}$
is the scaling dimension of the primary operator ${\cal O}$. The constant $\Cbdy$ is a normalization constant for which there is no canonical choice
at the linearized level. Note that the integral above diverges for $\Delta_{\cal O}\le d-2$, however, a universal finite term can still be extracted in this range. We return to this point in section \ref{discuss}.

In the following, we will show that the quantity $\dSO$ has the following four properties:
\begin{enumerate}
\item
$\dSO$ obeys a simple two-derivative wave equation \reef{gen:freefield} on the moduli space of causal diamonds $\Mdiamonds$, which was introduced in section~\ref{sec.2}.
\item
$\dSO$ reduces to a known `charge' associated with a spherical entangling surface in case that ${\cal O}$ is a conserved (higher spin) current \cite{deBoer:2015kda}.
\item
$\dSO$ can be interpreted as a resummation of all terms in the OPE of two operators of equal dimension
which contain ${\cal O}$ and all its conformal descendants. It is therefore a natural building block of contributions to correlation function where two operators fuse into the ${\cal O}$-channel. 
\item
In the case where the CFT has a holographic dual in the standard sense,
$\dSO$ has a very simple bulk description. If $\phi$ is the bulk scalar that
corresponds to ${\cal O}$, we define
\be 
\label{block}
\dSh x y=\frac{\Cblk}{8\pi G_\mt{N}} \int_{\tilde{B}(x,y)} \!\!\!\!\!d^{d-1} u \,\sqrt{h} \ \phi(u)\,, 
\ee
where $\tilde{B}(x,y)$ is the minimal surface whose boundary $\partial \tilde{B}(x,y)$ matches the maximal sphere at the waist of the causal diamond in the boundary CFT,
\ie the intersection of the past light-cone of $y$ with the future light-cone of $x$. 
We will show that $\dSh x y=\dS x y$ with an appropriate choice of the normalization constant $\Cblk$, which is determined by $\Cbdy$ and standard AdS/CFT parameters --- as we show explicitly in Appendix \ref{app:normalization}.
\end{enumerate}

We stress that the first three properties above do not rely on an underlying holographic construction and hence apply for generic CFTs. It is only point 4, which directly connects to the AdS/CFT construction and so hints at the interesting new perspective which these nonlocal observables may provide for holography.
Below, we will provide a more detailed explanation of each of these points and then discuss various other aspects of $\dSO$. 

However, before proceeding, we want to highlight that Eq.~\reef{gen:firstlaw} can be compactly re-expressed in terms of the conformal Killing vector $K^{\mu}$ which preserves the causal diamond --- see appendix \ref{dead} and Figure \ref{fig:KillingFlow}. In particular, using $K^\mu$, Eq.~\reef{gen:firstlaw} becomes\footnote{Recall for appendix \ref{dead} that $K^\mu$ is a timelike vector and hence our notation is $|K|=\sqrt{-\eta_{\mu\nu}K^\mu K^\nu}$.}
\be \label{gen:firstlaw5}
\dS x y= \Cbdy \int_{D(x,y)} \!\!\!\!\!d^d \xi\, \left( \frac{|K| }{2\pi} \right)^{\Delta_{\cal O}-d}\ 
\langle {\cal O}(\xi)\rangle\,,
\ee
where the factors of $2\pi$ arise from a standard choice of normalization for the vector. Of course, these factors could easily be absorbed by redefining the constant $\Cbdy$.

\subsection{Dynamics on the space of causal diamonds} \label{casual}

To show that $\dSO$ obeys a wave equation on the moduli space of causal diamonds is fairly straightforward. If we denote the generators of the conformal group by $L_i$, then
\be \label{intertwine}
\left(
L_i(x)+L_i(y)\right)\dS x y = \Cbdy \int_{D(x,y)} \!\!\!\!\!d^d \xi\, \left( \frac{(y-\xi)^2(\xi-x)^2 }{-(y-x)^2} \right)^{\half(\Delta_{\cal O}-d)}\,
\langle [L_i,{\cal O}(\xi)]\rangle
\ee
where $L_i(x)$ is the first order differential operator for the purely geometric action of the conformal group on the point $x^\mu$, 
and similarly for $L_i(y)$.\footnote{That is, $L_i(x)$ and $L_i(y)$ are given by the expressions in Eq.~\reef{eq:rotationGen} with $\Delta_\cO=0$.}
The fact that Eq.~(\ref{intertwine}) holds follows from the fact that the kernel that appears in Eq.~(\ref{gen:firstlaw}) can be interpreted formally as a three-point function of two primary
operators of dimensions zero and one primary operator of dimension $d-\Delta_{\cal O}$. Such a three-point function is conformally
invariant and as a result the action of $L_i(x)+L_i(y)$ on the kernel can be converted in the action of $-L_i(\xi)$ (with a contribution
from the non-trivial operator dimension of the third operator). A partial integration then yields Eq.~(\ref{intertwine}).
In fact, we could conversely have derived Eq.~(\ref{gen:firstlaw}) by insisting that it obeys the intertwining property (\ref{intertwine}),
and we can make this property more transparent by rewriting Eq.~(\ref{gen:firstlaw}) using the shadow operator formalism \cite{Ferrara:1972uq} as
\be \label{gen:firstlaw2shadow}
\dS x y = \Cbdy \int_{D(x,y)} \!\!\!\!\!d^d \xi\ \langle Y(x) Y(y) {\cal O}^{\ast}(\xi) \rangle\,
\langle {\cal O}(\xi)\rangle
\ee
where $Y$ represents a {\it formal} non-trivial primary operator of conformal dimension zero.

The action of second Casimir of the conformal group $\Casimir\equiv C^{ij}L_iL_j$ on $\dSO$ is obtained by applying Eq.~(\ref{intertwine}) twice. The left hand side of the equation then becomes
\be \label{wave1}
C^{ij} (L_i(x) + L_i(y))(L_j(x) + L_j(y)) \dSO\,.
\ee
Because $L_i(x) + L_i(y)$ represents the action of the conformal group on the moduli space, which is parametrized by 
pairs of (timelike separated) points $(x,y)$, these are also the Killing vectors on this space, and the Casimir operator $C^{ij} (L_i(x) + L_i(y))(L_j(x) + L_j(y)) $ is
the massless Klein-Gordon operator for the metric (\ref{eq.metriccosetd}).
On the right hand side, we get the combination\footnote{For non-scalar primaries ${\cal O}$, there is an extra contribution on the right-hand
side of the form $C_L\langle {\cal O}(\xi)\rangle$ with $C_L$ the second Casimir of the Lorentz representation of ${\cal O}$.}
\be
\langle C^{ij}[L_j,[L_i,{\cal O}(\xi)]]\rangle = \Delta_{\cal O}(d-\Delta_{\cal O}) \langle {\cal O}(\xi)\rangle
\ee
and therefore $\dSO$ obeys the following wave equation
\be \label{gen:freefield}
(\BoxM-m_{\cal O}^2)\,\dS x y=0\qquad \mbox{\rm with}\,\,\, m_{\cal O}^2\,L^2=\Delta_{\cal O}(d-\Delta_{\cal O}) \,,
\ee
where $\BoxM$ is the Klein-Gordon operator on the metric \eqref{eq.metriccosetd}. We conclude that the Casimir is represented on the space of causal diamonds $\Mdiamonds$ as $\Casimir = L^2\BoxM$. This can also be explicitly verified by acting with the Lorentz representation of $\Casimir$ on Eq.~\eqref{intertwine}. For our conventions and normalizations in this regard, see appendix \ref{app:conventions}.

\subsection{Operators with spin and conserved currents}
\label{sec:spin} 

Our construction can be easily generalized to the case where the primary operator is a traceless symmetric tensor of rank $\ell$
and scaling dimension $\Delta_{\cal O}$. In this case, 
conformal invariance again provides a natural candidate for a `first law'-like expression which takes the form
\be \label{gen:firstlaw2}
\dS x y  = \Cbdy \int_{D(x,y)} \!\!\!\!\!d^d \xi\, \left( \frac{(y-\xi)^2(\xi-x)^2 }{-(y-x)^2} \right)^{\half(\Delta_{\cal O}-d)}
\frac{s^{\mu_1}\cdots s^{\mu_{\ell}}\;\langle {\cal O}_{\mu_1 \ldots \mu_{\ell}}(\xi)\rangle}{ (-(y-\xi)^2(\xi-x)^2(y-x)^2)^{\ell/2} }\,
\,.
\ee
where
\be\label{eq:ConfKilling}
s^{\mu}=(y-\xi)^2\,(x-\xi)^{\mu}  -(x-\xi)^2\, (y-\xi)^{\mu} = -\frac{1}{2\pi} (y-x)^2\, K^{\mu}
\ee
with $K^{\mu}$, the conformal Killing vector introduced above --- see appendix \ref{dead} and Figure \ref{fig:KillingFlow}.\footnote{Note that $s^\mu$ and $K^\mu$ are both future-directed vectors within the causal diamond.}  Using this vector as in Eq.~\reef{gen:firstlaw5}, the above generalization can be written in the compact form:
\be \label{gen:firstlaw55}
\dS x y= \frac{\Cbdy}{(2\pi)^{\Delta_{\cal O}-d}} \int_{D(x,y)} \!\!\!\!\!d^d \xi\  |K|^{\Delta_{\cal O}-\ell-d}\,
K^{\mu_1}\cdots K^{\mu_{\ell}}\,
\langle {\cal O}_{\mu_1 \ldots \mu_{\ell}}(\xi)\rangle\,.
\ee
This expression in Eq.~\reef{gen:firstlaw2} follows from the shadow field formalism developed in \cite{SimmonsDuffin:2012uy} and the explicit result for the
three-point function of two scalars and one higher spin field, \eg  in \cite{Costa:2011mg}. From conformal symmetry arguments (or alternatively from explicit calculation --- see appendix \ref{app:conventions}), it follows again that the expression in Eq.~\eqref{gen:firstlaw2} satisfies a `spinning' wave equation on the space of causal diamonds:
\be \label{black}
\left(\BoxM-m_{\cal O}^2\right) \dS x y = 0 \qquad \text{with}\quad m_{\cal O}^2 \,L^2 =   \Delta_{\cal O}(d-\Delta_{\cal O}) - \ell(\ell+d-2)  \,.
\ee

\begin{figure}[t]
\centerline{\includegraphics[width=.27\textwidth]{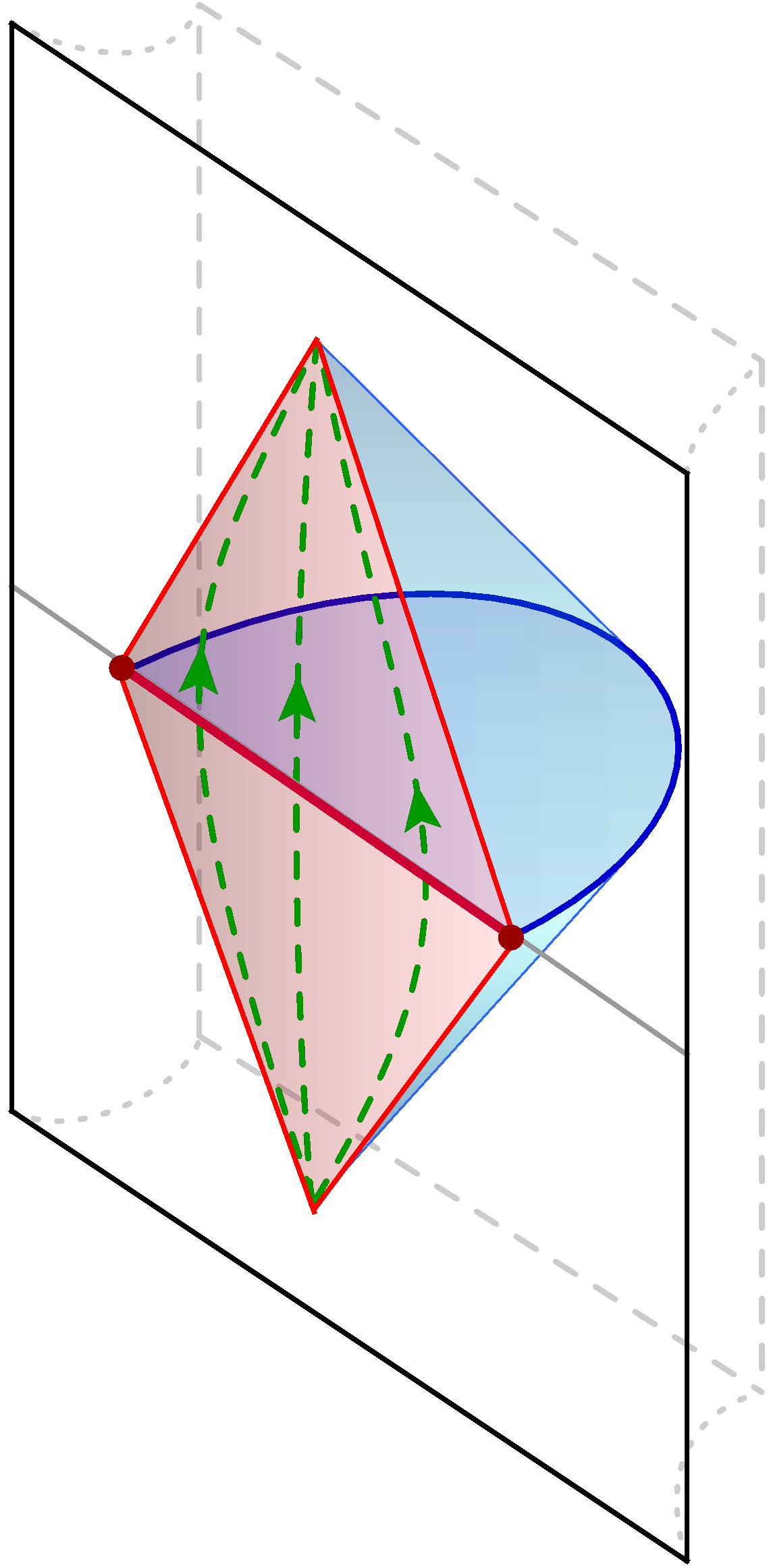}}
\setlength{\unitlength}{0.1\columnwidth}
\begin{picture}(0.3,0.4)(0,0)
\definecolor{Green}{rgb}{0,0.6,0.2}
\definecolor{Red}{rgb}{0.65,0.2,0.1}
\normalsize{
\put(5.85,3.3){\makebox(0,0){${\color{blue}\tilde{B}}$}}
\put(4.9,3.2){\makebox(0,0){${\color{Red}B}$}}
\put(4.8,1.55){\makebox(0,0){$x^\mu$}}
\put(4.83,3.6){\makebox(0,0){${\bf \color{Green}K}$}}	
\put(4.2,2.7){\makebox(0,0){${\color{red}D}$}}
\put(4.8,4.9){\makebox(0,0){$y^\mu$}}
}
\small{
}
\end{picture}
\caption{Domain of dependence $D(x,y)$ in the CFT$_2$ (red shaded) of a sphere $B(x,y)$, and the associated causal wedge in pure AdS$_3$ (blue shaded). The geodesic (blue) is the bulk minimal surface $\tilde{B}(x,y)$. Green arrows indicate the timelike Killing flow generated by $K^\mu$, which becomes null at the boundary of the domain of dependence (and vanishes at $x^\mu$ and $y^\mu$, see also Figure \ref{fig:flow}). 
}
\label{fig:KillingFlow}
\end{figure}
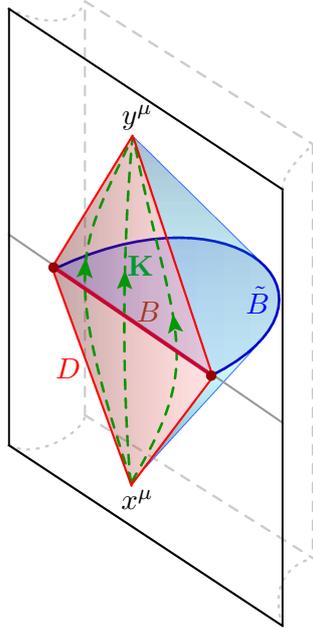

To illustrate the definition \eqref{gen:firstlaw2}, we turn to the second point in our list of properties above and show that it reproduces the known first laws \cite{deBoer:2015kda} when the operator is a conserved current. 
If the traceless symmetric tensor corresponds to a conserved current, then\footnote{Note that substituting Eq.~\reef{groc} into Eq.~\reef{black} yields
$m_{\cal O}^2L^2=-2(\ell-1)(\ell+d-2)$, which differs by a factor of two from the mass-squared reported in \cite{deBoer:2015kda}. However, as described above Eq.~\reef{eq.metricdSd}, restricting the submanifold of spheres on a fixed time slice requires `equating' the coordinates for the two tips of the causal diamond. This has the effect of reducing the mass.
Effectively one has $\BoxM \sim 2\,\nabla^2_\mt{dS}$  on this restricted moduli space studied in \cite{deBoer:2015kda}. In two dimensions the space ${\cal M}_\lozenge^{(2)}$ actually factorizes in two copies of dS$_2$ as in \eqref{eq.dS2dS2}. In this case one can make the above statement precise by noting that $\BoxM = 2 (\nabla^2_{{\rm dS}_2} + \bar{\nabla}^2_{\overline{\rm dS}_2})$ with each of the dS$_2$ spaces contributing $m^2_{\rm dS_2} L^2 = -\ell(\ell-1)$ and $m^2_{\overline{\rm dS}_2} = 0$, respectively (c.f., Eq.\ \eqref{slap1} and the discussion there).}
\be
\Delta_{\cal O} = \ell+d-2 \label{groc}\,.
\ee
Hence Eq.~\reef{gen:firstlaw55} can be written as
\bea
\label{gen:firstlaw3}
\dS x y  
&=& \frac{\Cbdy}{(2\pi)^{\ell-2}}\int_{D(x,y)}\!\!\!\!\! d^d \xi\  \frac{K^{\mu}J_{\mu}}{|K|^2} 
\labell{gen:firstlaw4}
\eea
where we have introduce the conserved current $J_{\mu}\equiv K^{\mu_2}\cdots K^{\mu_{\ell}}  
\langle {\cal O}_{\mu\mu_2 \cdots \mu_{\ell}}(\xi)\rangle$.\footnote{Current conservation follows here because $\langle {\cal O}_{\mu_1 \ldots \mu_{\ell}}(\xi)\rangle$ is both traceless and conserved and because $K^\mu$ is a conformal Killing vector \cite{deBoer:2015kda} --- see appendix \ref{dead}.} Now suppose we foliate the causal diamond by slices that are everywhere orthogonal to the
vector field $K^{\mu}$, and we also introduce a flow parameter in the direction of $K^{\mu}$
which we will call $\lambda$: $\partial_{\lambda}\equiv K^{\mu}\partial_{\mu}$. 
It is then clear that we can re-express the measure as $d^d \xi = d\lambda\ d^{d-1}\Sigma\, |K|$ where $d^{d-1}\Sigma\, |K|$ is the induced measure on each $(d-1)$-dimensional constant $\lambda$ slice. As a result, we obtain
\be
\label{gen:firstlaw5}
\dS x y = \frac{\Cbdy}{(2\pi)^{\ell-2}} \int_{D(x,y)}\!\!\!\!\! d\lambda\  d^{d-1}\Sigma \ n^{\mu}\,J_{\mu}
\ee
with $n^{\mu}=K^{\mu}/|K|$ being the timelike unit normal to the constant $\lambda$ slices. However, because $J_{\mu}$ is a conserved current, 
the integral over a slice of $n^{\mu} J_{\mu}$ does not depend on the slice. Hence 
\be
\label{gen:firstlaw6}
\dS x y = \frac{\Cbdy}{(2\pi)^{\ell-2}}  \int_{B(x,y)} \!\!\!\!\! d^{d-1}\Sigma\  n^{\mu}\, J_{\mu}\ \ \times\ \int d\lambda\,,
\ee
where $B(x,y)$ is a constant $\lambda$ slice, \eg the spherical region for which $D(x,y)$ is the domain of dependence. Note that the factor $\int d\lambda$ is in fact divergent but it can be absorbed into the normalization constant $\Cbdy$. Hence upon a redefinition of the normalization constant, the final result can be written as
\be
\label{gen:firstlaw66}
\dS x y = \frac{\tCbdy}{(2\pi)^{\ell-2}}  \int_{B(x,y)} \!\!\!\!\! d\Sigma^{\mu_1} \ 
K^{\mu_2}\ldots K^{\mu_{\ell}} \, 
\langle {\cal O}_{\mu_1 \ldots \mu_{\ell}}(\xi)\rangle\,.
\ee
Observe, that in fact, current conservation allows the $(d-1)$-dimensional surface defining the range of the remaining integral to be chosen as any Cauchy surface within the causal diamond   $D(x,y)$, \ie it need not be a constant $\lambda$ slice. Hence as claimed in the second point on our list above,
we have recovered precisely the first law for
conserved currents proposed previously in \cite{deBoer:2015kda}. In particular, the covariant version of the standard first law for entanglement entropy  is immediately recovered with the choice $\ell=2$, \ie $\mathcal{O}_{\mu_1\mu_2} = T_{\mu_1\mu_2}$.\footnote{Note another case which deserves special attention is $\ell=1$ and $\Delta_\cO=d-1$, which corresponds to ordinary conserved current, \ie $J_\mu=\langle\cO_\mu\rangle$. Na\"ively, the above arguments would suggest that the corresponding operator \reef{gen:firstlaw66} also satisfies the wave equation \reef{black} on the moduli space. However, an implicit assumption in the derivation of the wave equation is that current vanishes on the sphere $\partial B(x,y)$ and this is ensured in Eq.~\reef{gen:firstlaw66} by the vanishing of the conformal Killing vector on this surface --- see appendix \ref{dead}. However, there are no such factors of $K^\mu$ in the special case $\ell=1$ and so extra boundary term contributions would appear in the wave equation.}

\subsection{Connection to the OPE}
\label{sec:OPE}

The third point in the list of  features of Eq.~(\ref{gen:firstlaw}) 
is the connection to the operator product
expansion. In general, the OPE of two operators takes the form
\be
A(x)\, B(y) = \sum_i C_{AB}^{{\cal O}_i}\,
\frac{{\cal O}_i(y)}{(x-y)^{\Delta_A +\Delta_B-\Delta_{{\cal O}_i}}} + \mbox{{\rm conformal descendants}}
\ee
where on the right the sum is over all primary operators ${\cal O}_i$ and its conformal descendants. In two dimensions, where
the conformal group is infinite, we will take the sum to be over all quasi-primary operators and their descendants under the global
conformal group only. In principle, there is an infinite sum over conformal descendants on the right hand side, but this 
infinite sum can be repackaged as an integral of ${\cal O}_i$ smeared against a suitable kernel,
\be\label{eq:OPEansatz}
A(x)\, B(y) = \sum_i C_{AB}^{{\cal O}_i} \int_{D(x,y)} d^d\xi\; I_{AB{\cal O}_i}(x,y,\xi) \,{\cal O}_i(\xi)\,.
\ee
The kernel $I_{AB{\cal O}_i}(x,y,\xi)$ that appears here is completely fixed by conformal invariance. One can in principle
construct it by working out the relevant conformal Ward identities and solving for them. If one does this one recognizes
that the Ward-identities look exactly like those of a three-point function. In fact, this should not have come as a
surprise, as the 
shadow field identity (\ref{gen:firstlaw2shadow}) indeed implies that $I$ is proportional to a three-point function
\be
I_{AB{\cal O}_i}(x,y,\xi) \sim \langle A(x) B(y) {\cal O}_i^{\ast} \rangle\,.
\ee
This three-point function (for scalar operators) equals
\be \label{aux:3pt}
\langle A(x)\, B(y)\, {\cal O}_i^{\ast}(\xi) \rangle \sim (x-y)^{d-\Delta_{{\cal O}_i}-\Delta_A-\Delta_B}
(x-\xi)^{\Delta_{{\cal O}_i}-d-\Delta_A+\Delta_B}
(y-\xi)^{\Delta_{{\cal O}_i}-d+\Delta_A-\Delta_B}\, .
\ee
For the quantity $\dSO$ which appears in the first law, we imagine that there should not be any
special operators located at either $x$ or $y$, and indeed we recover the form of the first law in Eq.~(\ref{gen:firstlaw})
by taking $\Delta_A=\Delta_B=0$. Of course, in an actual conformal field theory, there is only one operator with
vanishing dimension, the identity operator, for which the three-point function above actually vanishes. One should
therefore view this is as a somewhat formal argument intended to explain the constraints imposed by conformal invariance alone.

Nevertheless, we notice from that Eq.~(\ref{aux:3pt}) that $(x-y)^{\Delta_A+\Delta_B}\langle A(x) B(y) {\cal O}_i^{\ast} \rangle$
also reproduces the kernel in Eq.~(\ref{gen:firstlaw}) as long as $\Delta_A=\Delta_B$. We can therefore use either
Eq.~(\ref{gen:firstlaw}), or its bulk counterpart (\ref{gen:bulk}), to compute the contribution of a particular operator and
all its conformal descendants to the OPE of two equal dimension scalar operators. 

For example, consider a four-point function 
\be
\langle A(x_1)\, B(y_1)\, C(x_2)\, D(y_2) \rangle 
\ee
of four scalar operators with $\Delta_A=\Delta_B$ and $\Delta_C=\Delta_D$. We can ask what the contribution to this
four-point function is when a particular operator ${\cal O}$ runs in the intermediate $(AB)-(CD)$-channel, also known
as a conformal block. Up to an overall normalization, we find that this conformal block equals
\be
(x_1-y_1)^{-\Delta_A-\Delta_B} (x_2-y_2)^{-\Delta_C-\Delta_C}
\langle \dS {x_1}{y_1} \, \dS{x_2}{y_2} \rangle.
\ee
We can now evaluate this two-point function using (\ref{gen:firstlaw}) and relate it to the integral of 
$\langle {\cal O}(\xi_1)\, {\cal O}(\xi_2) \rangle$ over two causal diamonds $D(x_1,y_1)$ and $D(x_2,y_2)$ on
the boundary.

 In the context of the AdS/CFT correspondence, a Euclidean version of this argument underlies the geodesic Witten diagram prescription of \cite{Hijano:2015zsa}.
Alternatively, we can use the bulk representation (\ref{gen:bulk}) which leads immediately to
an expression involving a double integral over two minimal surfaces connected by a bulk-bulk propagator,
reminiscent of the result in \cite{Hijano:2015zsa}.  Finally, the same quantity admits yet another
interpretation as the two-point function of $\dSO$ on the moduli space of causal diamonds. 
Notice that with all of the above we are working in Lorentzian signature (or mixed signature in case of the moduli
space of causal diamonds) and one has to be careful to precisely define
the types of correlators and Green's functions that appear. There is also a close relation to the `splines' introduced in \cite{Paulos:2012qa}.

\subsection{Holographic description}
\label{sec:bulk}

So far our discussion did not assume any special features of the CFT, however, we now turn to point 4 on our list which refers to the special case of holographic CFTs. In particular, we will be considering CFTs with a dual description in terms of weakly coupled gravity. In this setting, the scalar operator ${\cal O}$ in the boundary theory will be dual to a scalar field $\phi$ in the bulk and we wish to show that the following simple bulk expression provides an alternative definition of $\dSO$:
\be \label{gen:bulk} \dSh x y=\frac{\Cblk}{8\pi G_\mt{N}} \int_{\tilde{B}(x,y)} \!\!\!\!\!d^{d-1} u\, \sqrt{h}\; \phi(u) \,.
\ee
Here, as discussed above, 
$\tilde{B}(x,y)$ is the extremal surface reaching the asymptotic AdS boundary at the maximal sphere that bounds the causal diamond --- see Figure \ref{fig:KillingFlow}. Further, the measure $\sqrt{h}\, d^{d-1}u$ is simply the induced volume
element on $\tilde{B}$.
Now our claim, which we demonstrate below, is that 
\be\label{claim}
\dSh x y=\dS x y\,, 
\ee
with an appropriate choice of the normalization constant $\Cblk$.
Note that $\Cblk$ is fixed by standard AdS/CFT techniques once the normalization $\Cbdy$ in \eqref{gen:firstlaw} is given. In Appendix \ref{app:normalization}, we explicitly compute $\Cblk$ as a function of the CFT normalization $\Cbdy$, the dimension $d$ and the weight $\Delta_{\cal O}$ --- see Eq.~\eqref{eq:CbdyCblk} for the result. Note that it is natural to include an inverse factor of $8\pi G_\mt{N}$ in the definition of $\dSOh$, as this factor ensures that our new observable is dimensionless\footnote{Recall that $8\pi G_\mt{N}=\lp^{d-1}$ and we are assuming the usual `supergravity' convention where the bulk scalar $\phi(u)$ is a dimensionless field.} just as with its counterpart \reef{gen:firstlaw} in the boundary theory. The above holographic relation \reef{claim}  is in line with the general philosophy that minimal
surfaces should play a prominent role in the construction of these new boundary observables $\dS x y$, as is the case for holographic entanglement entropy.

To show the equality of Eqs.~(\ref{gen:firstlaw}) and (\ref{gen:bulk}), we can argue as follows: 
If we apply the conformal generator $L_i(x) + L_i(y)$ to the above expression, this has the
effect of an infinitesimal displacement of $\tilde{B}(x,y)$ in the direction of the Killing vector
field $L_i$. The field $\phi$ at this displaced location differs from the original value by an
amount $L_i\phi$, but the rest of the integrand remains unchanged because $L_i$ is a Killing vector field.
Therefore
\be \label{intertwine2}
(L_i(x)+L_i(y)) \int_{\tilde{B}(x,y)} \!\!\!\!\!d^{d-1} u\, \sqrt{h}\, \phi(u) = 
\int_{\tilde{B}(x,y)}\!\!\!\!\! d^{d-1} u \sqrt{h} \;L_i \phi(u) \,.
\ee
In case, this equation appears to be somewhat confusing, a simple one-dimensional version of this equation which illustrates the
idea is
$$
\int_{x+a}^{y+a} f(u)\, du = \int_x^y f(u+a)\,du  \,.
$$
We see that the bulk description of $\dSO$ enjoys a similar intertwining property as 
in Eq.~(\ref{intertwine}).

Applying the quadratic Casimir $\Casimir$ requires iterating Eq.~(\ref{intertwine2}) twice and we find
\be
C^{ij} (L_i(x) + L_i(y))(L_j(x) + L_j(y))\int_{\tilde{B}(x,y)}\!\!\!\!\! d^{d-1} u\, \sqrt{h} \, \phi(u) = 
\int_{\tilde{B}(x,y)}\!\!\!\!\! d^{d-1} u \,\sqrt{h}\; C^{ij} L_i L_j \phi(u) \,.
\ee
Now we can use the fact that in our conventions $C^{ij}L_iL_j$ is proportional to the d'Alembertian acting in the AdS spacetime, \eg \cite{vijay22}.\footnote{See appendix \ref{app:conventions} for details.} In particular, since $\phi(u)$ obeys a free massive field equation, we then have 
\be \label{wavy}
C^{ij} L_i L_j \phi(u) = - \Rads^2\nabla^2_{\text{\tiny AdS}} \phi(u) = -\mAdS^2\Rads^2\, \phi(u) = \Delta_{\cal O}(d-\Delta_{\cal O})\, \phi(u) \,, 
\ee
where we used the standard relation between the conformal dimension and the mass of the dual field, $\mAdS^2\Rads^2 = \Delta_{\cal O}(\Delta_{\cal O}-d)$. Hence we find that the holographic bulk expression in Eq.~\reef{gen:bulk} yields the same eigenvalue as found for the boundary expression in section \ref{casual}, \ie $\Casimir\,\dSh x y = \Delta_{\cal O}(d-\Delta_{\cal O})\,\dSh x y$, and the same wave equation (\ref{gen:freefield}) on kinematic space follows.

Hence we have shown that for scalar operators, $\dSO$ defined in Eq.~(\ref{gen:firstlaw}) for the boundary theory and $\dSOh$ defined in Eq.~(\ref{gen:bulk}) for the bulk theory obey the same wave equation of kinematic space. If we would in addition
show that both quantities obey the same boundary conditions for these equations, this would be sufficient to establish their equivalence up
to an overall normalization.  However, instead of studying the boundary conditions, there is a more direct argument to show the equivalence of Eqs.~(\ref{gen:firstlaw}) and (\ref{gen:bulk}). 

Inside the bulk causal domain attached to the boundary causal diamond, often referred to as the bulk causal
wedge, we can reconstruct the value of the field
using a bulk-boundary propagator which only involves the expectation value of the corresponding operator inside
the causal diamond \cite{Hamilton:2006az,Morrison:2014jha}. In other words, there exists a bulk-boundary propagator such that
\be
\phi(u) = \int_{D(x,y)}\!\!\!\!\! d^d \xi\, G_{b\partial}(u,\xi)\,\langle {\cal O}(\xi) \rangle
\ee
for any $u$ inside the causal wedge associated with $D(x,y)$. 
Inserting this expression into Eq.~(\ref{gen:bulk}), we find
\be
\dSh x y=\frac{\Cblk}{8\pi G_\mt{N}} \int_{D(x,y)} \!\!\!\!\!d^d \xi \left(\int_{\tilde{B}(x,y)} d^{d-1} u \,\sqrt{h}\, G_{b\partial}(u,\xi)
\right) \langle {\cal O}(\xi) \rangle\,.
\ee
The integral between brackets does not depend on the values of the field and we denote the result of this integral
by $H(x,y,\xi)$ resulting in
\be \label{gen:aux2}
\dSh x y=\frac{\Cblk}{8\pi G_\mt{N}} \int_{D(x,y)} \!\!\!\!\!d^d \xi\; H(x,y,\xi) \,\langle {\cal O}(\xi) \rangle\,.
\ee
This already takes the form of the first law and all that is left to do is to show that the kernel $H(x,y,\xi)$
agrees with that appearing in Eq.~(\ref{gen:firstlaw}). This can be seen as follows: The bulk-boundary propagator
$G_{b\partial}(u,\xi)$ is invariant under the isometries of AdS, implying
\be
(L_i(u)+L_i(\xi)) G_{b\partial}(u,\xi) = 0\,.
\ee
Combining the above with Eq.~(\ref{intertwine2}) shows that Eq.~(\ref{gen:aux2}) also obeys the intertwining property (\ref{intertwine}),
and as discussed below Eq.~(\ref{intertwine}), this uniquely fixes $H(x,y,\xi)$ up to an overall constant and hence it
must agree with the kernel appearing in the first law (\ref{gen:firstlaw}). 

One potential subtlety in the above analysis is that causal wedge reconstruction strictly speaking only applies
to the interior of the causal wedge, and to extend it to the boundary of the causal wedge requires us to make
an assumption that the field is continuous there. At the linearized level one could contemplate that there exist
solutions of the $\phi$ field equations with support outside and on the boundary of the causal wedge only. For example, 
one could assume that $\langle {\cal O}\rangle$ is zero everywhere inside the causal diamond and then discontinuously
jumps to a finite value at the boundary of the causal diamond. While one cannot, strictly speaking, exclude such
field configurations, they will tend to produce strange effects at higher orders and can for example produce a singular
energy-momentum tensor leading to a large back-reaction. For simplicity, we will in this paper simply ignore this issue
and restrict to continuous field configurations. An explicit calculation demonstrating the desired equality \reef{claim} for $d=2$ with smooth configurations is given in appendix \ref{exam}. 

It is interesting to observe that the bulk-boundary propagator for causal wedges is usually written in momentum space 
and behaves in such a way that a direct Fourier transform to position space is 
ill-defined \cite{Morrison:2014jha,claire78,Freivogel:2016zsb}.
However, after integrating the bulk operator over a bulk minimal surface, we apparently obtain the rather simple 
expression \eqref{gen:firstlaw} where the expectation value of the operator is smeared with a perfectly well-behaved kernel.
It would be good to have a better understanding of the origin of this simplification, but one could certainly say that this observation provides further evidence that the $Q({\cal O};x,y)$ are natural objects to study in the CFT.
 
\paragraph{Vector fields:} It is instructive to see what the bulk description is for an example of a non-scalar field. We briefly
describe the result for a (massless or massive) vector field, leaving the case of higher spin fields as an interesting exercise
for the reader. Given a bulk vector field $A_M$, we can always construct the $(d-1)$-form $\ast F$, with  the field strength $F_{MN}=\partial_M
A_N-\partial_N A_M$  corresponding to $A_M$. The appropriate bulk expression for $\dSO$ in this case turns out to be
\be \label{gen:bulkvector} Q_\mt{holo}({\cal O}_\mu; x, y)= \frac{\Cblk}{8\pi G_\mt{N}}\int_{\tilde{B}(x,y)}\!\!\! \ast F\,.
\ee
To show that this agrees with the first laws (\ref{gen:firstlaw2}) and (\ref{gen:firstlaw6}) for massive and massless vector fields, 
we can use the same group-theoretic argument that we used for scalar fields above. Interestingly, while the CFT counterpart (\ref{gen:firstlaw2})
diverges in the massless limit and becomes an infinite factor times (\ref{gen:firstlaw6}), the bulk expression (\ref{gen:bulkvector})
remains finite in the massless limit.

For massless vectors, the field equation reads $d\ast F=0$, and therefore we can continuously deform the bulk minimal
surface $\tilde{B}(x,y)$ without changing the value of $\dSOh$. In particular, we can deform it all the
way up to a spatial slice in the aymptotic AdS boundary. Then using the asymptotic behavior of a massless vector field in AdS directly, we find that Eq.~(\ref{gen:bulkvector}) agrees with Eq.~(\ref{gen:firstlaw66}) for a spin-one current. Further this argument can be applied for a vector field version of the derivation of the linearized Einstein equations from entanglement entropy \cite{eom2}, described in the introduction. For massive vectors, $d\ast F \sim m^2 *A$ and
this simple argument no longer applies. We return to these observations in the closing discussion section.

We finally notice that 
in writing Eq.~(\ref{gen:bulkvector}) we assumed the bulk action for the gauge field to be of Maxwell type. In 2+1 dimensions
it is also possible to have a topological theory with only a Chern-Simons term instead. In that case, the bulk description
should be replaced by Wilson loop of the gauge field.


\subsection{Euclidean signature} \label{euklid}

We can repeat much of the above logic in Euclidean signature, but there are some significant modifications. In this case, one might consider two distinct possibilities: The first would be the moduli space of pairs of (spacelike separated) points, which becomes $SO(1,d+1)/(SO(d)\times SO(1,1))$. The second distinct case would be the moduli space of $(d-2)$-dimensional spheres, which becomes  $SO(1,d+1)/(SO(1,d-1)\times SO(2))$. In either case, there is still a natural metric on the moduli space given by a suitable Wick rotation of Eq.~(\ref{eq.metriccosetd}). 

Similarly, as described in appendix \ref{dead}, the conformal Killing vector $K^\mu$ may be analytically continued to produce a conformal Killing vector of $\mathbb{R}^d$ which has fixed points either on a pair of points or on a $(d-2)$-sphere. This provides us two extensions of our new observables to Euclidean space through Eqs.~\reef{gen:firstlaw5} and \reef{gen:firstlaw55}. However, the causal diamonds are lost in Euclidean signature and so there is no natural finite domain with which to associate these observables.  As a result our analogous Euclidean expressions would now involve an integral over the entire Euclidean boundary. Of course, there is no obvious reason that such an integral should converge. However, we conjecture that it is possible to extract a universal finite term when the integrals are suitably regulated. This issue would not arise in the case of a conserved current where the integral in Eq.~\reef{gen:firstlaw66} is reduced to a Cauchy surface spanning the causal diamond. In Euclidean signature, when considering the $(d-2)$-spheres, this integral naturally continues to an integral over any $(d-1)$-dimensional surface whose boundary is the corresponding sphere. In the case of pairs of points, the natural domain would be an integral over closed $(d-1)$-dimensional surface enclosing one of the points.\footnote{This points to the necessity of introducing sources in the Euclidean framework, \ie otherwise such an operator would always evaluate to zero since the closed surface is contractible to a point. We will not pursue this issue further here but leave it as an interesting future project.}

The connection with the resummation of a local operator and its conformal descendants in the OPE remains valid given a pair of points. In fact, shadow operator formalism \cite{Ferrara:1972uq,SimmonsDuffin:2012uy} was originally developed in Euclidean signature. A similar discussion might be developed for the case of spheres, however, it would the OPE limit of $(d-2)$-dimensional surface operators, \eg see discussions in \cite{Hung:2014npa,Gomis:2009xg}.

In a holographic context, if we consider a sphere in the boundary theory, this again naturally defines a preferred extremal surface in the bulk. Hence the discussion of the holographic description of $\dSOh$ is essentially unchanged from that given in section \ref{sec:bulk}. On the other hand, given two spacelike separated points on the boundary, we must turn to a new class of natural minimal surfaces, namely, the geodesic connecting the two boundary points. Of course, integrating over this bulk surface in Eq.~\reef{gen:bulk} provides a natural construction of a bulk observable which is again entirely geometric in nature. The arguments
we gave in Lorentzian signature, which crucially relied on conformal invariance, can be repeated in Euclidean signature (at least formally) to show that the boundary and bulk descriptions of $\dSO$ agree in either case, and that it still obeys a wave equation on the corresponding moduli space. With the case of pairs of points, one makes direct contact with the geodesic Witten diagram prescription of
\cite{Hijano:2015zsa}, which was also derived in Euclidean signature --- as well as with the `splines' introduced in \cite{Paulos:2012qa}. Further in this case, for higher-spin symmetric traceless tensor fields, the natural bulk quantity
to consider is the contraction of the rank-$j$ tensor field with $j$ times the unit tangent vector along the geodesic. This object is
again quite distinct from its spherical or Lorentzian counterparts, for which we only worked out the vector field case --- we return to this point in section \ref{discuss}.

Of course, one could also consider the moduli space of spacelike separated pairs of points in Lorentzian signature and we discuss this possibility at length in appendix \ref{geom}. In this case, the coset geometry is $SO(2,d)/(SO(1,d-1)\times SO(1,1))$, which matches that for the moduli space of spheres in Eq.~\reef{eq:Mdiamonds}. In the appendix, by extending our considerations from Minkowski space to $\mathbb{R}\times S^{d-1}$ geometry, we show that the two moduli spaces are in fact identical. That is, the moduli space of spacelike pairs of points is the same geometric object as the moduli space of spheres or timelike pairs of points. It is interesting that in the $\mathbb{R}\times S^{d-1}$ geometry, a pair of spacelike separated points defines a region of finite volume, namely that enclosed by the past and future lightcones of both points. Further the conformal Killing vector $K^\mu$ is naturally extended to generate a flow on this region (with fixed points on the spacelike pair). Hence in this context, it would be natural to define nonlocal observables using Eqs.~\reef{gen:firstlaw5} and \reef{gen:firstlaw55} where the integration would now run over this new volume. In a holographic CFT, these observables would naturally have a gravity description analogous to Eq.~\reef{gen:bulk} except the bulk integral would run over the geodesic connecting the spacelike separated pair of points on the boundary. The latter construction would again connect directly to the discussion of geodesic Witten diagrams \cite{Hijano:2015zsa}. Of course, it would be interesting to fully explore the implications of this equivalence of these two moduli spaces in Lorentzian signature.
It would also be good to develop a better conceptual understanding of the peculiar differences between the various Lorentzian and Euclidean
versions of $\dSO$.

\subsection{Other fields}

We have so far discussed scalar fields and some aspects of higher spin fields described by symmetric traceless tensors. 
There are clearly many other types of fields one could contemplate studying carrying different representations of the
Lorentz group such as fermions or antisymmetric tensors. It would be interesting to study such fields as well, and to examine whether the natural
generalization of $\dSO$ remains a scalar on the space of causal diamonds or whether it can become a quantity
which carries nontrivial quantum numbers under the local $SO(d,d)$ Lorentz group on the generalized kinematic space. 

A natural starting point to explore such generalizations would be to put different fiducial operators $A$, $B$ at
the tips of the causal diamond and to write down a first law with a kernel of the form (\ref{aux:3pt}) for some
operator ${\cal O}$ which appears in the OPE of $A$ and $B$. The corresponding $\dSO$, perhaps better
denoted by $Q_{A,B}({\cal O})$, will then obey a modified field equation which can be obtained by repeating
the logic around Eq.~(\ref{wave1}). However, now $L_i(x)$ and $L_i(y)$ are no longer purely geometric but also involve
an internal piece due to the non-scalar nature of $A$ and/or $B$. It is however less obvious how to generalize the
bulk description (\ref{gen:bulk}) to this case, nor whether $Q_{A,B}({\cal O})$ can be extended in any natural
way beyond this linearized approximation. In Euclidean signature, the bulk geodesic connecting $A$ and $B$ could be
understood as the leading classical trajectory for a scalar particle connecting $A$ and $B$. However, the discussion of geodesic Witten diagrams
\cite{Hijano:2015zsa} suggests that integral along the geodesic should be weighted by a measure depending on the difference in the conformal weights of the operators $A$ and $B$. Further, if $A$ and $B$ carry
non-trivial Lorentz representations, one should presumably use classical trajectories for particles transforming precisely
under those representations. Once again, there are many interesting directions to explore and we have presumably only
uncovered the tip of the iceberg. 

\subsection{Two dimensions} \label{sec.genO}

In the remainder of the paper, our examination will focus primarily on two-dimensional CFTs. Hence to set the stage for the subsequent sections, we will explicitly illustrate ideas appearing in the previous discussion of our nonlocal observables for two dimensions. We will also show that certain straightforward generalizations and simplifications emerge for $d=2$. The latter seem to be closely related to the fact that light-cone coordinates (or complex coordinates in Euclidean section) provide a preferred framework in which to describe two-dimensional CFTs, \eg conserved currents naturally split into independent left- or right-moving  components. The preferential role of null coordinates for $d=2$ was also reflected in the discussion of the geometry of kinematic space in section \ref{sec:geometry}. In particular, we found that in this case, the metric on the moduli space of causal diamonds factorized into two copies of the metric on two-dimensional de Sitter space when using null coordinates --- see discussion below Eq.~\reef{eq.LCdef}.

Hence to begin, consider a general quasi-primary operator $\mathcal{O}$ with conformal weights ($h$, $\bar{h}$) in a two-dimensional CFT. Now we adopt the null coordinates introduced in Eqs.~\reef{eq.LCdef} and \reef{nulle} and then we may define the following observable:
\be \label{genfirst}
\dSt = 
\frac{\Cbdy}{2}  \int_{u}^v d\xi \left( \frac{(v-\xi)(\xi-u)}{(v-u)} \right)^{h-1} 
 \int_{\bar{u}}^{\bar{v}} d\bar{\xi} \left( \frac{(\bar{v}-\bar{\xi})(\bar{\xi}-\bar{u})}{(\bar{v}-\bar{u})} \right)^{\bar{h}-1} 
 \langle {\cal O}(\xi,\bar{\xi}) \rangle \,,
\ee 
where, as in Eq.~\eqref{gen:firstlaw}, the integration runs over the entire causal diamond. For general operators with $h\ne \bar h$, this expression is the two-dimensional version of Eq.~\reef{gen:firstlaw2} with $\Delta_{\cal O}=h+\bar h$ and $\ell=h-\bar h$. In 
the spinless case with $h = \bar h = \Delta_{\cal O}/2$, this formula agrees with the $d=2$ version of Eq.~\eqref{gen:firstlaw}.

When $\cal O$ is a component of a conserved current then either $\bar{h} = 0$ in which  $\cal O$ depends only on $\xi$, or $h = 0$ and $\cal O$ only depends on $\bar{\xi}$. In either of these cases, one of the integrals in Eq.~\reef{genfirst} becomes `trivial' and contributes only an overall (divergent) factor. In this case,  Eq.~\reef{genfirst} reduces to
\be
\label{eq.deltaSObarh0}
\dSO {\big |}_{\bar{h} = 0} = \tCbdy \int_{u}^v d\xi \left( \frac{(v-\xi)(\xi-u)}{(v-u)} \right)^{h-1} \langle{\cal O}(\xi)\rangle\,.
\ee
for $\bar{h} = 0$  --- with an analogous expression for $h=0$. Note we had to redefine the normalization constant above since  the integral over $\bar \xi$ yields a divergent result in the limit $\bar h\to0$, \ie $\tCbdy=\left.\Cbdy/\bar{h}\right|_{\bar{h} \rightarrow 0} $. 

\begin{figure}[t]
\center
\includegraphics[scale=0.7]{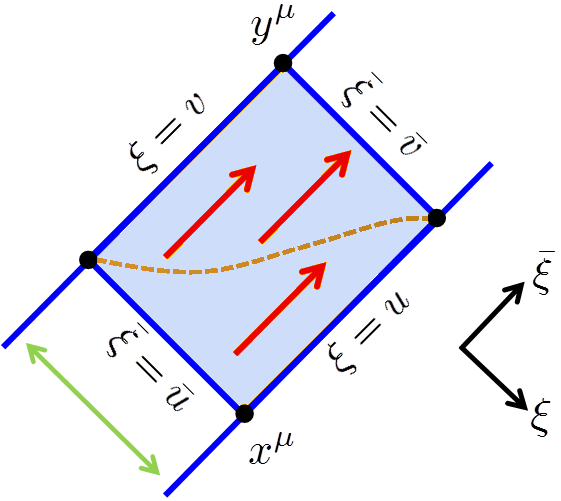}  
\caption{The nonlocal observable defined in Eq.~\reef{eq.deltaSObarh0} only involves right-moving modes and so is completely insensitive to the size and position of the causal diamond in the $\bar \xi$ direction.}    
\centering \label{holomorphic}
\end{figure}

Note that observables of the form given in Eq.~\reef{eq.deltaSObarh0} are completely independent of $\bar u$ and $\bar v$, \ie they are completely independent of the positions of the top-right and lower-left boundaries of the causal diamond in Figure \ref{holomorphic}. Since ${\cal O}(\xi)$ involves only right-moving modes, the nonlocal observable in Eq.~\reef{eq.deltaSObarh0} is only sensitive to expanding (or contracting) the causal diamond in the $\xi$ direction. In terms of the kinematic space, the result is completely independent of the position of the causal diamond in the second dS$_2$ factor in Eq.~\reef{eq.dS2dS2}, \ie the factor involving $\bar u$ and $\bar v$.

When the operator $\cal O$ in Eq.~\reef{eq.deltaSObarh0} is the right-moving component of the stress tensor $T_{\xi\xi}$ with $h=2$, we have
\be
\label{paste2}
Q(T_{\xi\xi}; v, u)  = \tCbdy \int_{u}^v d\xi \,\frac{(v-\xi)(\xi-u)}{(v-u)}\,  \langle T_{\xi\xi}\rangle\,.
\ee
Note that the above observable is completely independent of that constructed with the left-moving component $T_{\bar \xi\bar \xi}$, \ie independent of $Q(T_{\bar \xi \bar \xi}; \bar v, \bar u)$. The first law of entanglement \eqref{deltaS} (with $d=2$) actually corresponds to the sum of these two expressions
\be\label{paste3}  
\delta\see = Q(T_{\xi\xi}; v, u)  + Q(T_{\bar \xi \bar \xi}; \bar v, \bar u)\,,
\ee
along with the appropriate choice of the normalization factor, \ie we use $T_{tt}=T_{\xi\xi} + T_{\bar \xi \bar \xi}$ and $\tCbdy=2\pi$. 
Hence we observe that the entanglement entropy, at least in weakly excited states, naturally splits into right- and left-moving contributions. We will return to this important point  in the next section. 

\paragraph{Wave equations:} Following \cite{deBoer:2015kda}, it is immediate to see that Eq.~\eqref{genfirst} satisfies not just one but two wave equations on the kinematic space. This follows because the right- and left-moving parts of the integration kernel are both bulk-boundary propagators in two-dimensional de Sitter space, \ie in the separate dS$_2$ factors in Eq.~\eqref{eq.dS2dS2}. As a result, we find 
\be
\label{eq:deltaSOwave}
\begin{split}
 \left(\nabla^{2}_{{\rm dS}_2} -m_{{\rm dS}_2}^2\right) \dSO &= 0 \,,\qquad
 \text{with }\  m_{{\rm dS}_2}^2\, L^2 = - h(h-1) \,,\\
 \left(\overline{\nabla}^2_{\overline {\rm dS}_2} -\bar{m}_{\overline{\rm dS}_2}^2\right) \dSO &= 0 \,,\qquad
 \text{with }\ \bar{m}_{\overline{\rm dS}_2}^2\, L^2 = - \bar{h}(\bar{h}-1) \,,
\end{split}
\ee
where the d'Alembertians $\nabla^{2}_{{\rm dS}_2}$ and $\overline{\nabla}^2_{\overline {\rm dS}_2}$ only act on the copy of de Sitter space defined in Eq.\ \eqref{eq.dS2dS2} involving the right- and left-moving coordinates, respectively. 

Note that the wave equation \reef{gen:freefield} considered above for general dimensions corresponds to the sum of the two equations in Eq.~\reef{eq:deltaSOwave}, 
\be 
  \left( \nabla^2_\coset - m_{\cal O}^2 \right) \dSt = 2\left[\left( \nabla^{2}_{{\rm dS}_2}+ \overline{\nabla}^2_{\overline {\rm dS}_2}\right) - \left(m_{{\rm dS}_2}^2 +\bar{m}_{\overline{\rm dS}_2}^2\right) \right]\dSt = 0 \,, \label{slap1}
\ee
and hence $m_{\cal O}^2L^2=-2h(h-1)-2\bar h(\bar h-1)$.\footnote{Of course, we recover precisely the mass in Eq.~\reef{gen:freefield} upon substituting $\Delta_{\cal O}=h+\bar h$ and $\ell=h-\bar h$. Note that here we can see the reason why $m^2_{\cal O} L^2$ appears to be twice the mass that one would expect from the analysis on a fixed time-slice (as in \cite{deBoer:2015kda}): the mass on each copy of dS$_2$ in \eqref{eq:deltaSOwave} contributes {\it twice} (\ie with a factor of 2 in Eq.~\reef{slap1}) to the full mass $m^2_{\cal O}L^2$. } However, we see that Eq.~\reef{gen:freefield} is supplemented here by a second equation of the form
\be 
   \left[\nabla^{2}_{{\rm dS}_2}- \overline{\nabla}^2_{\overline {\rm dS}_2} - \Delta m^2  \right]\dSt = 0 
\qquad
 \text{with } \Delta m^2 L^2 =\bar h(\bar h-1) - h(h-1)\,.  \label{slap2}
\ee
We might note that in the case of a spinless equation (with $h=\bar h$), this second equation reduces to $\left( \nabla^{2}_{{\rm dS}_2}- \overline{\nabla}^2_{\overline {\rm dS}_2}\right)\dSO=0$. The importance of this constraint equation was emphasized in \cite{bartek66} and we return to this point in section \ref{sec:constraints}.

\paragraph{Relation to OPE:}
We wish to return to the arguments of section \ref{sec:OPE} in the special case of two dimensions, where we can make the necessary calculations more explicit. Let us consider the OPE of two general operators $A(u,\bar u)$ and $B(v,\bar v)$ inserted at the tips of the causal diamond, \eg see Figure \ref{holomorphic}. It is natural to write the analog of the resummation ansatz \eqref{eq:OPEansatz} for the OPE blocks in a factorized form for two dimensions: \be \label{eq:OPE}
A(u,\bar{u})\, B(v,\bar{v}) = \sum_i C^{{\cal O}_i}_{AB} \, \int_u^v d\xi\, \int_{\bar u}^{\bar v} d\bar{\xi} \; I_{AB{\cal O}_i}(u,v;\xi)\, \bar{I}_{AB{\cal O}_i}(\bar{u},\bar{v};\bar{\xi}) \,\mathcal{O}_i(\xi,\bar{\xi})\,.
\ee
As in general dimensions, the kernels $I_{AB{\cal O}_i}$ and $\bar{I}_{AB{\cal O}_i}$ are fixed by the global part of the conformal group. However, we can easily carry out this exercise in the case of $d=2$. Recall that for a quasi-primary $\mathcal{O}(z,\bar{z})$ we have
\be
[L_k,\mathcal{O}]=(k+1)\, h\, z^k \,\mathcal{O}+ z^{k+1}\,\partial_z \mathcal{O}\,,
\ee
where $L_{k}$ is the $k$-th element of the Virasoro algebra and $k=-1,0,1$ (for a primary this holds for all $k\in\mathbb Z$). Commuting $L_k$ through the OPE~\eqref{eq:OPE} therefore yields
\be
 \begin{split}
&\left[(k+1)h_A u^k + u^{k+1}\partial_u + (k+1) h_B v^k + v^{k+1} \partial_v\right] I_{AB{\cal O}_i}(u,v;\xi)  \\
 & \qquad\qquad = \left[(k+1)h_C \xi^k - \partial_\xi \xi^{k+1}\right] I_{AB{\cal O}_i}(u,v;\xi) \,.
 \end{split}
\ee
The solution to the three differential equations found by setting $k=-1,0,1$ in the above equation is
\be \label{eq.kernel}
I_{AB{\cal O}_i}(u,v;\xi) \sim \,(u-v)^{1-h_A-h_B-h_C} (u-\xi)^{-1-h_A+h_B+h_C} (\xi-v)^{-1+h_A-h_B+h_C}\, ,
\ee
where the symmetry analysis does not fix the normalization. Eq.~\eqref{eq.kernel} provides then the right-moving contribution. One can repeat the same argument for the left-moving factor and find the kernel $\bar{I}_{AB{\cal O}_i}(\bar{u},\bar{v},\bar{\xi})$ which has an analogous form. Let us add that having found a unique consistent solution of the Ward identities then justifies the validity of our factorized ansatz in Eq.~\reef{eq:OPE}.

Again, above we are only considering global conformal blocks, however, it would be interesting to extend this analysis to a resummation of entire Virasoro blocks.
The holographic description of such full conformal blocks has already been studied in \cite{Hijano:2015qja}.

\section{Interacting fields on $d=2$ moduli
space} \label{kintwo}

In the previous section, our considerations focused on new nonlocal observables $\dSO$ whose construction was motivated as a generalization of the  first law of entanglement \reef{deltaS}. A natural question then is whether for each observable $\dSO$, we can go beyond the linearized approximation implicit in the first law. That is, whether there is some nonlinear quantity equivalent to the full $\see$ in the first law, which can be defined for finite excitations away from the vacuum state. Of course, one would hope that such nonlinear quantities would still share at least some of the nice features of the corresponding observable $\dSO$, such as local 2-derivative equations of motion in the auxiliary space $\Mdiamonds$ or a natural appearance in the OPE. We start this line of investigation here by considering the full entanglement entropy $\see$ and asking whether there is a nonlinear extension of the wave equation \reef{black} for this quantity.

\subsection{Vacuum excitations}

When we apply the entanglement first law \reef{deltaS} as a diagnostic to characterize CFT states, \ie we apply the first law to all possible spheres, implicitly we are considering excited states which have a very small expectation value of the energy-momentum tensor everywhere. So our next step is to consider states where the expectation value $\langle T_{\mu\nu}\rangle$ may be finite. In particular, we will focus on two-dimensional CFTs, where an infinite number of excited states characterized only by the expectation value of the energy-momentum tensor can be obtained simply by acting with a local conformal transformation on the CFT vacuum on a plane.  The expectation value of the energy-momentum tensor for any of these states is given by the Schwarzian derivative, \ie with a local conformal transformation\footnote{For simplicity, we will continue to phrase our discussion here in terms of the null coordinates introduced in Eq.~\reef{eq.LCdef}. This is not entirely consistent with certain points in the following discussion where a Euclidean signature is implicit, \eg the path integral representation of ${\rm Tr} \rho_{\ssc A}^n$. 
However, one can easily Wick rotate to complex coordinates in Euclidean signature, \ie $z = x + i \, \tau$ and $\bar{z} = x - i \, \tau$ with $\tau=it$.}
\be
\label{eq.conftrafo}
w = f(z) \quad \mathrm{and} \quad \bar{w} = \bar f(\bar{z})\,,
\ee
we obtain right- and left-moving components of the stress tensor
\be
\label{eq.T}
\langle T_{zz} (z)\rangle  = \frac{c}{12} \left\{ \frac{f'''}{f'} - \frac{3 (f'')^{2}}{2 (f')^{2}}\right\} \quad \mathrm{and} \quad \langle T_{\bar z\bar z} (\bar{z}) \rangle = \frac{c}{12} \left\{ \frac{\bar{f}'''}{\bar{f}'} - \frac{3 (\bar{f}'')^{2}}{2 (\bar{f}')^{2}}\right\}.
\ee
Such functions $f(z)$ and $\bar f(\bar{z})$ may not exist globally, but we will ignore global issues in what follows,
in particular the fact that minimal surfaces can change discontinuously as, \eg in the BTZ black hole. Our discussion will
therefore only be valid for sufficiently short intervals or for excited states which are connected to the ground state
via a diffeomorphism. 

Now let us consider the entanglement entropy in the above states.\footnote{Further discussion of twist operators is presented in section \ref{discuss}.} First, we recall that the entanglement entropy can be evaluated using the replica trick, \eg \cite{Calabrese:2004eu,Doyon,Hung:2014npa,Hung:2011nu}: one begins with the R\'{e}nyi entropies
\be\label{eq:renyi}
S_n(A)=\frac{1}{1-n}\log {\rm Tr} \rho_{\ssc A}^n
\ee
which involves the reduced density matrix $\rho_{\ssc A}$ on the spatial domain $A$.
Here ${\rm Tr} \rho_{\ssc A}^n$ can be evaluated as a path integral on an $n$-fold cover of the original geometry on which the CFT lives. However an alternative description of this quantity comes in terms of twist fields, $\sigma_n$, which act in an $n$-fold replicated version of the CFT and implement twisted boundary conditions connecting the copies of the CFT along the entangling surface, \ie the boundary of $A$. In general dimensions then, the $\sigma_n$ are codimension-two surface operators with support on the boundary of $A$. However, two dimensions are special since $A$ will consist of a union of a set of disjoint intervals and the twist fields are local primary operators inserted the endpoints of each interval. In particular, 
in two-dimensional CFTs, the twist operators $\sigma_{\pm n}$ are local primaries with conformal weights \cite{Calabrese:2004eu,Doyon}
\be
\label{eq.htwists2}
h_{n} = \bar{h}_{n} = \frac{c}{24} \left(1 - \frac{1}{n^{2}} \right) \,.
\ee
Now we are interested in the $n\rightarrow 1$ limit in which the R\'{e}nyi entropy reduces to the entanglement entropy, \ie $\see=\lim_{n \rightarrow 1}S_n$.

In particular, the above discussion lets us relate the entanglement entropy of a single interval or of a single causal diamond to the two-point correlator of a pair of twist operators
\be 
\label{stormfront}
\see(w_1,w_2) =   \lim_{n \rightarrow 1}\, \frac{1}{1-n}\,\log\langle \sigma_n(w_1) \,\sigma_{-n}(w_2) \rangle \,,
\ee
where in the Minkowski vacuum of the CFT, the desired correlation function takes the form \cite{Calabrese:2004eu}
\be
 \langle \sigma_n(w_1)\, \sigma_{-n}(w_2) \rangle = \frac{a_{n}}{|w_1-w_2|^{2(h_n+h_{-n})}}
\ee
where $a_n$ is a numerical coefficient, with $a_{1} = 1$. Since the twist operators are primaries in a replicated CFT, this two-point function transforms  in a standard way under the conformal transformations~\eqref{eq.conftrafo}. Hence using Eqs.~\eqref{eq.htwists2} and \reef{stormfront}, the final result for the entanglement entropy in any of the above states~\eqref{eq.T} is given by
\be
\label{eq.S}
\see(u,\bar u;v, \bar v)= \frac{c}{12} \log{\frac{\left( f(u) - f(v) \right)^{2}}{\delta^{2} \, f'(u)f'(v)}} + \frac{c}{12} \log{\frac{\left( \bar{f}(\bar{u}) - \bar{f}(\bar{v}) \right)^{2}}{\delta^{2} \,\bar{f}'(\bar{u})\bar{f}'(\bar{v})}}\,,
\ee
where we introduced the short distance cut-off $\delta$. Further, ($u$, $\bar{u}$) and ($v$, $\bar{v}$) denote the tips of the causal diamond, as in figure \ref{holomorphic}, {\it after} the conformal transformation \eqref{eq.conftrafo}. As a simple example, if we choose $f(z)=z$ and $\bar{f}(\bar{z})  = \bar z$, we recover the standard result for a single interval
\be
\label{eq.Svac}
\see = \frac{c}{3}\, \log{\frac{| u - v |}{\delta}}\,,
\ee
where, however, the length of the interval is specified here by the distance between the past and future tips of the corresponding causal diamond.

A number of comments are in order here: First, Eq.~\eqref{eq.S} appeared already in \cite{Holzhey:1994we} and subsequently this approach was applied in \cite{Calabrese:2004eu} to obtain the entanglement entropy of a single interval both in a circular spatial domain and inin a thermal state (on an infinite line). Second, for holographic CFTs, Eq.~\eqref{eq.S} can be derived in full generality by using the Ryu-Takayanagi prescription and AdS$_{3}$ gravity --- see the discussion in the next section, as well as \cite{Roberts:2012aq}. Finally, we observe that the full entanglement entropy \eqref{eq.S} is the sum of independent contributions from the right- and left-moving modes 
\be
\see(u,\bar u;v, \bar v) = S_{\ssc R}(f;u,v) + S_{\ssc L}(\bar{f};\bar{u}, \bar{v})
\ee
with
\be
\label{eq.Sf}
S_{\ssc R}(f;u,v) = \frac{c}{12} \log{\frac{\left( f(u) - f(v) \right)^{2}}{\delta^{2} \, f'(u)f'(v)}}  \quad \mathrm{and} \quad 
S_{\ssc L}(\bar{f};\bar{u}, \bar{v})=\frac{c}{12} \log{\frac{\left( \bar{f}(\bar{u}) - \bar{f}(\bar{v}) \right)^{2}}{\delta^{2} \,\bar{f}'(\bar{u})\bar{f}'(\bar{v})}}\,.
\ee
That is, we interpret $S_{\ssc R}(f;u,v)$ as the contribution to the entanglement entropy for the right-moving modes in the state generated by the conformal transformation $w=f(z)$ since it is only sensitive to variations in the width of the causal diamond in the $\xi$ direction, as illustrated in figure \ref{holomorphic}. Of course, $S_{\ssc L}(\bar{f};\bar{u}, \bar{v})$ is the analogous contribution from the left-movers. Recall that the same split for $\delta\see$ in Eq.~\reef{paste3}. There it resulted since `holomorphicity' was a general feature of the observables constructed with conserved currents, \eg the stress tensor, as in Eq.~\reef{eq.deltaSObarh0}. With our analysis here, we see that this splitting survives at nonlinear level for the entanglement entropy.  

Furthermore, the right-moving contribution $S_{\ssc R}(f)$ turns out to solve the Liouville equation in the form
\be
\label{eq.liouville}
\delta^{2}\,\frac{\partial^{2}}{\partial u \, \partial v} S_{\ssc R}(f) =\frac{c}{6} \exp{\left(-\frac{12}{c}\, S_{\ssc R}(f) \right)} \,,
\ee 
and similarly with $S_{\ssc L}(\bar{f})$. To relate this result to our previous wave equations, we turn back to our motivation which was to find a nonlinear generalization of $\delta\see$, or rather of $Q(T_{zz}; v, u)$ in Eq.~\reef{paste3}. Having right- and left-moving contributions to the entanglement entropy, which are still sensibly defined in Eq.~\reef{eq.Sf} for states with finite energy densities, it is natural to define
\be
\DSR(u,v) =  S_{\ssc R}(f;u,v)\ -\  S_{\ssc R}(f(z)\!=\!z;u,v)\,.\label{rmove}
\ee
That is, we consider the finite difference between the right-moving contribution to the entanglement entropy for intervals in the state generated by $w=f(z)$ and that in the original vacuum state. Of course, there is a completely analogous difference $\DSL$ for the left-moving contribution. Note that these differences are UV finite and independent of the short distance cut-off $\delta$. Further,
for an infinitesimal conformal transformation \reef{eq.conftrafo}, \eg $f(z) = z + \epsilon\,g(z)$ and using Eq.~\reef{eq.T}, one can confirm that to leading order in $\epsilon$,
\be
\DSR(u,v) = Q(T_{zz}; v, u) + O(\epsilon^2)\,,
\label{horseD}
\ee
with $\tCbdy=2\pi$. That is, to leading order, the finite difference observable \reef{rmove} matches the linearized observable \reef{paste2} which yields the right-moving contributions to $\delta\see$ in the first law \reef{paste3}.

Quite remarkably, Eq.~\reef{eq.liouville} yields a local equation of motion for both $\DSR$ and $\DSL$ on, respectively, the right- and left-moving de Sitter factors in the kinematic space for two-dimensional CFTs, \eg see Eq.~\eqref{eq.dS2dS2},
\be\label{Dwave}
\nabla^{2}_{{\rm dS}_2}\, \DSR - V'(\DSR) = 0 \quad \mathrm{and} \quad \overline{\nabla}_{\overline{\rm dS}_2}^{2}\, \DSL - V'(\DSL) = 0\,,
\ee
where the nonlinear potential is given by
\be
V'(s) = - \frac{c}{6} \left[ 1 - \exp{\left( - \frac{12}{c} s \right)} \right] =  - 2 s + \frac{12}{c} s^{2} + \ldots\,. \label{gobb}
\ee
Hence if we linearize the wave equation \reef{Dwave} for, \eg $\DSR$, then the term $-2s$ is the above expression corresponds precisely to the mass term (with $L^2=1$) for $h = 2$ and $\bar{h} = 0$ needed to reproduce  Eq.~\eqref{eq:deltaSOwave}. Of course, we also recover the desired linearized wave equation for $\DSL$. Implicitly, we also have the equations
\be\label{Dwave2}
\overline{\nabla}_{\overline{\rm dS}_2}^{2}\, \DSR  = 0
\quad \mathrm{and} \quad
\nabla^{2}_{{\rm dS}_2}\, \DSL= 0\,,
\ee
which remain unchanged from their linearized counterparts.
Further, we might note that the higher order interactions in Eq.~\reef{gobb} are suppressed by inverse powers of the central charge at large $c$. Finally, notice that we have written the wave equations \reef{Dwave} as though they resulted from the variation of an action. In particular then, the underlying potential would be 
$V(s)=-\frac c6\,s -\frac 12 \exp(-12s/c)$, which has a single unstable extremum at $s=0$.

Our results here demonstrate that at least for the universal family of states characterized by~\eqref{eq.T}, the spatiotemporal organization of the entanglement entropy, or more precisely of its right- and left-moving contributions over the vacuum entanglement entropy, is governed by the Lorentzian structure of the moduli space of causal diamonds \eqref{eq.dS2dS2} and obeys a nonlinear and local propagation law on this space.

\subsection{Beyond vacuum excitations} \label{beyond}

In Eq.~\reef{rmove}, the vacuum state was chosen as the reference state in defining $\DSR$ and $\DSL$. However, this choice was arbitrary and we could easily consider the difference  
\be
\tDSR(u,v) =  S_{\ssc R}(f;u,v)\ -\  S_{\ssc R}(f_0;u,v)\,,\label{rmove2}
\ee
where the reference state is that generated from the vacuum by the conformal transformation $w=f_0(z)$. In this case, the corresponding wave equation derived from Eq.~\reef{eq.liouville} becomes
\be
\label{slugger}
\frac{\left( f_0(u) - f_0(v) \right)^{2}}{ f_0'(u)f_0'(v)} \,\frac{\partial^{2}}{\partial u \, \partial v} \tDSR -V'(\tDSR) = 0\,,
\ee 
with the same potential as in Eq.~\reef{gobb}. Inspecting the derivative term, we recognize this as the Laplacian on two-dimensional de Sitter space in transformed coordinates. That is, if we begin with dS$_2$ with null coordinates ($u_0$,$v_0$) --- \eg, see Eq.~\reef{eq.dS2dS2} ---  then the coordinate transformation $u_0=f_0(u)$ and $v_0=f_0(v)$ yields\footnote{Again, we have set $L^2=1$ here.}
\be
\label{transform}
ds^2_{{\rm dS}_2}=\frac{4\,du_0\,dv_0}{(u_0-v_0)^2}=\frac{ 4\,f_0'(u)f_0'(v)\,du\,dv}{\left( f_0(u) - f_0(v) \right)^{2}} \,.
\ee
Of course, the analogous discussion applies for the left-moving contribution with $\bar w=\bar f_0(\bar z)$ defining the reference state. Hence, with a new choice of reference state, one recovers precisely the same nonlinear wave equations and the only change is a coordinate transformation on the corresponding dS$_2$ geometry.
We should note that we are only examining the local geometry of the moduli space here and we have not concerned ourselves with any subtleties that may arise at the global level. 

The above results connect directly to the recent work of \cite{Asplund:2016koz}. They examined the linearized propagation of entanglement excitations in various nontrivial states, \eg for finite temperature or for finite spatial periodicity. Of course, both of these states in the CFT can be generated with an appropriate conformal transformation \cite{Calabrese:2004eu}, \eg $w=\exp(2\pi z/\beta)$ generates the thermal state from the flat space vacuum.\footnote{In this case, we are thinking of a conformal transformation of Euclidean space.} Hence these states fall into the universal class of states studied above and as described above, the auxiliary geometry describing the appropriate moduli space for these two examples will again be the direct product of two de Sitter spaces. Upon restricting to a fixed time slice as in \cite{Asplund:2016koz}, one then finds that $\delta \see$ propagates on the diagonal dS$_2$.

A natural interpretation emerging from the above discussion is that  $S_{\ssc R}(f)$ and $S_{\ssc L}(\bar{f})$ are proportional to conformal factors in the corresponding de Sitter factors of the kinematic space metric \reef{eq.dS2dS2}, \ie
\be
\label{eq.dsSfSbarf}
ds^{2}_\lozenge = 4 \exp\!\left( - \frac{12}{c} S_{\ssc R}(f;u,v) \right) du\, dv + 4 \exp\!\left( - \frac{12}{c} S_{\ssc L}(\bar{f};\bar u,\bar v) \right) d\bar{u}\, d\bar{v}\,.
\ee
This expression for the metric then directly connects the independent diffeomorphisms on each of the two-dimensional de Sitter spaces with the  conformal transformations \eqref{eq.conftrafo} in the CFT, as discussed above. In this interpretation, the short distance cut-off $\delta$ sets the curvature scale for each of the dS$_2$ geometries:
\be
{\cal R}_{{\rm dS}_2} = \frac{2}{\delta^{2}}\,,
\ee
where ${\cal R}_{{\rm dS}_2}$ is the Ricci scalar for each of the de Sitter geometries. Hence this approach makes a definite choice of the curvature of the moduli space, \ie $L=\delta$. Evaluating the above constant curvature condition on the right-moving part of the ansatz~\eqref{eq.dsSfSbarf} yields precisely the Liouville equation~\eqref{eq.liouville}. While this approach is very peculiar to two dimensions, the interpretation of the wave equation on the moduli space of causal diamonds as a constant curvature condition might well be more general. Indeed, as we discuss in section \ref{discuss}, a field equation identical to the one obeyed by $\delta \see$ can be derived in arbitrary dimensions by observing that $\Mdiamonds$ has constant scalar curvature.

\section{More interacting fields on $d=2$ moduli
space: higher spin case}
\label{sec.hs}

In the previous section, we have seen that entanglement entropy obeys nontrivial and local field equations on the moduli space which has the dS$_2\times$dS$_2$ geometry. In fact, the `fields' on the moduli space were right- and left-moving contributions to the entanglement entropy, \eg as defined in Eq.~\reef{rmove}. These results explicitly applied for a universal family of states, which were generated by a conformal transformation acting on the flat space vacuum or alternatively by exciting the vacuum state by the action of the stress tensor alone.
Moreover, the field equations \reef{Dwave}, as well as \reef{Dwave2}, are independent of the precise reference state. Rather all
dependence on the latter is encoded in a choice of coordinates, or implicitly on the form of the boundary conditions, on dS$_2\times$dS$_2$. In section \ref{sec:linearized}, we introduced the idea that in more general states where new operators (other than the stress tensor) acquire an expectation value, one would have to expand the discussion to consider new fields on the moduli space to account for these operators.
In this section, we wish to provide an explicit example of this generalization in the context of $d=2$ CFTs with higher spin symmetries. In particular, we will consider a theory with a conserved spin-three current. At the linearized level, we have seen  that there are new nonlocal observables \reef{eq.deltaSObarh0} associated with the right- and left-moving components of this current and that these satisfy linearized wave equations \reef{eq:deltaSOwave} on the moduli space. In the following, we will demonstrate that the latter extend to nonlinear equations where the fields corresponding to the spin-three current and to the entanglement entropy develop local interactions with each other on dS$_2\times$dS$_2$.

More specifically, we will consider a theory with only spin-two and spin-three fields, which can be described by a
$SL(3,\mathbb R)\times SL(3,\mathbb R)$ Chern-Simons theory (for a review with an emphasis on black hole solutions see
\eg \cite{Ammon:2012wc}). The most general solution of the field equations is described
by a flat gauge field subject to suitable boundary conditions, the latter encoding the expectation values of the stress-tensor
and spin-three currents in the dual CFT. In keeping with the general philosophy of this paper, we would like to probe
such backgrounds both with ordinary entanglement entropy as well as with a spin-three generalization thereof. The existence
and definition of such a generalization of entanglement entropy 
was proposed in \cite{Hijano:2014sqa} and some additional features were discussed in \cite{deBoer:2014sna}.
The spin-three entanglement entropy of \cite{Hijano:2014sqa} can be viewed as a generalization of the expressions for
ordinary entanglement entropy in higher spin theories in terms of Wilson lines originally proposed in 
\cite{Ammon:2013hba,deBoer:2013vca}. These two proposals were shown to be equivalent in \cite{Castro:2014mza} and were
tested against CFT computations in \cite{Datta:2014ska,Datta:2014uxa}, for more recent work see \cite{Chen:2016uvu}
and references therein.
In this section we will consider theories holographically dual to classical Chern-Simons theory, \ie we assume a large central charge (equivalently, a large Chern-Simons level $k$). 

We will be interested in computing the entanglement entropy in nontrivial states in Lorentzian signature and in particular
we will not be turning on any chemical potentials or any sources for the higher spin currents. In the presence of such
sources there are different types of boundary conditions depending on whether one takes a Lagrangian or Hamiltonian
point of view \cite{deBoer:2014fra} but we will not have to worry about this issue. Thanks to this the proposals of 
\cite{Ammon:2013hba,deBoer:2013vca} and \cite{Hijano:2014sqa} can be phrased as follows. 
In Chern-Simons theory one has two gauge fields $A$ and $\bar{A}$, one for each copy of the gauge group.
Entanglement entropy for the interval with endpoints $P,Q$ is computed by constructing the open Wilson loop
$\cW(P,Q)$ from $P$ to $Q$ for $A$, the open Wilson loop $\bar{\cW}(Q,P)$ from $Q$ to $P$ for $\bar{A}$, and then to evaluate
\be
S_R(P,Q)=c_R \log {\rm Tr}_R (\bar{\cW}(Q,P)\, \cW(P,Q))
\ee
with a suitable normalization and in a suitable representation $R$. Depending on the choice of representation $R$, 
different types of entanglement can be computed, as we will see below. Standard entanglement entropy is for example
obtained by taking $R$ to be the fundamental representation for pure gravity constructed with $SL(2,\mathbb R)\times SL(2,\mathbb R)$, and the adjoint representation for
the spin-three theory based on $SL(3,\mathbb R)\times SL(3,\mathbb R)$.

\subsection{Evaluation of Wilson loops}

The boundary condition on the Chern-Simons gauge fields $A$, $\bar{A}$ 
states that these should be gauge transforms of a suitable two-dimensional gauge field with a purely radial gauge
transformation. It is more convenient to write everything purely in terms of 2d data and write the radial dependence
explicitly. So from now on the open Wilson loops and gauge fields will be 2d and not 3d, and $S_R(P,Q)$ becomes
\be
S_R(P,Q)=c_R  \log {\rm Tr}_R (\bar{\cW}(Q,P) e^{\rho\Lambda_0} \cW(P,Q)  e^{-2\rho \Lambda_0} )
\ee
Here $\Lambda_0$ is the diagonal element of a $sl(2,\mathbb R)$ subalgebra of the gauge group. This $sl(2,\mathbb R)$
subalgebra (given by a choice of an embedding of $sl(2,\mathbb R)$ in the gauge group) is what sets the boundary conditions
for Chern-Simons theory and is also what determines the precise nature of the higher-spin symmetry of the dual CFT. 
Each inequivalent choice of $sl(2,\mathbb R)$ embedding describes a different higher spin theory. There is a preferred
$sl(2,\mathbb R)$ embedding in $sl(N,\mathbb R)$, the so-called principal embedding,
for which the fundamental representation of $sl(N,\mathbb R)$ is an irreducible representation of $sl(2,\mathbb R)$.
These give rise to the standard ${\cal W}_N$-algebras with one generator of spins $2,\ldots, N$ each, and this
is for $N=3$ the case that we will study.  

To regulate the divergent quantity $S_R(P,Q)$, we need to pick
a fixed and large value of $\rho$. 
The two-dimensional gauge fields are of the following form, where, as discussed above, we assume no sources have been turned on:
\be \label{dsgauge}
A=(\Lambda_+ + U(x^+))dx^+ ,\qquad \bar{A}=(\Lambda_- + \bar{U}(x^-))dx^-\, .
\ee
Here, $\Lambda_-,\Lambda_+$ together with $\Lambda_0$ are the three generators of the $sl(2,\mathbb R)$ subalgebra,
and $U(x^+)$ is short-hand notation for $\sum U_i(x^+) T_i$ where the sum runs over all generators $T_i$ of $sl(N,\mathbb R)$
which obey $[\Lambda_-,T_i]=0$. Similarly, $\bar{U}(x^-)=\sum_i \bar{U}_i(x^-)\bar{T}_i$ with 
$[\Lambda_+,\bar{T}_i]=0$. The particular form of the gauge fields in Eq.~(\ref{dsgauge}) is sometimes referred to
as Drinfeld-Sokolov gauge and the explicit form for the principal embedding in $sl(3,\mathbb R)$ will be given below.
In Eq.~(\ref{dsgauge}), $U$ and $\bar{U}$ contain the expectation values of the right- and left-moving conserved currents, respectively.

Since $A$ and $\bar{A}$ are flat, they are locally pure gauge\footnote{This need not be the case globally, because
the gauge field can have a non-trivial monodromy around the spatial circle. For example, this is  the case
for the higher spin analogues of conical defect and black hole geometries. Therefore, our analysis is strictly speaking 
restricted to sufficiently small intervals on the boundary if the spatial geometry is a circle, but should be valid
for arbitrarily large intervals in the planar case. It would be interesting to explore global aspects and possible
consequences for the field equations on dS${}_2\times$dS${}_2$ in more detail.}, $A=g^{-1}dg$ and ${\bar{A}}=\bar{g}^{-1}d\bar{g}$,
and 
\be
S_R(P,Q)= c_R \log {\rm Tr}_R ( \bar{g}^{-1}(P) \bar{g}(Q) e^{2\rho\Lambda_0} g^{-1}(Q) g(P)  e^{-2\rho \Lambda_0} )
\ee
We will now assume that when we decompose the representation $R$ in eigenvectors of $\Lambda^0$ there are unique
eigenvectors with a smallest and largest eigenvalue, and we will refer to these highest and lowest weight vectors as
$|\mu^+\rangle$ and $|\mu^-\rangle$. This assumption will hold in the cases that we will consider, and if it holds
the dominant contribution to $S_R(P,Q)$ is coming
from the matrix elements where we pick up the largest powers of $e^{\rho}$, which is precisely given by
the matrix elements between the highest and lowest weights. The dominant contribution is then
\be
S_R(P,Q) =  c_R \log \left[ e^{2\rho\eta} \langle \mu^- |  \bar{g}^{-1}(P) \bar{g}(Q) | \mu^+ \rangle 
          \langle \mu^+ |   g^{-1}(Q) g(P)     | \mu^- \rangle  \right]
\ee
with $\eta=\langle\mu^+|\Lambda_0|\mu^+\rangle-\langle\mu^-|\Lambda_0|\mu^- \rangle$. If we denote $\epsilon_R=e^{-\eta\rho}$
where $\rho$ is a fixed and large cutoff, we find that $S_R(P,Q)$ can be written as
\be\label{slap5}
S_R(P,Q) =  c_R \log \left[ \frac{ \langle \mu^- |  \bar{g}^{-1}(P) \bar{g}(Q) | \mu^+ \rangle }{\epsilon_R} \right] +
c_R \log\left[ \frac{
          \langle \mu^+ |   g^{-1}(Q) g(P)     | \mu^- \rangle  }{\epsilon_R} \right].
\ee
Hence we see that the result splits into separate right- and left-moving contributions.  Our experience in the previous section suggests that these separate terms will form a interesting basis to develop a local field theory on the moduli space.
Hence from now on, we will restrict to the right-moving contribution to $S_R(P,Q)$, 
which we denote by ${\cal S}_R(P,Q)$
\be \label{def:Sr}
{\cal S}_R(P,Q) \equiv c_R \log \left[ 
          \frac{\langle \mu^+ |   g^{-1}(Q) g(P)     | \mu^- \rangle}{\epsilon_R} \right]
\ee
which only depends on the right-moving coordinates $P^+$ and $Q^+$ and therefore naturally defines a function
on one of the two-dimensional de Sitter spaces. 

\subsection{Pure gravity example}

It is instructive to see how this works in the pure gravity case. There the two-dimensional gauge field is of the form
\be
A=\left( \begin{array}{cc} 0 & 1 \\ T(x^+) & 0 \end{array} \right)\,,
\ee
where $T(x^+)$ is the right-moving component of the stress tensor. Now we can write $A=g^{-1}\partial_+ g$ with
\be
g = \left( \begin{array}{cc} \partial_+\! \!\left(\frac{1}{\sqrt{\partial_+ f}}\right) & \ \frac{1}{\sqrt{\partial_+ f}} 
\\ \partial_+\!\!\left(\frac{f}{\sqrt{\partial_+f }} \right) &\  \frac{f}{\sqrt{\partial_+ f}} 
 \end{array}
\right)\,,
\ee
which indeed has the property that
\be
g^{-1}\partial_+ g = \left( \begin{array}{cc} 0 & 1 \\ T & 0 \end{array} \right)
\ee
and where, not surprisingly, $T$ is expressed as a Schwarzian derivative
\be
T=-\frac{1}{2} \{ f(x^+),x^+ \}  \,.
\ee
In the pure gravity case, the embedding of $sl(2,\mathbb R)$ in $sl(2,\mathbb R)$ is 
given by the identity map and in particular
\be
\Lambda_0=\left( \begin{array}{cc} 1 & 0 \\ 0 & -1 \end{array} \right)\,,\qquad
|\mu^+\rangle=\left( \begin{array}{c} 1 \\ 0 \end{array} \right)\quad
{\rm and}\quad
|\mu^-\rangle=\left( \begin{array}{c} 0 \\ 1 \end{array} \right)
\ee
where the highest and lowest weight states above correspond to $R$ being the fundamental representation.
To compute ${\cal S}_\mt{fun}(P,Q)$ in this case we therefore only need the 
$12$-matrix element of $g^{-1}(Q) g(P)$ and we get
\be\label{game8}
{\cal S}_\mt{fun}(P,Q) = c_\mt{fun} \log\frac{\left[ g^{-1}(Q)\, g(P) \right]_{12}}{\epsilon_\mt{fun}} = c_\mt{fun} \log \left[\frac{f(P^+)-f(Q^+)}{\epsilon_\mt{fun}\sqrt{\partial f(P^+) \,\partial f(Q^+)} } \right]
\ee
which is indeed in precise agreement with the results (\ref{eq.Sf}) obtained in the previous section, when we identify $c_\mt{fun}=c/6$, as well as $\epsilon_\mt{fun}=\delta$.

\subsection{Spin-three entanglement entropy}

We would now like to consider the spin-three case with the principal embedding, in which case the right-moving two-dimensional gauge field
takes the form
\be \label{a3}
A=\left(\begin{array}{ccc} 0 & 1 & 0 \\ T(x^+) & 0 & 1 \\ W(x^+) & T(x^+) & 0
\end{array}\right)dx^+\,,
\ee
where $T(x^+)$ and $W(x^+)$ are the right-moving components of the stress tensor and the spin-three current, respectively. Once again, we need to find a $g$ which obeys $g^{-1}dg=A$. Such a $g$ can be parametrized in terms of two
functions $\gamma_1$ and $\gamma_2$, but the equations are quite a bit more cumbersome compared to the
pure gravity case. To write $g$, it is convenient to first define
\be
\chi_1 = \frac{1}{(\gamma_1' \gamma_2'' - \gamma_2' \gamma_1'')^{1/3} }
,\qquad \chi_2 = \gamma_1 \chi_1, \qquad \chi_3=\gamma_2 \chi_1
\ee
and to parametrize $g$ as
\be
g=\left(\begin{array}{ccc}  \partial^2 \chi_1 - \theta \chi_1 & \partial \chi_1 & \chi_1 \\
  \partial^2 \chi_2 - \theta \chi_2 & \partial \chi_2 & \chi_2 \\
  \partial^2 \chi_3 - \theta \chi_3 & \partial \chi_3 & \chi_3 \end{array}\right)\,.
\ee
One can explicitly show that with this choice of $g$
\be
g^{-1}dg = \left(\begin{array}{ccc} 0 & 1 & 0 \\ T_1 & 0 & 1 \\ W & T_2 & 0\,,
\end{array}\right)
\ee
where $T_1$, $T_2$ and $W$ are lengthy expressions in terms of $\gamma_1$, $\gamma_2$ and $\theta$,
which can be viewed as generalizations of the Schwarzian derivative to the spin-three case. 
For a suitable choice of $\theta$ in terms of $\gamma_1$ and $\gamma_2$, which one
can algebraically determine, one gets $T_1=T_2$. We will however not need the explicit form of $\theta$
in what follows.

As an aside, we notice that there is an interesting 
action of $SL(3,\mathbb R)$ on $\gamma_1$ and $\gamma_2$, which follows from $g\rightarrow\epsilon g$,
and which leaves $T$ and $W$ invariant. It takes the form
\be
\gamma_1' = \frac{a + b\gamma_1 +c \gamma_2}{g + h\gamma_1 + i \gamma_2},\qquad
\gamma_2' = \frac{d + e\gamma_1 +f \gamma_2}{g + h\gamma_1 + i \gamma_2},\qquad
\ee
which is a direct generalization of the standard $SL(2,\mathbb R)$ action in the pure gravity case and which
presumable plays some sort of role in `W-geometry'.

With the explicit form of $g$ at hand we can now evaluate ${\cal S}_R(P,Q)$ for various
choices of representations $R$. From \cite{Ammon:2013hba,deBoer:2013vca} we know that
ordinary entanglement entropy is obtained by taking $R$ to be the adjoint representation. 
It is, however, a priori less clear which representation one should take in order to get
the spin-three generalization of entanglement entropy. We claim that the right quantity
(up to an overall normalization)
is obtained by taking a linear combination of the fundamental and adjoint representations
\be \label{ee3}
\see^{(3)} \sim {\cal S}_\mt{fun}-\frac{1}{2} {\cal S}_\mt{adj}\,,
\ee
where we put the normalization constants $c_\mt{fun}=c_\mt{adj}$. 

To see that Eq.~(\ref{ee3}) is the right quantity one can either consider its expansion around the
vacuum to first order in $\langle T(x^+)\rangle$ and $\langle W(x^+)\rangle$, and verify that it produces an observable of precisely the form in Eq.~(\ref{eq.deltaSObarh0}) with $h=3$ --- see also below. Alternatively, one can translate the original 
proposal of \cite{Hijano:2014sqa} (see also appendix~B of \cite{deBoer:2014sna})
in the present language and also arrive at Eq.~(\ref{ee3}). The relation of these papers to Eq.~(\ref{ee3})
can be summarized as follows: the highest weight of the fundamental representation minus 
one half the highest weight of
the adjoint representation is proportional to the $sl(3,\mathbb R)$ generator 
\be
U_0 = \frac{1}{3} \left( \begin{array}{ccc} 1 & 0 & 0 \\ 0 & -2 & 0 \\ 0 & 0 & 1 \end{array}\right).
\ee
which is precisely the generators used in the construction of \cite{Hijano:2014sqa}.

We notice that if we decompose the adjoint representation of $SL(3,\mathbb R)$ with respect to the $SL(2,\mathbb R)$ subgroup,
we obtain a three- and a five-dimensional representation which contain $T(x^+)$ and $W(x^+)$ as lowest weight
respectively. The Cartan generator of the five-dimensional representation is precisely $U_0$. This suggests
that to obtain a higher spin entropy in more general cases we should take linear combination of ${\cal S}_R$ 
with various representations $R$ in such a way the the corresponding highest weight is proportional to a Cartan generator
which is part of the same $SL(2,\mathbb R)$ representation as a particular higher-spin generator.

We are thus led to consider the following two quantities
\begin{eqnarray}
\see^{(2)} & = &  {\cal S}_\mt{adj} \,,\labell{deaf77} \\
\see^{(3)} & = &  {\cal S}_\mt{fun}-\frac{1}{2} {\cal S}_\mt{adj} \,,\nonumber
\end{eqnarray}
which we refer to as the spin-two and spin-three `entanglement entropies', respectively. That is, $\see^{(2)}$ is proportional to the ordinary entanglement entropy while $\see^{(3)}$ defines a new nonlinear observable related to the spin-three current, \ie it vanishes in states where $\langle W(x^+)\rangle=0$. We also note that the short distance regulators cancel out in our definition of $\see^{(3)}$ since $\epsilon_\mt{adj}=\epsilon_\mt{fun}^{\,2}$
--- see the definition of $\epsilon_R$ above Eq.~\reef{slap5}.  Hence our spin-three entanglement entropy is a completely UV finite observable.

Note that both expressions \reef{deaf77} carry an overall factor of the normalization constant $c_\mt{fun}$ since as above, we set $c_\mt{fun}=c_\mt{adj}$. However, we have not
fixed the precise normalization of $\see^{(2)}$ and $\see^{(3)}$ since we are mostly interested in the question whether
these obey local field equations on dS${}_2$ or not. It is in principle straightforward, by examining the first laws
for $\see^{(2)}$ and $\see^{(3)}$ and using the known relation between the generators in Eq.~(\ref{a3}) and the spin-two
and spin-three generators with their canonical normalization to determine the precise normalization factors.

\subsection{de Sitter field equations for higher spin entanglement entropy}

We can now explicitly compute $\see^{(2)}$ and $\see^{(3)}$ in the most general spin-two and spin-three background.
To simplify the final answers, we will denote $u=P^+$ and $v=Q^+$, and define
\bea
\Sigma_1 & = & (\gamma_2(u)-\gamma_2(v) ) \gamma_1'(u) - (\gamma_1(u)-\gamma_1(v) ) \gamma_2'(u)\,,
\nonumber \\
\Sigma_2 & = & (\gamma_2(u)-\gamma_2(v) ) \gamma_1'(v) - (\gamma_1(u)-\gamma_1(v) ) \gamma_2'(v)  \,,
\labell{killjoy0} \\
\Phi_1 & = & (\gamma_1'(u)\gamma_2''(u)-\gamma_2'(u)\gamma_1''(u))\,,
\nonumber \\
\Phi_2 & = & (\gamma_1'(v)\gamma_2''(v)-\gamma_2'(v)\gamma_1''(v))\,,
\nonumber
\eea
and with these we find that
\be\label{killjoy}
\see^{(2)} = c_\mt{fun}\,\log\left(-\frac{\Sigma_1\,\Sigma_2}{\epsilon^{\,2}_\mt{fun}\,\Phi_1\, \Phi_2}\right)\quad\ \ {\rm and}
\quad\ \ 
\see^{(3)} = \frac{c_\mt{fun}}6\,\log\left(-\frac{\Sigma_2^{\,3}\,\Phi_1}{\Sigma_1^{\,3}\, \Phi_2}\right).
\ee
Interestingly, these quantities obey the following local field equations, 
\be\label{toda}
\begin{split}
 \epsilon_\mt{fun}\,\frac{\partial \see^{(2)}}{\partial u \partial v} &= 2\,c_\mt{fun}\, e^{-\see^{(2)}/(2c_\mt{fun})} \cosh\left(3\see^{(3)}/c_\mt{fun}\right) \,,\\
 \epsilon_\mt{fun}\,\frac{\partial \see^{(3)}}{\partial u \partial v} &=  -c_\mt{fun}\,e^{-\see^{(2)}/(2c_\mt{fun})}\, \sinh\left(3\see^{(3)}/c_\mt{fun}\right) \,.
\end{split}
\ee
These equations are our spin-three extension of the Liouville equation appearing in Eq.~\reef{eq.liouville} in the previous section. In particular, they reduce to \reef{eq.liouville} if we set $\see^{(3)}=0$ and identify $\see^{(2)}$ with $S_{\ssc R}$. In passing, we observe that these equations \reef{toda}  are identical to the so-called Toda equations for $SL(3,\mathbb R)$ (for a summary of some
aspects of Toda theory and further references see, \eg \cite{Fateev:2007ab})
and reserve further comments for later. 

To see the de Sitter geometry of the kinematic space emerge in the field equations, we follow our previous approach in Eq.~\reef{rmove} and consider the following difference  
\be
\Delta \see^{(2)}=\see^{(2)}-\left.\see^{(2)}\right|_{\gamma_1(z)=z;\,\gamma_2(z)=z^2}\,.\label{extra1}
\ee
The second term corresponds to the usual vacuum entanglement entropy, \ie one can easily verify from Eqs.~\reef{killjoy0} and \reef{killjoy} that 
\be
\left.\see^{(2)}\right|_{\gamma_1(z)=z;\,\gamma_2(z)=z^2}=2\,c_\mt{fun}\,\log\frac{(u-v)^2}{2\,\epsilon_\mt{fun}}
\label{extra2}
\ee
and further that $\see^{(3)}=0$ with this choice of $\gamma_1$ and $\gamma_2$.
We can match this result with the expected flat space entanglement entropy \reef{eq.Svac} (more precisely, the right-moving contribution \reef{eq.Sf}) with $c_\mt{fun}=c/24$ and $\epsilon_\mt{fun}=\delta^2/2$.\footnote{Recall that below Eq.~\reef{game8}, we found $c_\mt{fun}=c/6$. However, note that in the conventions implicit in our calculations, the normalization factor $c_R$ for the representation $R$ which yields the ordinary entanglement entropy (\ie the fundamental representation for $SL(2,\mathbb R)$ and the adjoint  representation for $SL(3,\mathbb R)$) is given by the level $k$ of the Chern-Simons theory. Further, the relation between $c$ and $k$ is different for different gauge groups, \ie for $SL(2,\mathbb R)$, $c=6k$ and for $SL(3,\mathbb R)$, $c=24k$. Combined with our choice that $c_\mt{fun}=c_\mt{adj}$ here, this gives a precise explanation of the two different values found for $c_\mt{fun}$.} Now in terms of $\Delta \see^{(2)}$ and $\Delta \see^{(3)}= \see^{(3)}$, the field equations become
\be\label{newWave}
\begin{split}
&\nabla^{2}_{{\rm dS}_2}\,\Delta \see^{(2)} +4\,c_\mt{fun}  - 4 \,c_\mt{fun}\, e^{-\Delta\see^{(2)} /(2c_\mt{fun})} \cosh\left(3 \Delta\see^{(3)}/c_\mt{fun}\right) =0\,,\\
 &\nabla^{2}_{{\rm dS}_2}\,\Delta \see^{(3)} + 2\,c_\mt{fun}\, e^{-\Delta\see^{(2)} /(2c_\mt{fun})} \sinh\left(3 \Delta\see^{(3)}/c_\mt{fun}\right)=0 \,,
\end{split}
\ee
with $L^2=1$. It is therefore indeed true that spin-two and spin-three entanglement entropy obey local interacting field equations on dS${}_2$. Note that at the linearized level, these equations yield the expected masses --- see Eq.~\reef{eq:deltaSOwave} --- for the nonlocal observables \reef{eq.deltaSObarh0} associated with conserved currents with $h=2$ and 3. Further notice that these equations can be formulated in terms of extremizing the following action of an interacting field theory in dS$_2$
\be
\int d^2\xi\sqrt{-g}\left[-\frac12[\nabla(\Delta\see^{(2)})]^2-6[\nabla(\Delta\see^{(3)})]^2- V(\Delta \see^{(2)},\Delta\see^{(3)})\right]  \label{gobb2}
\ee
where
\be
 V(\Delta \see^{(2)},\Delta\see^{(3)}) = - 4\,c_\mt{fun}\,\Delta \see^{(2)} - 8 \,c_\mt{fun}^2\, e^{-\Delta\see^{(2)} /(2c_\mt{fun})} \cosh\left(3 \Delta\see^{(3)}/c_\mt{fun}\right)\,. \label{gobb3}
\ee
This action can again be related to the $SL(3,\mathbb R)$ Toda equations. 

The Toda equations \reef{toda} are widely believed to have to same relation to $W_3$-gravity as Liouville theory has to ordinary two-dimensional gravity (see, \eg \cite{deBoer:1991jc}). It therefore appears that we have
found the field equations of some higher spin theory of gravity on de Sitter space. Since $SL(3,\mathbb R)$ Toda theory is intimately related to the $W_3$-algebra, and in fact has the $W_3$-algebra as its symmetry, one suspects that there should be more direct argument to explain the appearance of Toda equations here, and it would be interesting to explore this further. 

In defining $\Delta \see^{(2)}$ and $\Delta\see^{(3)}$ above, we only considered subtracting the vacuum entanglement entropies. However, we could consider subtracting the result of other reference states as in section \ref{beyond}. In particular, if we choose states with $\gamma_2=(\gamma_1)^2$, these are all states where the spin-three current vanishes and Eq.~\reef{extra2} becomes
\be
\left.\see^{(2)}\right|_{\gamma_2=\gamma_1^{\,2}}=2\,c_\mt{fun}\,\log\frac{(\gamma_1(u)-\gamma_1(v))^2}{2\,\epsilon_\mt{fun}\,\gamma'_1(u)\gamma'_1(v)}\,.
\label{extra22}
\ee
As above, this matches the right-moving contribution to the entanglement entropy in Eq.~\reef{eq.Sf} with $\gamma_1(z)$ playing the role of the `holomorphic' function $f(z)$. Hence we could carry out the same analysis in section \ref{beyond} choosing any of these states as the reference state, \eg as in Eq.~\reef{rmove}. The results would be essentially the same, \ie the conformation transformation in the CFT would produce a coordinate transformation on the moduli space with $u_0=\gamma_{1,0}(u)$ and $v_0=\gamma_{1,0}(v)$.\footnote{Here, $w=\gamma_{1,0}(z)$ denotes the conformal transformation from the CFT vacuum in the flat space to the new reference state.} Of course, $\langle W(x^+)\rangle=0$ for all of these reference states. An interesting new direction to explore, however, would be to consider using a general reference state from the class defined by Eqs.~\reef{killjoy0} and \reef{killjoy}. That is, a reference state where $\gamma_1(z)$ and $\gamma_2(z)$ are completely independent functions and so the spin-three current has a nonvanishing expectation value. This may well reveal some `higher spin' structure in the geometry of the moduli space.

Let us mention once more that the results derived in the spin-three case are only valid for large $c$ and large Chern-Simons level $k$,
and one expects these results to receive $1/c$ corrections in the full quantum theory.

\subsection{First law from Wilson loops}

To conclude this discussion of higher-spin CFTs, we briefly describe the form of the `first law' in this formalism for a general $SL(N,\mathbb R)\times SL(N,\mathbb R)$ theory. Given the form
of ${\cal S}_R(P,Q)$ in Eq.~(\ref{def:Sr}), we can vary it by varying $g$, and with a bit of algebra, the general variation becomes
\be
\delta {\cal S}_R(P,Q) =c_R \int_Q^P dz 
\frac{             \langle \mu^+ |   g^{-1}(Q) g(z) \delta (g^{-1}(z) \partial g(z) ) g^{-1}(z) g(P)     | \mu^- \rangle  }{ 
          \langle \mu^+ |   g^{-1}(Q)  g(P)     | \mu^- \rangle  }.
\ee
However, since we have $g^{-1} \partial g = \Lambda^+ + U$ from Eq.~\reef{dsgauge}, the variation $ \delta (g^{-1}(z) \partial g(z) )$ is just
$\delta U$. In other words,
\be
\delta {\cal S}_R(P,Q) =c_R \int_Q^P dz 
\frac{             \langle \mu^+ |   g^{-1}(Q) g(z) \delta U g^{-1}(z) g(P)     | \mu^- \rangle  }{ 
          \langle \mu^+ |   g^{-1}(Q)  g(P)     | \mu^- \rangle  }.
\ee
This indeed has the form of some sort of kernel integrated against the local perturbation $\delta U$.
The usual first law is obtained by perturbing global AdS$_3$ background, which we get by taking as 
background $g^{-1} \partial g = \Lambda_+$, so $g=\exp(z \Lambda_+)$. Then
\be \label{hs:firstlaw}
\delta {\cal S}_R(P,Q) =c_R \int_Q^P dz 
\frac{             \langle \mu^+ |   e^{(z-Q)\Lambda_+} \delta U e^{(P-z)\Lambda_+}   | \mu^- \rangle  }{ 
          \langle \mu^+ |   e^{(P-Q)\Lambda_+}    | \mu^- \rangle  }.
\ee
From this one can see that the various components of $\delta U$ will be multiplied by $(z-Q)^a (P-z)^b/(P-Q)^c$
for suitable powers $a,b,c$ and with further work one can show that $a=b=c=h-1$ with $h$ being the conformal dimension of 
the relevant component of $\delta U$. This then is precisely the expression that appears in Eq.~(\ref{eq.deltaSObarh0}) but we have not attempted to determine the precise numerical coefficients which multiply
each of the components of $\delta U$ in Eq.~(\ref{hs:firstlaw}) as function of the representation $R$. But in explicit
examples, like the spin-three `entanglement entropy' (\ref{ee3}), this expression can easily be evaluated explicitly
and can be used to verify that only the spin-three currents appear in the first law for Eq.~(\ref{ee3}), as required.

\section{Dynamics and interactions: future challenges}  \label{loco9}

As we have seen in the previous section, there exist examples where nonlinear and local interactions in the space 
of causal diamonds occur naturally. This nonlinear dynamics was found to describe scale dependence of 
entanglement entropy (and its spin-three generalization) in states with general spin-two and spin-three excitations.

A very interesting question is whether this extends to higher dimensions (see also the discussion section), and whether other degrees of freedom can be included, \ie other fields on the moduli space associated with our new observables \reef{gen:firstlaw}. One might be tempted to look for local field equations just as in the spin-two and spin-three example, however, even in that case there was an issue with the
apparent locality: to write local equations we had to decompose entanglement entropy into left- and right-moving contributions and each of which obeyed a local field equation on a single dS${}_2$. It is not possible to capture this in terms of
a single local field equation obeyed by the sum of the left- and right-moving contributions.

When moving to scalar primaries as in Eq.~\reef{gen:firstlaw}, the situation becomes more complicated, since for these scalars in $d=2$, no simple separation in terms of
left- and right-moving modes exists, while scalars do obey the constraint (\ref{slap2}). As is familiar from several different examples such as exceptional and doubled field theory (for reviews of the latter see, \eg
\cite{Aldazabal:2013sca,Berman:2013eva,Hohm:2013bwa,Freidel:2015pka}), constructing interacting theories for constrained fields can be quite difficult. For example, it is in general not true that the product of two fields that obey the constraints 
still continues to satisfy them. Sometimes it is possible to modify the constraints in perturbation theory, but then
there is the potential issue that the theory becomes over-constrained. Keeping the constraints unaltered, one may have
to introduce explicit projection operators acting on products of fields, projecting the product back into the subspace
of fields which obey the constraints, thereby introducing nonlocalities into the theory. 

While we have not been able to find a compelling systematic framework to incorporate interactions, we believe this
is an important open problem with possibly many new applications, and as a prelude we describe below a preliminary
attempt at including interactions, which clearly demonstrates the sorts of issues one is running into. 

\subsection{Constraints}
\label{sec:constraints}

Even at the linearized level, there is already a challenge since we have identified a single wave equation in the moduli space $\Mdiamonds$ with the signature is $(d,d)$, \ie where there are $d$ timelike directions. Hence conventional techniques are unlikely to adequate to find physical solutions for this equation. An alternative perspective is that functions on the moduli space of the form in Eq.~\reef{gen:firstlaw} will only form a small subset of the solutions of Eq.~\reef{gen:freefield}. Of course, in section \ref{sec.genO}, we showed that in fact our wave equation was only one of two independent equations in $d=2$. The origin of these two equations can be traced to the structure of the the conformal group, which factorizes as  $SO(2,2) \simeq SL(2,\mathbb R)\times SL(2,\mathbb R)$. Each of these factors has an independent quadratic Casimir, which in turn produce  two independent wave equations on the moduli space, as shown in Eq.~\reef{eq:deltaSOwave}. The sum of these equations \reef{slap1} matches the wave equation \reef{gen:freefield} which we constructed for general dimensions, while their difference \reef{slap2} can be regarded as a supplemental constraint. 

Unfortunately, the conformal group is irreducible in higher dimensions and so the same structure does not appear in general. However, it was proposed in \cite{bartek66}, that one can identify constraints by examining $SO(2,2)$ subgroups of the full $SO(2,d)$ group. The reasoning will become apparent in section \ref{kongo}, where we consider the left- and right-moving Casimirs acting on the holographic version of our observables given in Eq.~\reef{gen:bulk}. In either case, the action of the Casimir acting on $\dSOh$ will produce the AdS d'Alembertian acting on the bulk scalar field, using Eqs.~\reef{eq:bulkCas} and \reef{four}, and hence their difference vanishes. This calculation is then easily lifted higher dimensions by considering AdS$_3$ submanifolds within the full AdS$_{d+1}$ bulk geometry. The form and action remains unchanged for the quadratic Casimirs of the left- and right-moving $SL(2,\mathbb R)$ factors in the $SO(2,2)$ group acting on the AdS$_3$ slice and hence their difference again vanishes when acting on Eq.~\reef{gen:bulk}. Hence in this holographic framework, one is able to identify additional operators which annihilate the nonlocal observables \reef{gen:bulk}.   We observe that implicitly a key ingredient here was the intertwining property \reef{intertwine2} which carries the action of the conformal generators on the boundary observable to the scalar field appearing inside the bulk integral. It then turns out that the difference of the `Casimirs' is trivial when acting on the scalar.

The latter observation allows us to extend this construction of constraints
to the nonlocal observables \reef{gen:firstlaw} for general CFTs. Here again the conformal generators satisfy an intertwining property \reef{intertwine}. Hence the idea is to find (combinations of) generators which are trivial in the representation acting on a scalar primary --- see Eq.~\reef{eq:rotationGen}. Motivated by the holographic discussion, we can identify a large number of 
such trivial operators, which can be elegantly written in pure CFT language as
\be \label{cons0}
\Gamma_{abcd}\,|{\cal O}(x)\rangle\equiv J_{[ab} J_{cd]}\, |{\cal O}(x)\rangle=0\,,
\ee
where $J_{ab}$ are the generators of $SO(2,d)$ defined in appendix~\ref{sec:ConvGeneral} (with indices running over embedding space coordinates $a,b,c,d=-,0,\ldots,d$) and ${\cal O}(x)$ is an arbitrary scalar primary. 
In fact, Eq.~\reef{cons0} identifies a family of ${{d+2}\choose{4}}=(d+2)(d^3-d)/24$ `trivial' operators, which then outnumbers the number of independent $SO(2,2)$ subgroups. One might think of this extended family of constraints as being associated with a $SO(k,4-k)$ subgroup, where $k$ corresponds to the number of timelike directions in the four-plane spanned by the $a,b,c,d$ directions in the embedding space. Clearly, this set of operators is closed under conformal transformations, \ie $\Gamma_{abcd}$ forms an antisymmetric tensor representation of the conformal group.

The operators identified with $SO(2,2)$ subgroups then emerged from the four-planes spanned by $X^-,X^{\mu},X^{\nu},X^d$, where $\mu,\nu=0,\cdots,d-1$ correspond to the spacetime directions of the CFT. The corresponding operators can be written in terms of the conformal generators as: 
\be \label{cons2}
12\,\Gamma_{-\mu\nu d}\equiv \Sigma_{\mu\nu} =  2 \{ M_{\mu\nu},D\} - \{ P_{\mu},Q_{\nu} \} + \{ Q_{\mu},P_{\nu} \}\,.
\ee
Again acting on any scalar primary, we have $\Sigma_{\mu\nu}|{\cal O}(x)\rangle=0$. That is, substituting for the generators in Eq.~\reef{cons2} with the expressions in Eq.~\reef{eq:rotationGen}, one finds that the above combination of generators simply vanish, \ie $\Sigma_{\mu\nu}=0$. As we will see below, when we substitute the representation of the generators acting on functions on the moduli space, these operators are nontrivial and hence $\Sigma_{\mu\nu}\,\dSO=0$ becomes an nontrivial constraint on the nonlocal observables.

With $k=1$, we can consider the four-planes spanned by $X^- \pm X^d,X^{\mu},X^{\nu},X^{\rho}$ in the embedding space 
(in the notation of Eq.~\eqref{eq.metricaux}), for which we find constraints of the form
\be \label{cons1}
\Gamma_{\mu\nu\rho-}+\Gamma_{\mu\nu\rho d}= M_{[\mu\nu} P_{\rho]}
\qquad{\rm and} \qquad
\Gamma_{\mu\nu\rho-}-\Gamma_{\mu\nu\rho d}= M_{[\mu\nu} Q_{\rho]} \,.
\ee
One can readily verify that these operators vanish identically on scalar primaries, using the explicit representation given in \eqref{eq:rotationGen}. 
The final case (\ie $k=0$) comes from considering a four-plane spanned by $X^{\mu},X^{\nu},X^{\rho},X^{\sigma}$, for which we obtain
\be
\label{cons3}
\Gamma_{\mu\nu\rho\sigma}=M_{[\mu\nu} M_{\rho\sigma]}\,.
\ee
Again, given the expressions (\ref{cons2}--\ref{cons3}), the identities $\Gamma_{abcd}|{\cal O}(x)\rangle=0$ may not look terribly familiar, however, they follow from conformal invariance and one can readily confirm that they hold for any scalar primary in any CFT by substituting for the conformal generators using Eq.~\reef{eq:rotationGen}.  

Having identified the family \reef{cons0} of trivial operators acting on scalar primaries, we again make use of the intertwining property \reef{intertwine} satisfied by the conformal generators to write
\be \label{intertwine3}
\Gamma_{abcd}(x,y)\,\dS x y = \Cbdy \int_{D(x,y)} \!\!\!\!\!d^d \xi\, \left( \frac{(y-\xi)^2(\xi-x)^2 }{-(y-x)^2} \right)^{\half(\Delta_{\cal O}-d)}\,
\langle\Gamma_{abcd}(\xi)\, {\cal O}(\xi)\rangle=0\,.
\ee
Hence Eq.~\reef{intertwine3} identifies a nontrivial set of constraints which any physical solution of the wave equation \reef{gen:freefield} must satisfy. That is, any solution corresponding to the `smeared' expectation value of a scalar primary, as in Eq.~\reef{gen:firstlaw}, must satisfy the additional constraints: 
\be\label{bead}
\Gamma_{abcd}(x,y)\,f(x, y) =0\,.
\ee
Acting on functions on the moduli space, the operators above take the form
\beq
\label{locked}
\begin{split}
 \Gamma_{abcd}(x,y)&=(J_{[ab}(x)+J_{[ab}(y))\, (J_{cd]}(x)+J_{cd]}(y))\\
 &=J_{[ab}(x)J_{cd]}(y)+ J_{[ab}(y) J_{cd]}(x)\,.
\end{split}
\ee
where $J_{ab}(x)$ and $J_{ab}(y)$ are given by the differential operators in Eq.~\reef{eq:rotationGen} with $\Delta_\cO=0$. To illustrate this, we show the explicit form of Eq.~\eqref{cons2}
when acting on a function $f(x,y)$ on the moduli space $\Mdiamonds$:
\begin{eqnarray}
\frac{1}{2} \Sigma_{\mu\nu}(x,y)\, f(x,y)& = & (x^2-y^2) \left( \frac{\partial\ }{\partial x^{\nu}} \frac{\partial\ }{\partial y^{\mu}} - 
\frac{\partial\ }{\partial x^{\mu}} \frac{\partial\ }{\partial y^{\nu}} \right)f(x,y)\nonumber \\
& & - 2 \left( (x_{\mu}-y_{\mu}) \frac{\partial\ }{\partial x^{\nu}} - (x_{\nu}-y_{\nu}) \frac{\partial\ }{\partial x^{\mu}} \right) 
\left( y^{\alpha} \frac{\partial\ }{\partial y^{\alpha}} \right)f(x,y) \labell{sigmai}\\
& & + 2 \left( (x_{\mu}-y_{\mu}) \frac{\partial\ }{\partial y^{\nu}} - (x_{\nu}-y_{\nu}) \frac{\partial\ }{\partial y^{\mu}} \right) 
\left( x^{\alpha} \frac{\partial\ }{\partial x^{\alpha}} \right)f(x,y)\,. \nonumber 
\end{eqnarray}
It is also interesting to consider this constraint in the center of mass coordinates of Eq.~\eqref{eq:coordsDdiamond}, with which the same operator takes the form
\be\label{walk}
\begin{split}
\frac{1}{2} \Sigma_{\mu\nu}(c,\ell)\,f(c,\ell)
&= 4\left( (c\cdot\ell) \,\delta_{[\mu}^\alpha \,\delta_{\nu]}^\beta + \ell_{[\mu} \,\delta_{\nu]}^\alpha \, c^\beta - \ell_{[\mu} \, \delta_{\nu]}^\beta \, c^\alpha \right)\, \frac{\partial\ }{\partial c^\alpha} \frac{\partial\ }{\partial \ell^\beta}\,f(c,\ell) \\
 &\qquad + 4 \left( \ell_{[\mu} \, \delta_{\nu]}^\alpha \, \ell^\beta  \right) \left(\frac{\partial\ }{\partial c^\alpha} \frac{\partial\ }{\partial c^\beta} -\frac{\partial\ }{\partial \ell^\alpha} \frac{\partial\ }{\partial \ell^\beta} \right)f(c,\ell) \,.
\end{split}
\ee

As a consistency check, we note that in two dimensions there is only one non-trivial constraint and that it reduces to the spinless constraint equation, \ie Eq.~\eqref{slap2} with $h=\bar h$:
\be\label{slap3}
\begin{split}
\text{for } d=2: \quad \Sigma_{01}(u,v,\bar u,\bar v) &=4(u-v)^2\partial_u\partial_v
-4(\bar u-\bar v)^2\partial_{\bar u}\partial_{\bar v}\\
&= 4L^2 \left( \nabla^{2}_{{\rm dS}_2}- \overline{\nabla}^2_{\overline {\rm dS}_2}\right) \,,
\end{split}
\ee
where we are using the null coordinates defined in Eq.~\reef{nulle}.

As a further confirmation of these conclussions, consider inserting a point-like source for the operator ${\cal O}$ at a point $\xi^\mu$ in  $\mathbb{R}^{1,d-1}$, which is timelike separated from the causal diamond $\lozenge=(x^\mu,y^\mu)$, \ie timelike separated from both $x^\mu$ and $y^\mu$. This source generates an expectation value $\langle{\cal O}\rangle$ inside the causal diamond and it follows, for example, from
the shadow field representation (c.f., Eq.~\eqref{gen:firstlaw2shadow}) that in this case the $x^\mu$ and $y^\mu$ dependencies are captured by 
\be
Q({\cal O};x,y) \sim \langle 1(x)\, 1(y)\, {\cal O}(\xi)\rangle  \sim \left( \frac{-(x-y)^2}{(x-\xi)^2 (\xi-y)^2} \right)^{\Delta_{\cal O}/2} 
\ee 
--- see also Eq.~\eqref{eq:2dResult} for the two-dimensional version of this formula.
Now one can verify that $\Sigma_{\mu\nu}(x,y)$ indeed yields zero when acting on this expression, for all values of $\Delta_{\cal O}$ and all choices of $\xi^\mu$. 

As commented above, Eq.~\reef{bead} produces $(d+2)(d^3-d)/24$ additional constraints in higher dimensional CFTs. However, not all of these constraint equations are independent. In particular, there are relations which show that the  $\Sigma_{\mu\nu}$ constraints are sufficient to ensure all constraints of the form \eqref{bead} will be satisfied.
One can show this by using simple but tedious algebra to express all the combinations in Eq.~\eqref{cons0} in terms of $\Sigma_{\mu\nu}$ as follows:
\be\label{algebra}
\begin{split}
 \Gamma_{\mu\nu\rho d}(x,y) &= \frac{1}{2(x^2-y^2)} \left(  (y^2-1) x_{[\mu}- (x^2-1) y_{[\mu} \right) \Sigma_{\nu\rho]}(x,y) \,,\\
 \Gamma_{\mu\nu\rho -}(x,y) &= \frac{1}{2(x^2-y^2)} \left((y^2+1) x_{[\mu}- (x^2+1) y_{[\mu}   \right) \Sigma_{\nu\rho]}(x,y) \,,\\
  \Gamma_{\mu\nu\rho\sigma}(x,y) &= \frac{4}{x^2-y^2} \, x_{[\mu} \, y_{\nu}\,\Sigma_{\rho\sigma]}(x,y) \,,
\end{split}
\ee
where these relations are to be understood to hold with the $\Gamma_{abcd}(x,y)$ expressed as in Eq.~\reef{locked}, \ie the operators are represented as acting on functions $f(x,y)$ on the moduli space.
Therefore the $\Sigma_{\mu\nu}$ alone form a sufficient set of constraints. Hence we expect that these together with the field equation and initial data on a codimension-$d$ surface determine the value of the physical solutions (corresponding to the nonlocal observables) everywhere on the space of causal diamonds. 

However, let us observe that the number of constraints is still larger than what one might have na\"ively expected: $\Sigma_{\mu\nu}$ has $d(d-1)/2$ components,\footnote{We note that \cite{bartek66} also found $d(d-1)/2$ constraint equations, however, their constraints have a slightly different form from that given in Eq.~\reef{sigmai}, \eg their constraints would be independent of the $c^\mu$ coordinates which appear in Eq.~\reef{walk}.} while we would a priori expect $d-1$ independent constraint equations.  That is, we might expect that the total number of equations, \ie the wave equation and the constraints combined, would equal $d$, the number of timelike directions. Hence, we conjecture that further analysis will show that the sufficient set of constraints can be further reduced to $\Sigma_{0i}$, which would give the desired number of equations.

It is relatively straightforward to establish that there are no algebraic relations amongst the $\Sigma_{\mu\nu}$, \ie relations of a form similar to those given in Eq.~\reef{algebra}.\footnote{One approach is to consider these operators on a specific submanifold of the moduli space where their explicit form simplifies, \eg the submanifold $\ell^\mu=R\,\delta^\mu_0$.} However, one can still consider differential relations between the constraints. For example, one can show that 
\be\label{jacobi}
J_{[ab}(x,y)\, \Gamma_{cdef]}(x,y)\, f(x,y) = 0 
\ee 
for all sets of six indices. Note here we are saying that these combinations of operators vanish when acting on any function on the moduli space.\footnote{Substituting Eq.~\reef{locked}, as well as $J_{ab}(x,y)=J_{ab}(x)+J_{ab}(y)$, one finds that the result contains either $J_{[ab}(x)\, J_{cd]}(x)=0$ or the analogous expression for $y$.} Hence Eq.~\reef{jacobi} yields $d+2\choose 6$ relations amongst the $\Gamma_{abcd}$, and implicitly then, amongst the $\Sigma_{\mu\nu}$ through Eq.~\reef{algebra}. For example, with $d=4$, Eq.~\reef{jacobi} provides 1 additional relation, whereas our discussion above suggested we should be able to find 3 extra relations. We have preliminary results on a set of further relations between the $\Sigma_{\mu\nu}$ operators, which may allow us to reduce them to a set of $(d-1)$ independent constraints. However, the full structure is intricate and we hope to report on these issues elsewhere --- see also further discussion in section \ref{discuss}.

Notice that there is an interesting similarity between the constraints that appear here and those that feature in doubled and exceptional field theories, \eg
\cite{Aldazabal:2013sca,Berman:2013eva,Hohm:2013bwa,Freidel:2015pka}, in that both are expressed in terms of a set of second order differential operators. It would be interesting to explore whether the techniques developed in the context of these theories could be of relevance for understanding dynamics on the moduli space of causal diamonds as well. We expect that there will be further relations, however, we defer a more detailed analysis of the constraints, which is presumably essential in order to properly formulate interactions, to future work.

\subsection{Holographic dynamics in AdS$_3$} \label{kongo}

To illustrate some of issues one encounters while attempting to generalize $\dS x y$ to nonlinear order, we consider possible 
nonlinear generalisation of the decoupled dS$_2\times$dS$_2$  wave equations, Eqs.~\eqref{eq:deltaSOwave}, 
for operators which are not conserved currents. We will seek guidance in holography, that is, we want to define 
holographically $\dSh x y$ by insisting that it obeys a local wave equation on the space of causal diamonds even if the corresponding bulk probe scalar $\phi$ interacts nonlinearly in AdS$_{3}$. 

We assume that the equation of motion for the scalar $\phi$ reads
\be
\label{eq.eomphi}
\nabla_{\mathrm{AdS}_{3}}^{2} \phi  = V'(\phi)\,.
\ee
Eventually we will specialize to
\be
\label{eq.Vprime}
V'(\phi) = \mAdS^2 \phi + \zeta \phi^{2}
\ee
and we will work perturbatively in the bulk coupling constant $\zeta$. We use the standard holographic result
\be
\label{eq.massinAdS3}
\mAdS^2\,\RAdS^2 = 4 h(h-1),
\ee
with $\bar{h} = h = \frac{1}{2}\Delta_{\cal O}$.

Let us briefly recapitulate the group theory behind dynamics on the space $\Mdiamonds$. As explained in section \ref{sec:bulk}, the second Casimir has a natural action on the bulk charge $\dSh x y$ defined as an integral over the bulk geodesic $\gamma(x,y)$:
\be \label{eq:bulkCas}
\begin{split}
{\cal C}_2\, \dSh x y &\equiv C^{ij} (L_i(x) + L_i(y))(L_j(x) + L_j(y))\left[ \frac{\Cblk}{8\pi G_\mt{N}} \int_{\gamma(x,y)}d\kappa \, \phi \right]\\
& = 
\frac{\Cblk}{8\pi G_\mt{N}}\int_{\gamma(x,y)}d\kappa \, {\cal C}_2 \phi \,,
\end{split}
\ee	
where we now work with the explicit representation of ${\cal C}_2$ on AdS$_3$ of the form
\be
{\cal C}_2 \, \phi \equiv -4\left[ L_0^2 - \frac{1}{2} \left( L_1 L_{-1}+L_{-1} L_1\right)\right] \phi\,.
\ee
with AdS$_{3}$ isometry generators 
\be\label{eq:AdS3Ln}
L_{-1} = \partial_{\xi}, \quad L_{0} = - \frac{1}{2}\,z \, \partial_{z} - \xi \, \partial_{\xi} \quad \mathrm{and} \quad L_{1}= z \, \xi \, \partial_{z} + \xi^2 \, \partial_{\xi} -  z^{2} \, \partial_{\bar{\xi}}\, ,
\ee
where $\xi = x - t$ and $\bar{\xi} = x+ t$. This immediately yields
\be
{\cal C}_{2}\,\phi= -\RAdS^2\, \nabla^{2}_{\mathrm{AdS}_{3}}\,\phi\,.\label{four}
\ee
Note that we used `right-moving' generators $L_n$ above. One can similarly define `left-moving generators' $\bar L_n$ by exchanging $\xi$ and $\bar \xi$ in their definition, which would lead to the same Laplacian on AdS$_3$.
Using these results, Eq.\ \eqref{eq:bulkCas} reads
\be
{\cal C}_2 \dSh x y =\frac{\Cblk}{8\pi G_\mt{N}} \int_{\gamma(x,y)}d\kappa \, \left( -\RAdS^2\, \nabla^{2}_{\mathrm{AdS}_{3}}\,\phi \right) \,.
\ee
Hence it is clear that the dynamics of the AdS$_3$ scalar field $\phi$ is intimately linked to the dynamics of $ \dSh x y$. If $\phi$ satisfies a linear wave equation, so will $ \dSh x y$. 

However, if we assume that $\phi$ interacts nonlinearly as in \eqref{eq.Vprime}, we find the following identity on the moduli space $\Mdiamonds$:
\be
\label{eq:nonlocal}
\BoxM \dSh x y = -\frac{4 h(h-1)}{L^2}\,\dSh x y - \frac{\Cblk\,\zeta\, \RAdS^2}{8\pi G_\mt{N}\, L^2} \int_{\gamma(x,y)}d\kappa \, \phi^{2}\,,
\ee
where we used $\BoxM = \frac{1}{L^2} \,{\cal C}_2$ on the space causal diamonds. 
It is clear that the last term in this expression, $ \int_{\gamma} d \kappa \,  \phi^{2}$ is not a 
local functional of $ \dSh x y$. One may for example notice that while local functionals of $\dSh x y$ will no longer
obey the constraints, the additional quadratic term in \eqref{eq:nonlocal} still does because it is the integral of a 
scalar quantity over a minimal surface. As a result, the equation of motion in the space of causal diamonds becomes nonlocal. In the next two subsections we will examine possible remedies. First, we will study whether there are any quadratic interaction terms that can be consistently added to Eqs.~\eqref{eq:deltaSOwave}. Subsequently, we will look for  natural quadratic modifications of the holographic definition of $\dSh x y$ given by Eq.~\eqref{gen:bulk} which will induce simple nonlinear dynamics in 
dS$_2\times$dS$_2$.

\subsection{Allowed quadratic local interaction terms on the space of causal diamonds}

The simplest possible solution to the nonlocality encountered in Eq.~\eqref{eq:nonlocal} would be a nonlinear modification of the dS wave equations. We will now give an argument that there is no straightforward and consistent nonlinear extension
of the field equations at the quadratic 2-derivative level.

To quadratic order in $\dS x y$, we can try to supplement Eqs.~\eqref{eq:deltaSOwave} 
by the following general set of local (i.e. at most 2-derivative) interaction terms
\be
\label{eq.deltaSOwave2}
\begin{split}
\left(\nabla^2_{\rm dS_2} - m^{2}_{{\rm dS}_2} \right) \dSO &= \lambda \left\{ \alpha_{1} \dSO^{2} +  \alpha_{2} \left(\nabla_{\rm dS_2}  \dSO \right)^{2} +  \alpha_{3} \left(\bar{\nabla} _{\overline{\rm dS}_2} \dSO \right)^{2} \right\}\,, \\ 
\left(\bar{\nabla}^2_{\overline{\rm dS}_2} - \bar{m}^{2}_{\overline{\rm dS}_2} \right) \dSO &=\lambda \left\{ \bar{\alpha}_{1} \dSO^{2} +  \bar{\alpha}_{2} \left(\bar{\nabla}_{\overline{\rm dS}_2}  \dSO \right)^{2} + \bar{\alpha}_{3} \left(\nabla_{\rm dS_2}  \dSO \right)^{2} \right\} \,,
\end{split}
\ee
where $\alpha_{i}$ and $\alpha_{j}$ are numbers independent of $\lambda$. The solutions to these equations will also have
an expansion in $\lambda$ of the form $\dSO = \dSO^{(0)} + \lambda \dSO^{(1)} + {\mathscr{O}}(\lambda^2)$. If the equations
\eqref{eq.deltaSOwave2} are consistent, we should be able to consistently solve for $\dSO^{(1)}$ given an initial solution
$\dSO^{(0)}$ of the free wave equation. A trivial set of consistent solutions of \eqref{eq.deltaSOwave2} is of the form
$\alpha_2=\bar{\alpha}_2=2 a$, $\alpha_1=-a m^{2}_{{\rm dS}_2}$, $\bar{\alpha}_1=-a \bar{m}^{2}_{\overline{\rm dS}_2}$, for
some constant $a$. For these values one finds the solution $\dSO^{(1)} = a (\dSO^{(0)})^2$, but this can be removed
using a field redefinition
\be
\label{eq.fieldredef}
\dSO \rightarrow \dSO + \lambda \, a\, \dSO^{2} .
\ee
To avoid such trivial solutions of \eqref{eq.deltaSOwave2} we can for example choose 
\be
\bar{\alpha}_{2} = - \alpha_{2},
\ee
but for simplicity we keep our notation for $\alpha_{1}$ and $\alpha_{2}$ unaffected. To examine whether \eqref{eq.deltaSOwave2}
is an consistent set of equations for $\dSO^{(1)}$, we notice that a necessary condition is the 
``integrability'' condition  
\be
[\nabla^2_{\rm dS_2} - m^{2}_{{\rm dS}_2},\bar{\nabla}^2_{\overline{\rm dS}_2} - \bar{m}^{2}_{\overline{\rm dS}_2} ]
\dSO^{(1)}=0.
\ee
Expanding the commutator and using the field equations \eqref{eq.deltaSOwave2} converts this into
\be \label{ic}
\begin{split}
0 & =  2( \bar{m}_{\overline{\rm dS}_2} ^{2} \alpha_{1} - m_{\rm dS_2}^{2} \bar{\alpha}_{1} ) (\dSO^{(0)})^{2} \\
&\quad+ 2 (\bar{m}^2_{\overline{\rm dS}_2}  \alpha_{2} - \bar{\alpha}_{1}) (\nabla_{\rm dS_2} \dSO^{(0)})^{2} 
    + 2 (m^2_{\rm dS_2} \alpha_{2} + \alpha_{1}) (\bar{\nabla}_{\overline{\rm dS}_2} \dSO^{(0)})^{2}\\
&\quad + 4 \alpha_{2} ( \nabla_{\rm dS_2} \bar{\nabla} _{\overline{\rm dS}_2}\dSO^{(0)} )^{2} + \alpha_{3} \bar{\nabla}^{2}_{\overline{\rm dS}_2} (\bar{\nabla}_{\overline{\rm dS}_2} \dSO^{(0)})^{2} - \bar{\alpha}_{3} \nabla^{2}_{\rm dS_2} (\nabla_{\rm dS_2}  
\dSO^{(0)})^{2}.
\end{split}
\ee
A consistent, nontrivial interactive generalization of Eqs.~\eqref{eq:deltaSOwave} to quadratic terms exists if and only
if the above expression vanishes for completely generic $\dSO^{(0)}$ of the form given by Eq.~\eqref{genfirst} with some nonzero $\alpha_{i}$ or $\bar{\alpha}_{j}$.

To test it, we consider several sample forms\footnote{Note that we do not insist on $\langle {\cal O} \rangle$ being the expectation 
value in an actual CFT state. However, quite remarkably, for $\langle \cal O \rangle$ coming from insertion(s) of $\cal O$ at a point 
outside but causally affecting the causal diamond, one can fulfill the integrability condition
 by fixing only one of the interaction terms 
leaving the rest arbitrary. Precisely in these cases the expectation value of $\langle \cal O \rangle$, but also $\dSh x y$
are holomorphically factorized, which was a crucial feature in the derivation of the Liouville equation for entanglement entropy.
One might speculate that holomorphic factorization will be an important ingredient in understanding the role of interactions
in two dimensions.} of $\langle {\cal O} \rangle$ and evaluate the integrability condition \label{ic} using $\dSO^{(0)}$ obtained from Eq.~\eqref{genfirst}. 
Perhaps unsurprisingly, we need to make all $\alpha_{i}$ and $\bar{\alpha}_{j}$ vanish in order to satisfy the
integrability condition \eqref{ic} for generic $\langle {\cal O} \rangle$. As a result, the only consistent 2-derivative set of local equations for $\dSO$ up to quadratic order in the amplitude are the free wave equations \eqref{eq:deltaSOwave} or the trivial 
modifications obtained from them using the field redefinition \eqref{eq.fieldredef}. 


\subsection{Quadratic modifications of the holographic definition of $\dSO$}

Let us now sketch an attempt to extend our definition of $\dSh x y$ beyond the linearized approximation, and in particular let us specify to the case of quadratic interactions as in Eq.\ \eqref{eq.Vprime} and work perturbatively in the coupling $\zeta$. To this end, we start with the most general ansatz for a charge $\dSh x y$ which is quadratic in the bulk field $\phi$, contains two derivatives acting on $\phi$ and a double integral over the bulk geodesic. Since points in the space of causal diamonds are represented by minimal surfaces, having a double integral over the same bulk
surface has a chance of corresponding to local interactions in the space of causal diamonds.
 We will show that these requirements are insufficient. 

Before writing the ansatz, we need a convenient parametrization of the geodesic $\gamma$. On a constant time slice ($t=0$) this is a semi-circle of the form
\be
(x-x_0)^2 + z^2 = R^2\,.
\ee
The AdS$_3$ isometry generators are then given by $L_n$ and $\bar L_n$, c.f., Eq.\ \eqref{eq:AdS3Ln}.
We are going to substitute 
\be
x = - r \tanh \kappa\,,\qquad z=\frac{r}{\cosh\kappa}\,,
\ee
so that $r=R$ describes a geodesic centered at $x=x_0$, and the latter is affinely parametrized by $\kappa \in [-\infty,\infty]$. 
The AdS$_3$ metric in this coordinate system is
\be
ds^2 = \RAdS^2 \left[ d\kappa^2 + \frac{\cosh^2\kappa}{r^2}(dr^2-dt^2)\right]\,.
\ee
In these variables, the symmetry generators \eqref{eq:AdS3Ln} read
\bea
L_{-1} & = & \frac{1}{2}\partial_t - \frac{1}{2}\tanh \kappa\, \partial_r -\frac{1}{2r} \partial_{\kappa} \,,\\
L_0 & = & -\frac{1}{2}(t-r\tanh\kappa) \partial_t  -\frac{1}{2}(r-t\tanh\kappa) \partial_r +\frac{t}{2 r}\partial_{\kappa} \,,\\
L_1 & = & \frac{1}{2}\left(r^2+t^2 - 2 t r \tanh\kappa\right)\partial_t 
-\frac{1}{2} \left((r^2+t^2)\tanh\kappa - 2 t r\right)\partial_r +\frac{r^2-t^2}{2 r} \partial_{\kappa}\,,
\eea
and the barred generators are obtained by sending $t\rightarrow -t$. 
The AdS$_3$ wave equation now reads
\be \label{eq:wavekappa}
\RAdS^2\,\nabla_{\text{\tiny AdS}_3}^2 \phi \equiv \left[\partial_\kappa^2+2\tanh\kappa \, \partial_\kappa + \frac{r^2}{\text{sinh}^2\kappa} (\partial_r^2 - \partial_t^2) \right] \phi = \mAdS^2 \, \phi + \zeta \, \phi^2 \,.
\ee

The last observation needed before we can write our ansatz for the nonlinear $\dSh x y$ is that the generators simplify when evaluated on the geodesic $t=0$ and $r=R$:
\be
\begin{split}
L_{-1}\big{|}_\gamma = \frac{1}{2}(\partial_t-\tanh\kappa\,\partial_r ) &-\frac{1}{2R} \partial_{\kappa} \,,\qquad
L_1\big{|}_\gamma = \frac{R^2}{2}(\partial_t-\tanh\kappa\,\partial_r )+\frac{R}{2}\, \partial_{\kappa} \,,\\
&L_0\big{|}_\gamma = \frac{R}{2} (\tanh\kappa\, \partial_t  -\partial_r) \,.
\end{split}
\ee
From this it is clear that, while the combination $L^-|_\gamma  \equiv \left(R L_{-1} - R^{-1} L_1 \right)_\gamma = -\partial_\kappa$ parametrizes derivatives along the geodesic, there are two independent derivative operators, which act as derivatives orthonormal to the geodesic: on the one hand, we have simply $L_0|_\gamma$, on the other hand $L^+|_\gamma \equiv \left( R L_{-1}  + R^{-1} L_1\right)_\gamma = R (\partial_t -\tanh \kappa \, \partial_r  )$.

We can now write the general two-derivative ansatz for a quadratic charge $\dSh x y$ as the linearized solution known from \eqref{gen:bulk} plus the following double integral:
\be\label{eq:ansatzN}
\begin{split}
\dSh x y &= \frac{\Cblk}{8\pi G_\mt{N}}\int_{\gamma(x,y)} d\kappa \, \phi + \int_{\gamma(x,y)} d\kappa \int_{\gamma(x,y)} d\kappa' \bigg\{ K_0 \, \phi \, \phi' + K_1\, (L^+\phi)(L^+\phi') \\ 
&\qquad+ K_2\, (L^+\phi)(L_0\phi') + K_3\, (L_0\phi)(L^+\phi') + K_4\, (L_0\phi)(L_0\phi') \\
&\qquad + K_5\, \phi (L_0L_0\phi')+ K_6 \, \phi (L_0L^+\phi') + K_7 \, \phi (L^+L^+\phi') +K_8 \, \phi (L_0L^-\phi')\bigg\} \,,
\end{split}
\ee
where $K_i \equiv K_i(\gamma(\kappa),\gamma(\kappa'))$ is a bilocal kernel along the two integrals over the geodesic, and we also abbreviate $\phi\equiv \phi(\gamma(\kappa))$ and $\phi'\equiv \phi(\gamma(\kappa'))$. The idea behind this ansatz is that it makes manifest some of the desired symmetry properties. At the same time the ansatz is completely general (within our assumptions) for the following reason: we have distributed the two orthonormal transverse derivatives over the $\phi$'s in all possible ways. Further, we have not used any $L^-$ generators as 'outermost' derivatives because they would reduce to pure $\kappa$-derivatives along the geodesic, as noted above -- but $\partial_\kappa$ can always be integrated by parts and absorbed into the definition of the kernel $K_i$. Finally, we used the equations of motion (see below) to remove some other combinations (such as $\phi \,L_0L_0 \phi'$) that one could have written in \eqref{eq:ansatzN}. We note that in a time-independent setup, one can show that the kernels $K_{i\geq 5}$ provide nothing new and can be absorbed into $K_{i\leq 4}$.

Giving the ansatz \eqref{eq:ansatzN}, the goal is to determine the kernels $K_i$ such that the quantity thus defined satisfies a nonlinear wave equation of the form 
\be \label{eq:BoxQh}
\BoxM \dSh x y = m^2_{\cal O} \, \dSh x y + \zeta_{\cal O} \, {\dSh x y}^2\,,
\ee
where we expect as before $m^2_{\cal O} L^2 = - \mAdS^2\RAdS^2$ and an analogous identity relating $\zeta_{\cal O}$ to $\zeta$.  To evaluate the left hand side of \eqref{eq:BoxQh} explicitly, we write\footnote{See appendix \ref{app:conventions} for details.}
\be \label{eq:BoxMAdS}
 \BoxM = \frac{1}{L^2}\,{\cal C}_2 = \frac{2}{L^2} \left[ \left(-L_0^2 +\frac{1}{4} ((L^+)^2 - (L^-)^2) \right) + \left(-\bar L_0^2 +\frac{1}{4} ((\bar L^+)^2 - (\bar L^-)^2) \right) \right] \,.
\ee 
The round brackets make it clear that the full Casimir written above factorizes into a holomorphic and an anti-holomorphic part. If we write $L_n$ and $\bar L_n$ in terms of AdS$_3$ derivative operators, then the two parts act as the same operator (i.e., each of them is proportional to the AdS$_3$ Klein-Gordon operator). For our present purposes we can therefore replace all $\bar L_n$ in \eqref{eq:BoxMAdS} by $L_n$. The operator \eqref{eq:BoxMAdS} is then written in a form that makes it easy to act on the ansatz \eqref{eq:ansatzN} and manipulate the resulting expression purely by using group theoretic commutators between the $L_n$, and  the equations of motion of $\phi$ which can now be stated as
\be
\nabla_{\text{\tiny AdS}_3}^2 \phi \equiv  \frac{1}{\RAdS^2} \,\left[4\,L_0^2 - ((L^+)^2 - (L^-)^2) \right] \phi = \mAdS^2\,\phi + \zeta\,\phi^2\,.
\ee
Commuting the $L_n$ through the ansatz and demanding a result of the form of the right hand side of Eq.\ \eqref{eq:BoxQh} yields a set of differential equations for the kernels $K_i$. We find that these differential equations have no non-trivial solution. An ansatz of the form \eqref{eq:ansatzN} is therefore not consistent with the nonlinear dynamics described by \eqref{eq:BoxQh}. 

It will be a very interesting future problem to investigate this issue more closely. What nonlinear form of $\dSh x y$ does satisfy nonlinear dynamical equations on the space of causal diamonds? One can start by including higher derivative terms in the ansatz \eqref{eq:ansatzN}. One quickly finds that only an infinite number of derivatives leads to a consistent set of differential equations for the kernels. The resulting solution is hence highly nonlocal. We believe that all these facts might be a hint that a nonlocal completion of the generalized first law might suffer from similar nonlocal behavior as does the general modular Hamiltonian in the familiar case of entanglement entropy. We hope that the space of causal diamonds might provide a useful new perspective for reorganizing (or perhaps resumming) such objects in an illuminating way. 

Another approach that would be interesting to explore in this context involves integrals not just over bulk minimal surfaces, but over codimension one spatial slices connecting the minimal surface $\tilde B$ to the boundary interval $B$ (c.f., Figure \ref{fig:KillingFlow}). This approach has recently been taken in 
\cite{Beach:2016ocq} to compute perturbations of entanglement entropy at second order in perturbation theory around the vacuum state. It would be interesting to use the second order results of 
\cite{Beach:2016ocq} as a starting point to learn about the general structure of higher order interactions also for scalar primaries: 
if there is interesting dynamics on the space of causal diamonds then the second order expansion of entanglement entropy should
tell us about the three-point function between $\delta \see$ and two other operators $Q({\cal O};x,y)$, and higher order
terms about higher-point functions involving at least one $\delta \see$. If $\delta \see$ somehow couples universally
to the other degrees of freedom $\dS x y$, just like gravity couples universally to all fields in AdS, this should
allow us to completely construct essentially the full interacting theory on the space of causal
diamonds, and, by working backwards, also tell us what the right 
nonlinear extension of $Q({\cal O};x,y)$ should be. It is tempting to speculate that the integrals over spatial slices 
which appear in the second order expansion of entanglement entropy 
need to be upgraded to integrals over the entire bulk causal wedge to describe the nonlinear extension
of $Q({\cal O};x,y)$.
We leave the investigation of these interesting possibilities for future work.

\section{Discussion} \label{discuss}

With the goal of extending the `holographic' structure presented in \cite{deBoer:2015kda} to a dynamical framework, in this paper, we extended our discussion to consider all spherical regions throughout the $d$-dimensional spacetime of the CFT, rather than focusing on those in a fixed time slice. In this context, it is also useful to think in terms of the causal diamonds associated with each of the spheres. Then one readily shows that the moduli space of all causal diamonds is described the coset geometry  $\Mdiamonds\equiv SO(2,d)/[SO(1,d-1)\times SO(1,1)]$, which investigated in some detail in section \ref{sec.2}. 
Motivated by the first law of entanglement \reef{deltaS}, we constructed families of nonlocal observables in CFT, which involved integrating the expectation value of a standard local primary operator over a causal diamond, as in Eqs.~\reef{gen:firstlaw} and \reef{gen:firstlaw2}. One of the nice features of these observables was that they satisfy a simple two-derivative wave equations, \reef{gen:freefield} and \reef{black}, on the moduli space $\Mdiamonds$. In sections \ref{kintwo} and \ref{sec.hs}, we showed that in two-dimensional CFTs, these linear wave equations could be extended to nonlinear equations with local interactions, at least for particular observables evaluated in a certain universal class of states. Hence these CFT observables can be described in terms of local dynamics on the moduli space of causal diamonds. Another nice feature of our new observables is that for holographic CFTs, they have a simple bulk description \reef{gen:bulk} involving a integral of the dual field over the extremal bulk surface reaching the asymptotic AdS on the sphere in the boundary theory. In many earlier works, \eg \cite{Balasubramanian:2013lsa,diff1,diff2,diff3,Czech:2015qta,Lin:2014hva}, these extremal Ryu-Takayanagi surfaces were found to serve as useful probes of the bulk geometry. Then here, we are beginning to see that they also provide interesting probes of the configuration of the matter fields in the bulk. While we have presented a number of compelling results in this paper, the program of describing general CFTs in terms of nonlocal observables on the moduli space of causal diamonds, and also formulating holography in this framework for holographic CFTs, still faces a number of technical challenges.

Two features of the moduli space, which seem rather surprising at first, are that this new space is 2$d$-dimensional and has signature $(d,d)$. Of course, recognizing the space of causal diamonds and the space of timelike separated pairs of points makes clear that the dimension of the moduli space must be $2d$, \ie twice the number of coordinates needed to specify a single point. However, this represents quite a departure from the framework studied in \cite{deBoer:2015kda}, which had a character more akin to standard holography. In particular, for spheres on a fixed time slice, there was a single `holographic' direction associated with the size of the spheres. 

\paragraph{Too Many Times:} On the other hand, coming to grips with the $(d,d)$ signature of the moduli space presents a greater challenge. In particular, as noted in section \ref{sec:constraints}, the wave equations, \reef{gen:freefield} and \reef{black}, that we have identified are quite unconventional since they involve $d$ timelike directions, \ie the $\ell^\mu$ directions in Eq.~\reef{eq:dsspheres2}. A related comment would be that a natural set of initial conditions would come from the value of the observables on infinitesimal causal diamonds, \ie from the submanifold where $\ell^\mu\to0$. From the discussion of section \ref{sec:CausalDi}, \ie Eq.~\reef{eq:DistFinal}, this submanifold lies on the time infinity of the moduli space, however, by definition, it is a codimension $d$ surface. Hence it seems clear that the wave equation by itself is insufficient to produce full solution. Rather, it must be supplemented by additional constraint equations, as discussed in part in section \ref{sec:constraints}.

The case of $d=2$ is special and in fact two independent (conventional) wave equations emerged very naturally, as shown in Eq~\reef{eq:deltaSOwave}. The sum of these equations \reef{slap1} matches the wave equation identified for general dimensions, and hence their difference \reef{slap2} can be regarded as a supplemental constraint. These two equations \reef{eq:deltaSOwave} appeared because the moduli space factorized into the product of two de Sitter geometries for $d=2$ CFTs, as discussed below Eq.~\reef{eq.dS2dS2}. However, an alternative perspective is that, as shown in section \ref{casual}, the wave equation results from acting on the new observables with the quadratic Casimir of the conformal group. In this regard, $d=2$ is special because the conformal group factorizes as  can be seen to $SO(2,2) \simeq SL(2,\mathbb R)\times SL(2,\mathbb R)$ and hence each factor produces an independent quadratic Casimir. The sum of the two Casimirs yields that for the full group and hence generates the expected wave equation, while their difference yields the constraint equation \reef{slap2}. 

In higher dimensions, the conformal group is irreducible and however, as proposed in \cite{bartek66}, one can still focus on $SO(2,2)$ subgroups (as well as other subgroups) of the full $SO(2,d)$ group. This approach gives rise to an elaborate system of constraints (\ref{cons2}--\ref{cons3}), as discussed in section \ref{sec:constraints}. However as discussed there, there are various (algebraic and differential) relations amongst the $\Gamma_{abcd}(x,y)$ operators constructed there, \eg see Eqs.~\reef{algebra} and \reef{jacobi}. While we were unable to prove it, it seems that a natural conjecture is that the $\Sigma_{0i}$ form a sufficient set of constraints to identify the physical solutions of the wave equation \reef{gen:freefield}. It is interesting to note then that the operators appearing in these constraints are second order differential operators in the timelike directions on the moduli space, \ie $\ell^\mu$. This becomes clearer if we consider Eq.~\reef{walk} on the submanifold where $\ell^\mu=R\,\delta^\mu_0$:
\be\label{walk2}
\begin{split}
\frac{1}{2} \Sigma_{0i}(c,\ell)\,f(c,\ell)&=2R^2\left(
\frac{\partial\ }{\partial \ell^0} \frac{\partial\ }{\partial \ell^i} -
\frac{\partial\ }{\partial c^0} \frac{\partial\ }{\partial c^i} 
\right)f(c,\ell)\\
&\qquad +2R \left(c^k \, \frac{\partial\ }{\partial c^k} \frac{\partial\ }{\partial \ell^i}-c^k \, \frac{\partial\ }{\partial \ell^k} \frac{\partial\ }{\partial c^i}\right) f(c,\ell) \,,
\end{split}
\ee
where $k$ is only summed over $k=1,\cdots,d-1$. For comparison purposes, we also consider
\be\label{walk2x}
\frac{1}{2} \Sigma_{ij}(c,\ell)\,f(c,\ell)
= -2Rc^0\left(\frac{\partial\ }{\partial c^i} \frac{\partial\ }{\partial \ell^j}-
\frac{\partial\ }{\partial c^j} \frac{\partial\ }{\partial \ell^i}\right) f(c,\ell)\,.
\ee
Here we see these latter constraints are only first order in `time' derivatives. 

While the appropriate initial value problem is not entirely clear, our proposal above was that initial data would be specified on the codimension $d$ submanifold where $\ell^\mu\to0$. Then, combining the constraints $\Sigma_{0i}(x,y)\,f(x,y)=0$ with Eq.~\reef{gen:freefield}, we have $d$ second-order wave equations which would propagate the physical solutions out across the moduli space.\footnote{Note that the distinction between different types of constraints, \eg as in Eqs.~\reef{walk2} and \reef{walk2x}, should be done covariantly with the projection operators: $P^{\parallel}_{\mu\nu}=\ell_\mu\ell_\nu/\ell^2$ and $P^\perp_{\mu\nu}=\eta_{\mu\nu}-P^{\parallel}_{\mu\nu}$. That is the two classes of constraints would be replaced by $\Sigma_{0i}\to P^{\parallel \sigma}_{\mu} P^{\perp\rho}_{\nu}\,\Sigma_{\sigma\rho}$ and $\Sigma_{ij}\to P^{\perp\sigma}_{\mu} P^{\perp\rho}_{\nu}\,\Sigma_{\sigma\rho}$. This covariant description reinforces the idea that all $d(d-1)/2$ components of $\Sigma_{\mu\nu}$ play a role in describing the physical observables on the moduli space.}  In this context, we can regard the constraints $\Sigma_{ij}(x,y)\,f(x,y)=0$ as imposing constraints on the initial data, \ie the values of $f$ and its first `time' derivatives on the codimension $d$ initial value surface. Further, the additional constraints discussed towards the end of section \ref{sec:constraints} would verify that the $\Sigma_{ij}$ constraints are consistent with the propagation produced by the $\Sigma_{0i}$ equations. This intriguing structure is then reminiscent of the constraint equations appearing in gauge theories or gravity and it may be hinting that there is a hidden gauge symmetry underlying the present equations. However, the full structure of the constraints and the associated initial value problem is intricate and remains to be understood. We hope to return to these issues in future work.

Let us note in this context that one might anticipate some simplifications when we restrict attention to stationary configurations. In particular, it seems that in such a case, we would only need to consider spherical domains on a fixed time slice. The moduli space of such balls\footnote{Equivalently we can consider the space of maximally symmetric minimal surfaces of codimension-one in Euclidean AdS${}_d$.} again reduces the $d$-dimensional de Sitter geometry studied in \cite{deBoer:2015kda}, as discussed in section~\ref{sec.2}. Hence one may expect that the problem reduces to solving the standard Lorentzian wave equation on this geometry. However, it turns out that when evaluated on a stationary configuration, our nonlocal observables \reef{gen:firstlaw} do not satisfy the na\"ive wave equation on the dS$_d$ space. In the case of holographic CFTs, there is a simple intuition for this fact: stationary bulk solutions do also not obey the Euclidean AdS${}_d$ equations of motion. That is, performing the time integral in the first law with stationary sources yields a stationary kernel which is not appropriate for a free wave propagation on Euclidean AdS$_d$. It would be interesting to fully investigate this in a more general context.




\paragraph{Time Evolution:} One of our motivations here was to move extend the construction of \cite{deBoer:2015kda}, which focused on fixed time slices, to a new framework which could describe the time dynamics of the CFT. Hence we must observe that describing the time evolution of the CFT remains to be understood in the current framework. As discussed in section \ref{sec:CausalDi}, according to the metric \reef{eq:dsspheres2}, the displacements $dc^\mu$ of the centre of the causal diamond are all spacelike while the displacements $d\ell^\mu$ deforming the causal diamond are all timelike. But in particular then, translating  the causal diamonds in time corresponds to a spacelike motion on the moduli space!

This `unusual' feature becomes readily evident in Eq.~\reef{goshwoggle}, however, we can gain some insight into the time evolution as follows: Choose a fixed time foliation of the original flat spacetime and consider spheres restricted to these time slices. Following the discussion of section \ref{sec:geometry}, this amounts to choosing the coordinates for the tip and the tail of the corresponding causal diamonds as $x^0 = t+R$, $y^0=t-R$ and $\vec{x}=\vec{y}$. With this restriction, the metric \reef{eq.metriccosetd} becomes
\be
\label{newmetric}
ds_{\mathrm{dS}_{d+1}}^{2} = \frac{L^2}{R^{2}} \left(- dR^2 +dt^2+ d\vec{x}^{2} \right)\,.
\ee
Hence we have identified a submanifold of the full coset with the geometry of $(d+1)$-dimensional de Sitter space.\footnote{One might note that the above de Sitter geometry is a somewhat unusual choice in the context of the present paper because it is not a totally geodesic submanifold. This can be seen since we can regard the new submanifold \reef{newmetric} as a coset itself: dS$_{d+1}=SO(1,d+1)/SO(1,d)$. However, the $SO(1,d+1)$ isometries of the submanifold do not form a subgroup of $SO(2,d)$, the isometries of the full coset. As a result, as is easily verified, our nonlocal operators do not satisfy a simple wave equation on the dS$_{d+1}$ geometry.} However, this submanifold clearly exposes the somewhat surprising feature noted above, namely, the sphere radius $R$ plays the role of time while the CFT time $t$ appears as a {\it space-like} coordinate.

Hence this key issue remains an open question for this new moduli space approach, \ie how to construct a natural description of the real-time dynamics of the underlying CFT using this framework.

\paragraph{K\"ahler-like structure:} In a two-dimensional CFT, we saw that right- and left-moving contributions \reef{eq.Sf} to the entanglement entropy  had an interesting interpretation as conformal factors for the two de-Sitter factors of the moduli space in Eq.~\eqref{eq.dsSfSbarf}. With this interpretation the Liouville equations \eqref{eq.liouville} were equivalent to demanding a positive constant curvature for the conformally rescaled de-Sitter metrics.
One problem with this interpretation, however, is that the sum $\see=S_{\ssc R}(f)+S_{\ssc L}(\bar{f})$ itself does not appear in the geometry and we need to able to split it into right- and left-moving components to write Eq.~\eqref{eq.dsSfSbarf}. As a consequence, it is not easy to generalize this structure to higher dimensions. 

Interestingly, it is possible to identify a different mode in the metric on the moduli space such that demanding 
constant scalar curvature gives rise to a field equation which is identical to the field equation obeyed by $\delta \see$, suggesting
a close relation between the two. To write down this mode, we first point out that 
the metric on the space of causal diamonds \eqref{eq.metriccosetd} can be obtained from the following K{\"a}hler-like structure:
\be
\label{eq.Kaehler}
V = 2L^2 \,\log\!\left[- (x-y)^{2} \right]
\ee
via
\be
h_{\mu \nu} = \frac{\partial^{2}}{\partial x^{\mu} \partial y^{\nu}} V\,.
\ee
This is reminiscent of the findings of Refs.~\cite{Czech:2015qta,Czech:2015kbp}, where the role of the potential $V$ was 
played by the entanglement entropy in a two-dimensional CFT. It is clear though that in higher dimensions  this direct association is no longer true, albeit one might still try to  express Eq.~\eqref{eq.Kaehler} in terms of the entanglement entropy through its leading divergent term, \ie the area law contribution. 

To make the connection with K\"ahler geometry more transparent, we will temporarily relabel $y^{\mu}$ as $x^{\bar{\mu}}$ and define $g_{\mu\bar\nu} \equiv \tfrac{1}{2} h_{\mu\bar\nu}$ and $g_{\mu\nu} \equiv 0$. 
The Ricci tensor for this type of K\"ahler metric takes a simple form 
\be
{\cal R}_{\mu\bar{\nu}} = -\partial_{\mu}\partial_{\bar{\nu}} \log\det g_{\alpha\bar{\beta}} = \frac{d}{L^2}g_{\mu\bar{\nu}} \,, \qquad {\cal R}_{\mu\nu} = 0 \,,
\ee
where we used a specific property of the metric \eqref{eq.metriccosetd} that
\be \label{auxjj}
\det g_{\alpha\bar{\beta}} = \frac{ (2L^2)^d}{(x-\bar{x})^{2d}}\, .
\ee
The moduli space is therefore a constant curvature space with Ricci scalar
\be
{\cal R}=\frac{2d^2}{L^2}\, .
\ee
Another property of K\"ahler metrics is the simplicity of the scalar Laplacian
\be
\nabla^2_\coset = 2 g^{\alpha\bar{\beta}} \partial_{\alpha} \partial_{\bar{\beta}}\,.
\ee

Let us now try to look for variations of the K\"ahler potential that do not change the value of the
scalar curvature and see whether these variations obey an interesting equation. By explicitly varying
the scalar curvature ${\cal R}=2g^{\mu\bar{\nu}}R_{\mu\bar{\nu}}$, we find \begin{eqnarray}
\delta {\cal R} & = & -2 g^{\mu\bar{\lambda}} \delta g_{\bar{\lambda}\rho} g^{\rho\bar{\nu}} \, R_{\mu\bar\nu}- 2 g^{\mu\bar{\nu}} \partial_{\mu}\partial_{\bar{\nu}} \left( g^{\alpha\bar{\beta}}\delta g_{\alpha\bar{\beta}}
\right)
\nonumber \\
& = & -\frac{2d}{L^2} g^{\mu\bar{\lambda}} \delta g_{\bar{\lambda}\rho} g^{\rho\bar{\nu}}  g_{\mu\bar{\nu}}
- \nabla^2_\coset \left(g^{\alpha\bar{\beta}}\delta g_{\alpha\bar{\beta}}\right) \nonumber\\
& = & (-\frac{2d}{L^2} -\nabla^2_\coset) \left(g^{\alpha\bar{\beta}}\delta g_{\alpha\bar{\beta}}\right) \,,
\end{eqnarray}
where in the second line we used Eq.~(\ref{auxjj}). Using the explicit form of $g_{\alpha\bar{\beta}}$ in terms of
$\delta V$ we can finally write the requirement $\delta {\cal R}=0$ as 
\be
\left( \nabla^2_\coset +\frac{2d}{L^2}\right)\nabla^2_\coset \delta V = 0\,.
\ee
If we therefore were to take $\delta V=\delta \see$, this equation would indeed be satisfied. 

It would be interesting to explore this intriguing potential connection between $\delta \see$ and $\delta V$ further. 
If correct one could speculate that it might even be valid at the nonlinear level, and that constant scalar curvature
on the moduli space of causal diamonds yields the full nonlinear equation for entanglement entropy valid in generic 
gravitational backgrounds but in the absence of other sources. It is also intriguing to notice that for space-like separated
points, $V$ itself is proportional to the geodesic distance between the two points, so that the constant curvature
condition may have a natural meaning in that case as well. To test these ideas, one could for example check whether
they apply to entanglement entropy in explicitly known non-trivial gravitational backgrounds such as black holes.
We hope to return to these issues at some point in the future.

\paragraph{Generalized twist operators:} One open question is to provide a nonlinear generalization of observables introduced in section \ref{sec:linearized}. Motivated by considerations of entanglement entropy, we are drawn to consider twist operators with regards to this issue. Recall that as was briefly reviewed in section \ref{kintwo}, the entanglement entropy, as well as the R\'enyi entropies, can be evaluated in terms of twist operators in an $n$-fold replicated version of the CFT --- see also \cite{Calabrese:2004eu,Doyon,Hung:2014npa,Hung:2011nu}. Further in higher dimensional CFTs, \ie for $d\ge3$, the twist operators $\sigma_n$ are codimension-two surface operators with support on the entangling surface. In \cite{Hung:2014npa,Smolkin14}, it was argued that an effective twist operator $\tilde{\sigma}_n$ is defined if one considers correlation functions where the twist operator only interacts with other operators which are all from a single copy of the  replicated CFT. In particular, one finds
\be \label{aux12}
\tilde{\sigma}_n = e^{-(n-1)\hm}
\ee
where $\hm$ is the modular Hamiltonian. This expression should apply for general geometries but, of course, the special case of a spherical entangling surface (in the CFT vacuum) is of interest here, where $\hm$ is given by the local expression in Eq.~\reef{deltaS}. This expression is particularly useful to investigate the limit $n\to1$, which then yields
\be\label{tangle88}
\tilde{\sigma}_n \simeq 1-(n-1)\hm+\cdots\,.
\ee
In particular, this demonstrates that the modular Hamiltonian is the only nontrivial contribution in the OPE limit of the twist operator which survives in the $n \rightarrow 1$ limit. Ref.~\cite{alex} suggested augmenting the twist operators with (the exponential of) a charge term which had the form of one of our new observables \reef{gen:firstlaw66} with a spin-one conserved current. A similar extension \cite{Hijano:2014sqa} involving higher spin observables \reef{eq.deltaSObarh0} was considered in the context of two-dimensional CFTs of the form discussed in section \ref{sec.hs}.

Given these considerations, it is tempting to generalize Eq.~\reef{aux12} to a family of `generalized twist operators' based on our nonlocal observables, \eg
\be\label{woggle3}
\tilde{\sigma}({\cO}) = e^{-\mu\, \dSO }\,.
\ee
We have included a numerical coefficient $\mu$ so that the linearized observable would emerge in a `first law'-like expression with the limit $\mu\to0$.\footnote{We have distinguished $\mu$ from the index $n$ in Eq.~\reef{aux12} since we need not consider the replicated CFT in defining $\tilde{\sigma}({\cO})$, \ie it can be defined in a single copy of the CFT.} 
However, it is not immediately clear whether one can meaningfully construct the power series in $\mu$ implicit in the above definition of $\tilde{\sigma}({\cO})$. We hope to return to study this question and other issues for this possible nonlinear generalization of our nonlocal observables in the future.

\paragraph{Universal constant?:} As noted in section \ref{sec:linearized},
the integral in Eq.~\reef{gen:firstlaw} diverges for $\Delta_{\cal O}\le d-2$ unless the expectation value vanishes at the boundaries of the causal diamond. That is, if $\langle\cO\rangle$ is nonvanishing somewhere, then Eq.~\reef{gen:firstlaw} diverges for causal diamonds over some region of the moduli space. However, we still expect that a universal finite term can be extracted from this expression in this situation. Examining Eq.~\reef{eq:delSbdy}, where $\dSO$ is evaluated for a constant expectation value, we see that the result remains finite for $\Delta_{\cal O}< d-2$. In fact, divergenes only arise for $\Delta_\cO =d-2,\ d-4, \cdots$. Hence our calculation has implicitly analytically continued the expression to produce a finite result in the range $\Delta_{\cal O}< d-2$. We expect that the same universal result could be produced if we explicitly introduced a short distance cut-off and focused on the cut-off independent constant term in the final result. Further we expect for the special values of $\cO$ where Eq.~\reef{eq:delSbdy} corresponds to the appearance of a logarithmic divergence whose coefficient would yield the universal contribution. These considerations would then put these universal contributions on the same footing as the constant $F$ in the $F$-theorem \cite{fthem1,fthem2,prince1,prince2}. However, there are subtleties defining $F$ using entanglement entropy \cite{mutual} and so as in that case, one might ask if a more robust definition of $\dSO$ for the cases where Eq.~\reef{gen:firstlaw} contains divergences.

Using the usual AdS/CFT dictionary, \eg Eqs.~\reef{slap} and \reef{dash}, it is straightforward to see that analogous divergences appear in the holographic definition in Eq.~\reef{gen:bulk}. That is, the integral over the extremal surface in $\dSOh$ will diverge for $\Delta_{\cal O}\le d-2$. Of course, the result in Eq.~\reef{eq:deltaSlimitHolo} for a constant expectation value indicates that these divergences can again be avoided by a suitable analytic continuation or with a suitable regulator, \ie the results there precisely match those in Eq.~\reef{eq:delSbdy}. Hence the equivalence $\dSO=\dSOh$ survives for operators with $\Delta_{\cal O}\le d-2$. However, the question of whether the wave equation \reef{gen:freefield} applies in this regime still requires more careful investigation.\\

It is clear from the discussion above that our studies here have left open a variety of interesting questions and we hope to continue to study these in future research. 


\section*{Acknowledgments}
We would like to thanks Nele Callebaut, Alejandra Castro, Bartek Czech, Ben Freivogel, Diego Hofman, Veronika Hubeny, Aitor Lewkowycz, R.\ Loganayagam, Markus Luty, Miguel Paulos, Guilherme Pimentel, Mukund Rangamani, James Sully, Erik Tonni and Claire Zukowski for useful discussions and comments. 
This work is part of the research programme of the Foundation for Fundamental Research on Matter (FOM), which is part of the Netherlands Organisation for Scientic Research (NWO). Research at Perimeter Institute is supported by the Government of Canada through the Department of Innovation, Science and Economic Development and by the Province of Ontario through the Ministry of Research \& Innovation. FMH is grateful to Perimeter Institute and UC Davis for hospitality while this work was in progress. RCM is also supported in part by research funding from the Natural Sciences and Engineering Research Council of Canada,  from the Canadian Institute for Advanced Research, and from the Simons Foundation through the ``It from Qubit" Collaboration.

\appendix

\section{Geometric details}
\label{app:geometric}

In this appendix, we consider various geometric details which are useful for the discussions in the main text. In particular, in the first section, we discuss the details of the derivation of the precise form of the metric \eqref{eq.metriccosetd} on the moduli space of causal diamonds. In the second section, we discuss the moduli space for pairs of spacelike separated points, which arises naturally in a number of instances, \eg two dimensions.
Finally, in the last section, we elaborate on the form and properties of the conformal Killing vector which can be constructed to preserve the form of any given causal diamond.

\subsection{Derivation of metric on the space of causal diamonds}
\label{app:metric}

In the following, we present further details in the derivation of the metric \eqref{eq.metriccosetd} on the moduli space of causal diamonds. Our approach is to continue working in the embedding space introduced in section \ref{sec:geometry}, make a general ansatz compatible with the required symmetries, and subsequently impose conditions which fix the free parameters.

We remind the reader that the metric needs to be of the form \eqref{eq.metricpartialcoset}, which we reproduce here for convenience:
\be
\label{eq.metricpartialcoset2}
ds^{2}_\coset = L^2 \left( - \langle  d T, d T \rangle + \langle  d S, dS \rangle\right)\,,
\ee
where the vectors $T^b$ and $S^b$ still need to be fully determined, subject to the conditions in Eqs.~\reef{eq.condTSnorm} and \reef{eq.condTSorth}, \ie
\beqa
&&\langle T, T \rangle  \,= -1\,, \quad
\langle S, S\rangle \, = 1\,, \quad
\langle T, S\rangle  \,= 0\,,
\labell{eq.condTSnorm2}\\
&&\qquad\quad\langle S ,X\rangle \big|_{z \rightarrow 0} = \;\langle T, X\rangle \big|_{z \rightarrow 0} = 0 \,. \labell{eq.condTSorth2}
\eeqa
The form of the metric  \reef{eq.metricpartialcoset2} was derived in section \ref{sec:geometry} by demanding $SO(1,d-1)\times SO(1,1)$ invariance. 

Let us start with the observation that for any metric of the form
\be
\label{eq.fibration}
ds^{2} = N\, da^{2} + 2 N_{i} \,da \, dm^{i} + g_{i j} \, dm^{i} dm^{j}\,,
\ee
where $a$ is a Killing coordinate, \ie none of the metric components depends on $a$, one obtains its $SO(1,1)$ coset by taking
\be
\label{eq.moddingoutU1}
ds^{2} = \left( g_{i j} - N_{i} N_{j} / N^{2} \right) dm^{i} dm^{j}\,,
\ee
where $m^{i}$ are the coordinates on the final coset. In order to obtain the metric on the space of causal diamonds, we thus need to parametrize $T^{b}$ and $S^{b}$ in terms of the corresponding $m^{i}$-coordinates, which in our case are simply $x^{\mu}$ and $y^{\mu}$ specifying the tips of a causal diamond. We then need to evaluate Eq.~\eqref{eq.metricpartialcoset2}. The corresponding Killing coordinate will be that associated with the $SO(1,1)$ boost and this will allow us to use Eq.~\eqref{eq.moddingoutU1} to explicitly write out the desired coset metric.

As it turns out, the following parametrization of $T^{b}$ and $S^{b}$ does the job for us:
\be
T^{b} = (T^{-1}, \tau_{x} \, x^{\mu} + \tau_{y} \, y^{\mu}, T^{d}) \quad \mathrm{and} \quad S^{b} = (S^{-1}, \sigma_{x} \, x^{\mu} + \sigma_{y} \, y^{\mu}, S^{d})\,.
\ee
In order to demonstrate this, let us start with the conditions~\eqref{eq.condTSorth2}, which by taking their two independent linear combinations can be recast as
\be
C^{(0)}_{x}-2 w x + C^{(2)}_{x} \, w^{2} = 0 \quad \mathrm{and} \quad C^{(0)}_{y}-2 w y + C^{(2)}_{y} \, w^{2} = 0\,,
\ee 
where
\bea
C^{(0)}_{x} &=&-  \frac{\sigma_{y} (- T^{-1} + T^{d}) + \tau_{y} (S^{-1} - S^{d})}{\tau_{y} \sigma_{x} - \tau_{x} \sigma_{y}}\,, \nonumber \\
C^{(2)}_{x} &=& - \frac{-\sigma_{y} (T^{-1} + T^{d}) + \tau_{y} (S^{-1} + S^{d})}{\tau_{y} \sigma_{x} - \tau_{x} \sigma_{y}}\,, \nonumber \\
C^{(0)}_{y} &=& \frac{\sigma_{x} (T^{-1} - T^{d}) + \tau_{x} (-S^{-1} + S^{d})}{\tau_{y} \sigma_{x} - \tau_{x} \sigma_{y}}\,, \nonumber \\
C^{(2)}_{y} &=& \frac{\sigma_{x} (T^{-1} + T^{d}) - \tau_{x} (S^{-1} + S^{d})}{\tau_{y} \sigma_{x} - \tau_{x} \sigma_{y}}\,.
\eea
Clearly, neither $T^{b}$ nor $S^{b}$ can depend on $w^{\mu}$. As a result, demanding conditions~\eqref{eq.condTSorth2} amounts to solving a set of 4 \emph{independent} equations:
\be
\label{eq.condspherefinal}
C_{x}^{(0)} = x^2 \quad \mathrm{and} \quad C_{x}^{(2)} = 1 \quad \mathrm{and} \quad C_{y}^{(0)} = y^2 \quad \mathrm{and} \quad C_{y}^{(2)} = 1\,.
\ee
Together with the three normalization conditions \eqref{eq.condTSnorm2}, Eqs.~\eqref{eq.condspherefinal} allow to solve for 7 out of 8 parameters specifying $T^{b}$ and $S^{b}$ vectors (up to an irrelevant discrete choice of the vectors' orientations). The remaining real parameter corresponds to the boost freedom. Let us then solve Eqs.~\eqref{eq.condspherefinal} together with Eqs.~\eqref{eq.condTSnorm2} for $T^{d} = 0$. The solution reads
\bea
T^{-1} &=& \frac{x^2 - y^{2}}{\sqrt{-(x-y)^{2}(1-x^{2})(1-y^{2})}}\,, \nonumber \\
\tau_{x} &=& -\frac{-1+y^{2}}{\sqrt{-(x-y)^{2}(1-x^{2})(1-y^{2})}}\,, \nonumber \\
\tau_{y} &=&  \frac{-1+x^{2}}{\sqrt{-(x-y)^{2}(1-x^{2})(1-y^{2})}}\,,\nonumber \\
T^{d} &=& 0\,, \nonumber \\
S^{-1} &=& \frac{-1 + x^2 y^{2}}{\sqrt{-(x-y)^{2}(1-x^{2})(1-y^{2})}}\,,\nonumber \\
\sigma_{x} &=& \tau_{x}\,, \nonumber \\
\sigma_{y} &=& -\tau_{y}\,, \nonumber \\
S^{d} &=& \frac{(-1+x^{2})(-1+y^{2})}{\sqrt{-(x-y)^{2}(1-x^{2})(1-y^{2})}}\,.
\eea
We will regenerate the missing parameter by evaluating the metric~\eqref{eq.metricpartialcoset2} by performing a boost in the $(T,S)$-plane,
\be
\label{eq.boost}
(T')^{b} = \cosh{a} \, T^{b} + \sinh{a} \, S^{b}\,, \qquad (S')^{b} = \cosh{a} \, S^{b} + \sinh{a} \, T^{b}\,,
\ee
which preserves the conditions in Eqs.~\eqref{eq.condTSnorm2} and \eqref{eq.condTSorth2}. It is then a matter of tedious and rather unilluminating calculation to recast the metric in the form~\eqref{eq.fibration} and identify the corresponding $g_{i j}$ and $N_{i}$. After using Eq.~\eqref{eq.moddingoutU1}, we are led to the desired metric on the $SO(2,d)/[SO(1,d-1)\times SO(1,1)]$ coset:
\be
ds^{2}_{\coset} = h_{\mu \nu} dx^{\mu} dy^{\mu} =  \frac{4 L^2}{(x-y)^{2}} \left( -\eta_{\mu \nu} + \frac{2(x_{\mu}-y_{\nu})(x_{\nu}-y_{\nu})}{(x-y)^{2}}\right) dx^{\mu} dy^{\nu}\,.\label{jolly1}
\ee

\subsection{Conformal Killing Vectors} \label{dead}

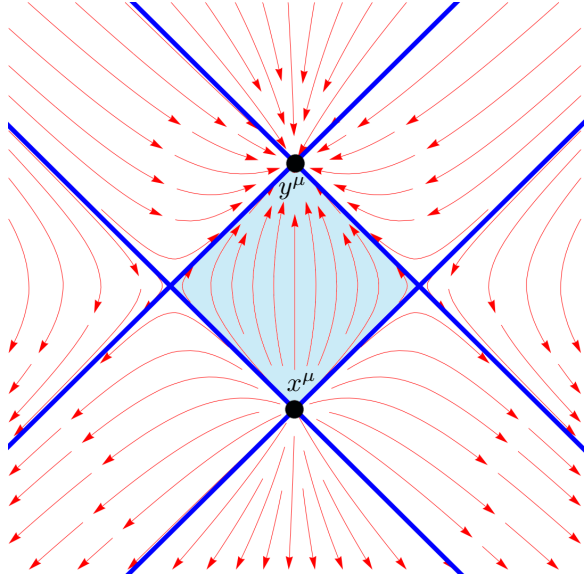
\begin{figure}[t]
\centerline{\includegraphics[width=.5\textwidth]{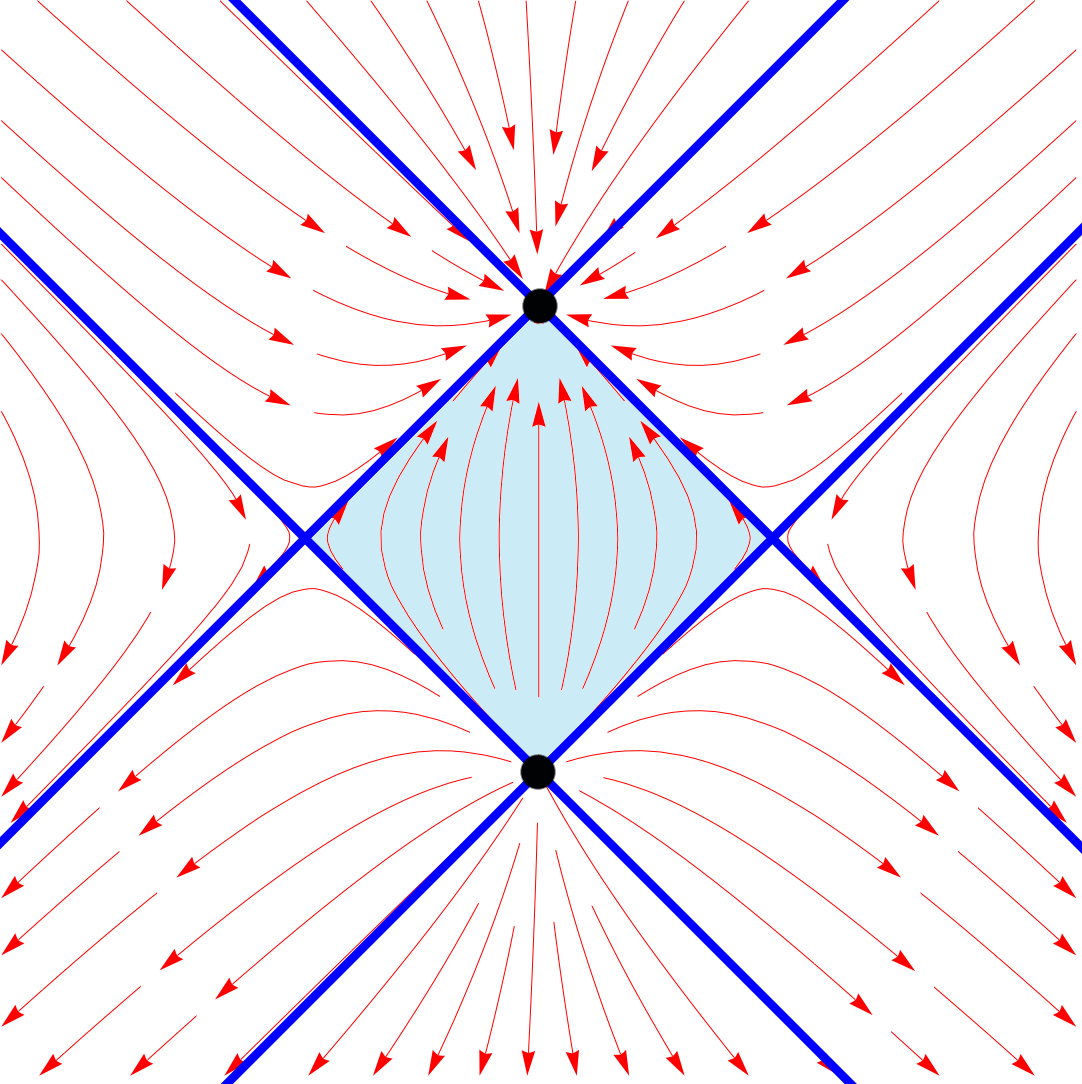}}
\setlength{\unitlength}{0.1\columnwidth}
\begin{picture}(0.3,0.4)(0,0)
\small{
\put(4.97,3.8){\makebox(0,0){${y^\mu}$}}
\put(5.05,2.1){\makebox(0,0){${x^\mu}$}}
}
\end{picture}
\caption{Flow lines of the conformal Killing vector $K^\mu$. The causal diamond $\lozenge(x^\mu,y^\mu)$ is shaded blue. 
}
\label{fig:flow}
\end{figure}

Given a causal diamond in Minkowski space, which is defined by the positions of the future and past tips $(y^\mu,x^\mu)$, there is a conformal Killing vector which preserves the diamond:\footnote{As usual, our notation here is that $(y-x)^2=\eta_{\mu\nu}(y-x)^\mu(y-x)^\nu$.}
\be
\label{Killer}
K^\mu(w)\,\partial_\mu=-\frac{2\pi}{(y-x)^2}\left[(y-w)^2\,(x^\mu-w^\mu)-(x-w)^2\,(y^\mu-w^\mu)\right]\,\partial_\mu\,.
\ee
From this expression, one can easily see that the vector vanishes at  $w^\mu=x^\mu$ and $w^\mu=y^\mu$, and when both $(y-w)^2=0$ and $(x-w)^2=0$, \ie when Eq.~\reef{eq.spherecond} is satisfied. Hence the tips of the causal diamond and also the maximal sphere at the waist of the causal diamond are  fixed points of the flow defined by $K$. Further, one sees that $K$ is null on the boundaries of the causal diamond, \ie when either $(y-w)^2=0$ or $(x-w)^2=0$. Finally, one can also observe that within the rest of the causal diamond $K$ is timelike and future directed. Figure \ref{fig:flow} illustrates the Killing flow both inside and outside of the causal diamond for a cross-section of the diamond. 

Working with standard `Cartesian' coordinates $w^\mu=(t,\vec{x})$ in Minkowski space, if one chooses the frame where $y^\mu=(R,\vec{x}_0)$ and $x^\mu=(-R,\vec{x}_0)$ then the conformal Killing vector takes a recognizable form, \eg \cite{eom2}
\be
\label{Killer2}
K^\mu\,\partial_\mu=\frac{\pi}{R}\left[(R^2-|\vec{x}-\vec{x}_0|^2-t^2)\,\partial_t-2t\,(x^i-x^i_0)\,\partial_i\right]\,.
\ee
Given this expression, one sees that the perturbation of the entanglement entropy in Eq.~\reef{deltaS} can be written in a covariant form as
\be
\label{deltaH}
\delta\see=  \int_{B} d\Sigma^\mu \ \langle T_{\mu\nu}\rangle\ K^\mu   \,,
\ee
where the integration runs over 
$|\vec{x}-\vec{x}_0|^2\le R^2$ on the $t=0$ time slice.
However, in this form, we can regard the integrand is a conserved current which allows us to move the surface of integration to be any Cauchy surface spanning the associated causal diamond.  That is, if we define $J_\mu\equiv
\langle T_{\mu\nu}\rangle\ K^\mu$, it follows that $\nabla^\mu J_\mu=0$ because the stress tensor is conserved and traceless, \ie $\nabla^\mu \langle T_{\mu\nu}\rangle=0=\langle T^\mu{}_\mu\rangle$, and because $K$ is a conformal Killing vector, \ie $\nabla^\mu K^\nu+\nabla^\nu K^\mu=\frac2d\, \nabla\cdot K\, \eta^{\mu\nu}$. Of course, similar statements apply for the higher spin observables constructed in section \ref{sec:spin}.

We might note that in two dimensions using the null coordinates introduced in Eqs.~\reef{eq.LCdef} and \reef{nulle}, the conformal Killing vector takes a particularly simple form:
\be
K^{\xi} = 2\pi\,\frac{(v-\xi)(\xi-u)}{(v-u)} \quad \mathrm{and} \quad K^{\bar{\xi}} = 2\pi\,\frac{(\bar{v}-\bar{\xi})(\bar{\xi}-\bar{u})}{(\bar{v}-\bar{u})}\,.
\ee
This allows us to re-express the observables \reef{genfirst} for $d=2$ CFTs as
\be
\label{eq.deltaSOK}
\dSt = \frac{\Cbdy}{2} \int_u^v  d\xi \ \bigg( \frac{K^{\xi}}{2\pi}\bigg)^{h-1}\
\int_{\bar u}^{\bar v}d\bar \xi\ \bigg(\frac{ K^{\bar{\xi}}}{2\pi}\bigg)^{\bar h-1}\, O(\xi,\bar{\xi})\,.
\ee

In the context of the AdS/CFT correspondence, the conformal Killing vector \reef{Killer} extends to a proper Killing vector of the AdS geometry as follows: We describe the AdS geometry with Poincar\'e coordinates
\be
\label{eq.AdSmetric2}
ds^{2} = G_{MN}\,dW^M dW^N=\frac{\Rads^2}{z^{2}} \left(dz^{2} + \eta_{\mu \nu} \, dw^{\mu} dw^{\nu}\right)\,,
\ee
where we have introduced a $(d+1)$-dimensional vector notation, \eg we denote the bulk coordinates as $W^M=(w^\mu,z)$. Hence we indicate the tips of the causal diamond in the boundary with $Y^M=(y^\mu,0)$ and $X^M=(x^\mu,0)$. With this notation, the bulk Killing vector becomes
\be
\label{Killer3}
K^M\,\partial_M=-\frac{2\pi}{(Y-X)^2}\left[(Y-W)^2\,(X^M-W^M)-(X-W)^2\,(Y^M-W^M)\right]\,\partial_M\,,
\ee
where our notation here is that $(Y-X)^2=G_{MN}(Y-X)^M(Y-X)^N$. With this expression, one can easily verify that the tips of the causal diamond in the boundary are fixed points of the Killing flow, as is the extremal surface where $(Y-W)^2=0$ and $(X-W)^2=0$. Further one can see that the Killing vector becomes null on the boundaries of the causal wedge in the bulk.

One can also consider the analytic continuation of Eq.~\reef{Killer} to Euclidean signature, which follows by simply replacing the Lorentzian inner product there by $(y-x)^2=\delta_{\mu\nu}(y-x)^\mu(y-x)^\nu$. As discussed in section \ref{euklid}, there are two distinct moduli spaces to consider in Euclidean signature and  associated conformal Killing vectors arise from different choices of the vectors $x^\mu$ and $y^\mu$. If we choose real vectors, then $x^\mu$ and $y^\mu$ now define a pair of spacelike points and these points are the only fixed points of the flow defined by $K^\mu$.\footnote{That is, in Euclidean signature, the only solution of $(x-w)^2=0$ is $w^\mu=x^\mu$ and hence we cannot simultaneously solve $(y-w)^2=0$ and $(x-w)^2=0$. Note that if we were considering spacelike separated points but in Lorentzian signature, there would be the simultaneous solution of these two equations would define a spacelike hyperbola --- see the following section.} Hence this conformal Killing vector generates the $SO(1,1)$ symmetry in the coset $SO(1,d+1)/(SO(d)\times SO(1,1))$, which corresponds to the moduli space of pairs of points discussed in section \ref{euklid}.

The second distinct moduli space in Euclidean space is the space of all $(d-2)$-dimensional spheres, which is described by the coset  $SO(1,d+1)/(SO(1,d-1)\times SO(2))$. In this case, the associated conformal Killing vector results from choosing `complex' vectors $x^\mu$ and $y^\mu$. In particular, using the notation of Eq.~\reef{eq:coordsDdiamond}, we choose
\bea
y^\mu&=&c^\mu+\ell^\mu=c^\mu + i R\,n^\mu\labell{weird}\\
x^\mu&=&c^\mu-\ell^\mu=c^\mu - i R\,n^\mu = (y^*)^\mu\nonumber
\eea
where $n^\mu$ is an arbitrary unit vector in $R^d$. The conformal Killing vector then becomes
\be
\widetilde K^\mu(w)\,\partial_\mu=-\frac{2\pi i}{(y-y^*)^2}\left[(y-w)^2\,((y^*)^\mu-w^\mu)-(y^*-w)^2\,(y^\mu-w^\mu)\right]\,\partial_\mu
\label{KillerB}
\ee
where we have introduced an extra overall factor of $i$ to produce a real vector.
Since $w^\mu$ correspond to real positions, we cannot satisfy the equations $w^\mu=y^\mu$ or $w^\mu=(y^*)^\mu$. On the other hand, the equations $(y-w)^2=0$ and $(y^*-w)^2=0$ can be simultaneously solved by setting
\be
(c-w)^2-R^2=0\qquad{\rm and}\qquad n\cdot(c-w)=0\,.
\label{sphe}
\ee
That is, the flow of the new vector $\widetilde K^\mu$ has a fixed point on a $(d-2)$-sphere of radius $R$ centred at $w^\mu=c^\mu$ and lying in the $(d-1)$-dimensional hyperplane defined by $n\cdot(c-w)=0$. Hence this new Killing vector generates the $SO(2)$ symmetry in the coset describing the moduli space of $(d-2)$ dimensional spheres in $\mathbb{R}^d$.

\subsection{Moduli space of spacelike separated pairs of points} \label{geom}

Here we would like to consider the analog of our generalized kinematic space \reef{eq:Mdiamonds} for pairs of spacelike separated points in a $d$-dimensional CFT  (with Lorentzian signature). Recall that $\Mdiamonds$ was the moduli space of all causal diamonds, or equivalently of all spheres, or equivalently of all timelike separated pairs of points. Considering the space of spacelike separated points arises naturally in a number of instances, \eg upon analytically continuing to a Euclidean signature, as discussed briefly in section \ref{euklid}. In fact, in two dimensions, a causal diamond can be defined either in terms of a pair of timelike separated points or a pair of spacelike separated points.\footnote{For example in Figure \ref{fig:2dcoords}, the causal diamond can be defined in terms of the extreme points at the left and right corners, \ie $(\xi,\bar \xi)=(v,\bar u)$ and $(u,\bar v)$.} Hence it seems that $d=2$ is a special case where the two moduli spaces are equivalent, \ie the space of timelike separated pairs of points is the same geometric object as the space of spacelike separated pairs of points. Our final conclusion here is that in fact this equivalence extends to CFTs {\it in arbitrary dimensions!}

\begin{figure}[t]
\centerline{\includegraphics[width=.4\textwidth]{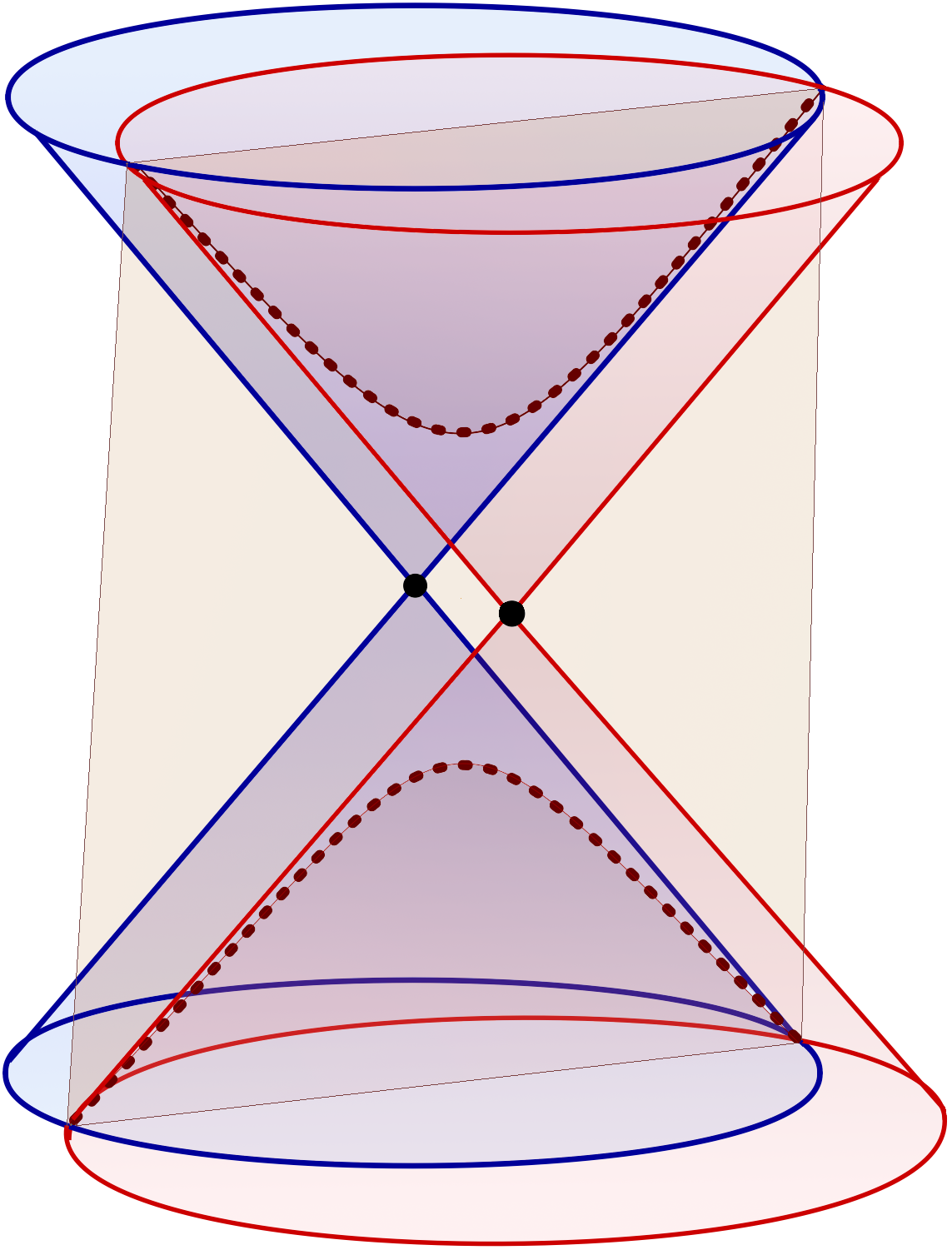}}
\caption{Illustration of the one-to-one correspondence between spacelike separated points and the space of spacelike hyperbolas: the intersection of lightcones of two spacelike separated points forms a spacelike hyperbola (dashed maroon curve) which lies in a timelike codimension-one hyperplane (shaded in yellow).
}
\label{fig:cones}
\end{figure}

To understand this new moduli space, we begin by considering the intersection of the lightcones from a pair of  spacelike separated points. 
As illustrated in Figure \ref{fig:cones}, the intersection of the lightcones defines a spacelike hyperbola lying in a fixed timelike hyperplane (of codimension one). Hence in analogy to the previous discussion of kinematic space, we may say that  the moduli space of pairs of spacelike separated points is equivalent to the moduli space of spacelike hyperbola. There is no obvious analog of the causal diamonds since for spacelike separated points, the two lightcones do not enclose a finite-volume region anywhere, as can be seen in the figure. 

Next we would like understand the coset structure of this moduli space by turning to the embedding space introduced in section \ref{sec:geometry}. However, it is easiest to think in terms of a construction of the moduli space of spacelike hyperbolae in a $d$-dimensional CFT. A bit of thought shows that such a hyperbola will be described by choosing a pair of orthogonal unit vectors, $T^b$ and $S^b$, satisfying precisely the same conditions given in Eqs.~\reef{eq.condTSnorm} and \reef{eq.condTSorth}. 
This construction is again easily illustrated with the Poincar\'e patch coordinates \reef{poinc} where a convenient choice of the unit vectors is 
\bea
T^b&=&(1,0,0,\ldots,0)\quad\longrightarrow\ \  \eta_{\mu\nu}\,w^\mu w^\nu=-1\,,\nonumber\\
S^b&=&(0,0,1,\ldots,0)\quad\longrightarrow\ \ w^1=0\,.
\labell{example2}
\eea
The expressions on the right denote the surfaces in the asymptotic geometry that are picked out by the orthogonality constraints \reef{eq.condTSorth}, \ie $S^b$ selects a particular timelike codimension-one hyperplane in the boundary while $T^b$ selects a spacelike hyperboloid. The intersection of these two surfaces then yields the (codimension-two) hyperbola 
\be\label{wacked}
(w^0)^2 - \sum_{i=2}^{d-1}(w^i)^2 =1\qquad \text{on the hyperplane }w^1=0\,.
\ee 
   
Now following the discussion of section \ref{sec:geometry}, a particular pair of unit vectors, $T^b$ and $S^b$, specifies a particular hyperbola in the boundary geometry. We sweep out the rest of the moduli space by acting with $SO(2,d)$ transformations, \ie Lorentz transformations in the embedding space. However, the coset structure of the resulting moduli space of hyperbolae is then determined by the symmetries preserved by any particular choice of the unit vectors. However, since the constraints on the present unit vectors are precisely the same as in section \ref{sec:geometry}, these symmetries are also the same and hence we arrive at the same coset as given in Eq.~\reef{eq:Mdiamonds}, namely,  
\be \frac{SO(2,d)}{SO(1,d-1)\times SO(1,1)} \,. \label{wacked2}
\ee

At first sight, this result may seem rather counterintuitive. Spacelike and timelike separated pairs of points are by definition very different kinds of objects in Minkowski space and yet we found that in a $d$-dimensional CFT, the moduli spaces of such pairs are described by the same coset structure irrespective of whether the separation is spacelike or timelike. Further in the language of the embedding space, the two spaces are being described by precisely the same family of orthogonal unit vectors, \ie pairs satisfying Eqs.~\reef{eq.condTSnorm} and \reef{eq.condTSorth}. Of course, this indicates that not only do we have two moduli spaces described by the same coset geometry \reef{wacked2} but that in fact we are considering one and the same moduli space from two different perspectives!

In order to develop a better understanding of this counterintuitive result consider the following: The first point to note is that our intuition about spacelike and timelike separated pairs of points is firmly rooted in flat Minkowski space. However, recall that in the embedding space, the the Poincar\'e patch coordinates \reef{poinc} only cover a portion of the AdS hyperboloid \reef{eq.hyperboloid} and some $SO(2,d)$ transformations will take us out of this region, \ie pairs of points maybe mapped beyond the corresponding Minkowski space in the asymptotic boundary. 
Hence it is more appropriate to think of working on global coordinates for the AdS geometry or transforming the CFT to the `cylindrical' background $\mathbb{R}\times S^{d-1}$ 
(with $R$ being the time direction).\footnote{With this transformation, we are actually extending the original Minkowski space to a geometry where the conformal group acts properly everywhere.}

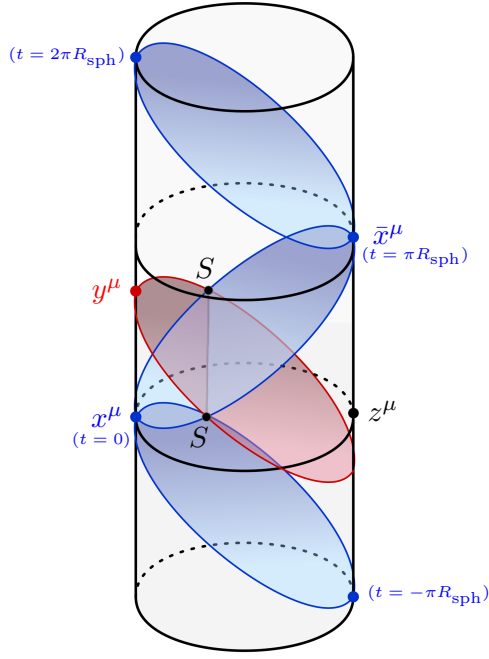
\begin{figure}[t]
\centerline{\includegraphics[width=.2\textwidth]{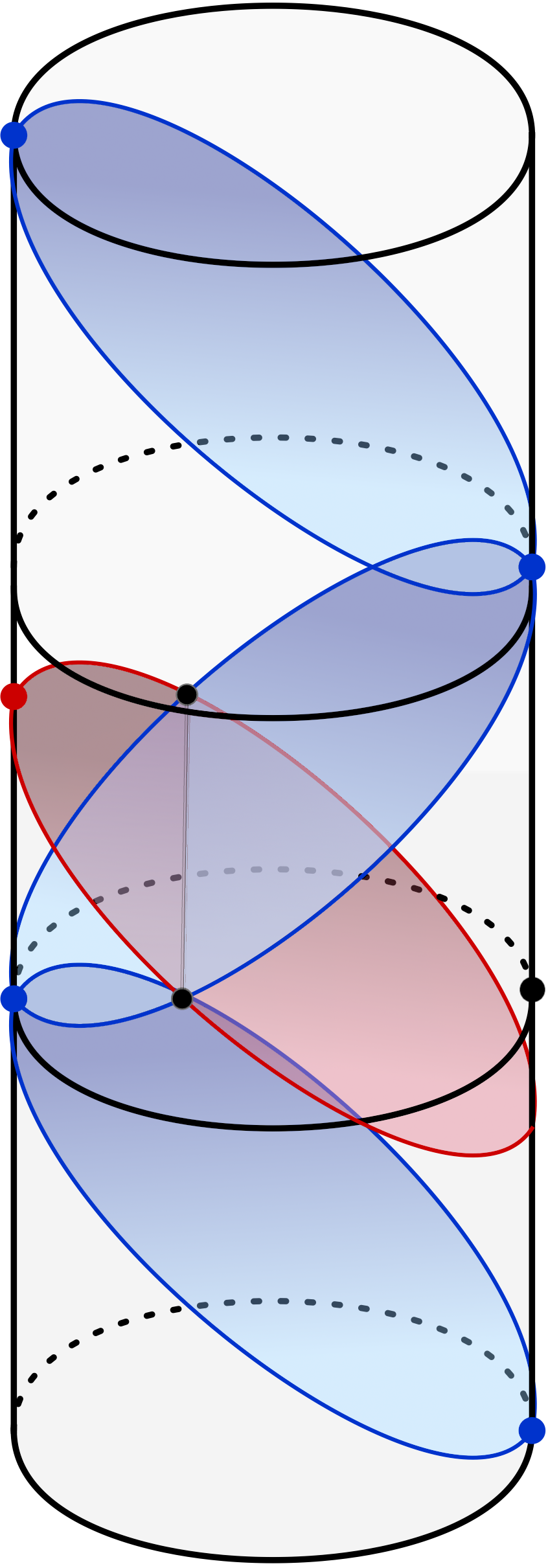}}
\setlength{\unitlength}{0.1\columnwidth}
\begin{picture}(0.3,0.4)(0,0)
\normalsize{
\put(3.8,3.6){\makebox(0,0){${\color{red}y^\mu}$}}
\put(3.8,2.5){\makebox(0,0){${\color{blue}x^\mu}$}}
\put(6.2,2.55){\makebox(0,0){${z^\mu}$}}
\put(4.6,2.32){\makebox(0,0){${S}$}}
\put(4.65,3.8){\makebox(0,0){${S}$}}
\put(6.25,4.1){\makebox(0,0){${\color{blue}\bar x^\mu}$}}
}
\tiny{
\put(3.75,2.3){\makebox(0,0){${\color{blue}(t=0)}$}}
\put(3.45,5.65){\makebox(0,0){${\color{blue}(t=2\pi\Rs)}$}}
\put(6.48,3.9){\makebox(0,0){${\color{blue}(t=\pi\Rs)}$}}
\put(6.6,.93){\makebox(0,0){${\color{blue}(t=-\pi\Rs)}$}}
}
\end{picture}
\caption{Illustration of the CFT on cylindrical background $\mathbb{R}\times S^1$. The point $z^\mu$ is the antipodal point from the point $x^\mu$ on the constant time slice containing this point and $z^\mu$ has the maximal spacelike geodesic distance $\pi \Rs$ from $x^\mu$. Blue lines are past and future lightcones of $x^\mu$. The point $\bar{x}^\mu$ corresponds to the position where the future light cone of $x^\mu$ first self-intersects. The sphere $S$ (indicated by black points) can be described as the intersection of past lightcone of $y^\mu$ either with future lightcone of $x^\mu$, or alternatively with past lightcone of the antipodal point $\bar{x}^\mu$. In the former case, $S$ is characterized by a pair of timelike separated points, in the latter case by a pair of spacelike separated points. 
}
\label{fig:cylinder1}
\end{figure}

In the latter geometry, there are limits to how far apart the pairs of points can be.\footnote{As in flat space, we measure the separation between points in $\mathbb{R}\times S^{d-1}$ as the (minimal) proper distance along geodesics connecting the points.} In particular for spacelike separated points, the maximum separation is $\pi \Rs$ where $\Rs$ is the radius of curvature of the $S^{d-1}$, \ie maximally separated pairs are antipodal pairs on the $(d-1)$-sphere --- see Figure \ref{fig:cylinder1}. Similarly, the maximal separation for  a timelike  pair is $2\pi\Rs$. For example, if the two points lie at the same pole on the sphere, then with this maximal time separation, the lightcones from these two points intersect at a point on the the opposite pole and hence the corresponding sphere has the maximal angular size, \ie the sphere's proper size has actually shrunk to zero but the `enclosed' ball covers the entire $S^{d-1}$. In fact, as illustrated in the figure, the null cones of these two maximally (timelike) separated points actually coincide.\footnote{In the embedding space, the two points considered here are actually coincident points on the boundary of the AdS hyperboloid \reef{eq.hyperboloid}. It is only when we consider the universal cover of the AdS hyperboloid (as we do implicitly here) that the points are separated. In particular, if we had been more precise we should have replaced the $SO(2,d)$ group in the numerator of
\reef{wacked2} by a suitable infinite cover in this case.} 
This leads to the observation that because of the compact structure of the $S^{d-1}$, when we choose any single point in the $\mathbb{R}\times S^{d-1}$, by following the past and future null cones, we actually specify two families of preferred points. The first being points lying at the same pole of the sphere at $t=2\pi n\Rs$ where $n$ is any integer (and we have assumed the initial point lies at $t=0$, \ie $n=0$). The second family is points on the opposite pole lying at $t=2\pi (n+\half) \Rs$ where $n$ is again any integer.

This insight then allows us to understand the equivalence of the two spaces discussed above in very concrete terms. Consider the two timelike separated points designated $x$ and $y$ shown in Figure \ref{fig:cylinder1}. The future lightcone of $x$ and the past light cone of $y$ intersect on the sphere designated $S$. However, now consider the point $\bar x$ where the future lightcone of $x$ (first) converges to a point on the opposite pole of the sphere. The pair $\bar x$ and $y$ is now a spacelike separated pair of points. The past and future lightcones from these two points intersect at the {\it spheres}, $S$ and $\tilde S$, respectively. Now, in an appropriate conformal frame, where $\bar x$ and $y$ are spacelike separated points in flat Minkowski, these two spheres become the two branches of the corresponding spacelike hyperbola discussed above.\footnote{Each branch is topologically a $(d-1)$-sphere when we include the point at infinity.} However, the key point here is that in the $\mathbb{R}\times S^{d-1}$ conformal frame, we can specify spheres either in terms of the intersection of the past and future lightcones of a pair of timelike separated points or in terms of the intersection of the past light cones from two spacelike separated points. Hence we recognize that moduli spaces of spacelike and timelike pairs in fact provide two different perspectives of the same geometric object!

Given that the moduli spaces of spacelike and timelike pairs (on $\mathbb{R}\times S^{d-1}$) are the same, it is interesting that the discussion in section \ref{sec:CausalDi} implies that the limit in which a timelike separated pair approaches a null separated pair of points is a limit that takes us to timelike infinity in the moduli space --- see footnote \ref{call8}. This is a consistency check in that it shows that there is no trajectory on the moduli space that carries one between timelike separation to spacelike separation. Of course, it would be interesting to further explore the implications of this equivalence.

\section{Conventions for symmetry generators}
\label{app:conventions}

\subsection{General definitions}
\label{sec:ConvGeneral}

\paragraph{Spinless case:} 
Given the conformal symmetry generators $L_i(x)$, we define the second Casimir as the object $\Casimir\equiv C^{ij}L_i(x)L_j(x)$  (where $i,j=1,\cdots,(d+1)(d+2)/2$) which acts on scalar primaries $\mathcal{O}$ in the CFT with dimension $\Delta_\mathcal{O}$ such that:
\be 
[\Casimir ,{\cal O}(x)] = \Delta_{\cal O}(d-\Delta_{\cal O}) {\cal O}(x)\,.
\ee
In this appendix we discuss various realizations of $\Casimir$ on objects which carry a representation of the conformal group: fields in AdS$_{d+1}$ and functions on the moduli space of causal diamonds. 

The conformal algebra in $d$ dimensions is isomorphic to the group $SO(2,d)$ Lorentz group of the embedding space \eqref{eq.metricaux}. We write the action of $SO(2,d)$ generators on primaries as  (see, \eg \cite{Fradkin:1996is})
\be \label{eq:rotationGen}
\begin{split}
  M_{\mu\nu}|{\cal O}( x)\rangle &= i (x_\mu\partial_\nu - x_\nu \partial_\mu)|{\cal O}( x)\rangle \,,\\
   P_\mu|{\cal O}( x)\rangle &= i\partial_\mu |{\cal O}( x)\rangle \,,\\
   Q_\mu|{\cal O}( x)\rangle &= i \left( x^2 \partial_\mu -2x_\mu x^\nu \partial_\nu -2 x_\mu\Delta_\mathcal{O} \right) |{\cal O}( x)\rangle \,,\\
  D|{\cal O}( x)\rangle &= i\left( x^\mu \partial_\mu +\Delta_{\cal O} \right) |{\cal O}( x)\rangle \,,
\end{split}
\ee 
where $|{\cal O}( x)\rangle={\cal O}(x)|0\rangle$ and the vacuum state $|0\rangle$ is annihilated by all of the generators.
The $SO(2,d)$ Lorentz generators $J_{ab} = -J_{ba}$ are hence represented by 
\be \label{Jab}
\begin{split}
 J_{\mu\nu} = M_{\mu\nu}  \,, \quad J_{\mu-} = \frac{1}{2} (P_\mu+Q_\mu) \,, \quad  J_{d\mu} = \frac{1}{2} (P_\mu-Q_\mu) \,,\quad J_{-d} = D \,.
\end{split}
\ee
These satisfy the algebra
\be \label{eq:ConfAlg}
[J_{ab},J_{cd}] = i\, \eta^{{}_{(2,d)}}_{bc} \, J_{ad} -i\, \eta^{{}_{(2,d)}}_{ac} \, J_{bd} - i \, \eta^{{}_{(2,d)}}_{bd} \,  J_{ac} + i \, \eta^{{}_{(2,d)}}_{ad} \, J_{bc}\ \,,
\ee
where $\eta^{{}_{(2,d)}}_{ab} = \text{diag}(-1,-1,1,\ldots,1)$ is the embedding space metric. 
In terms of these Lorentz generators, we can represent the action of the Casimir on operators by $\Casimir \equiv - \frac{1}{2} J^{ab}J_{ab}$, which acts as a differential operator whose eigenfunctions are the primary states:
\be \label{eq:CasimirDef}
\begin{split}
\Casimir |{\cal O}(x)\rangle &\equiv - \frac{1}{2} J^{ab}(x)J_{ab}(x) |{\cal O}(x)\rangle = \Delta_{\cal O} (d-\Delta_{\cal O}) \, |{\cal O}(x)\rangle\,.
\end{split}
\ee

Since $SO(2,d)$ acts on the AdS$_{d+1}$ hyperboloid in embedding space as standard Lorentz transformations, the above generators can also be represented as isometry generators of AdS$_{d+1}$. This representation is given in embedding space coordinates by $J_{ab} = i (X_a\partial_b-X_b\partial_a)$. In particular, the AdS$_{d+1}$ Laplacian is represented by the combination
\be 
\RAdS^2 \, \nabla^2_{AdS}  = \frac{1}{2} J^{ab}J_{ab} =- \Casimir \,.
\ee

Similarly, the action of the Casimir is represented on the moduli space of causal diamonds. 
Using the explicit representation \eqref{eq:rotationGen}, it is straightforward to verify the following relation between the second Casimir as a differential operator acting on the space of causal diamonds, and the scalar Laplacian on the same space: 
\be \label{eq:CasimirDiff}
\begin{split}
 \Casimir f(x,y) \equiv -\frac{1}{2} ( J^{ab}(x)+J^{ab}(y))( J_{ab}(x)+J_{ab}(y) ) f(x,y) = L^2 \, \BoxM\, f(x,y) \,,
\end{split}
\ee
where $f(x,y)$ is any function on the space of diamonds $\lozenge=(x^\mu,y^\mu)$, and $\BoxM$ is the Laplacian on the moduli space of diamonds \eqref{eq.metriccosetd}.

As an application of this, one can explicitly check that the kernel in Eq.~\eqref{intertwine} is an eigenfunction of the Casimir as represented by Eq.~\eqref{eq:CasimirDiff}:
\be 
 \Casimir \left( \frac{|y-\xi||\xi-x| }{|y-x|} \right)^{\Delta_{\cal O}-d}
 = \Delta_{\cal O}(d-\Delta_{\cal O}) \left( \frac{|y-\xi||\xi-x| }{|y-x|} \right)^{\Delta_{\cal O}-d} \,.
\ee

\paragraph{Generalization with spin:} It is straightforward to generalize the above discussion to the case of primary operators with symmetric-traceless indices, ${\cal O}_{\mu_1\cdots\mu_\ell}$. In this case, the eigenvalues of the conformal Casimir are
\be 
[\Casimir,{\cal O}_{\mu_1\cdots\mu_\ell}(x)] = \left[ \Delta_{\cal O} (d-\Delta_{\cal O}) - \ell(\ell+d-2) \right]{\cal O}_{\mu_1\cdots\mu_\ell}(x) \,.
\ee
One can explicitly verify that Eq.~\eqref{eq:CasimirDiff} then still holds for tensors instead of functions $f(x,y)$. Most importantly, we find that the kernel in our proposal  \eqref{gen:firstlaw2} for the `first law'-like expression with spin satisfies
\be 
L^2 \, \BoxM \left( G^{\mu_1\cdots\mu_\ell}(x,y;\xi) \, T_{\mu_1 \cdots \mu_\ell}(\xi)\right) =\left[ \Delta_{\cal O} (d-\Delta_{\cal O}) - \ell(\ell+d-2) \right] G^{\mu_1\cdots\mu_\ell} (x,y;\xi)\, T_{\mu_1 \cdots \mu_\ell} (\xi)
\ee
where $T_{\mu_1 \cdots \mu_\ell}$ (\eg $ T_{\mu_1 \cdots \mu_\ell}= \langle {\cal O}_{\mu_1\cdots\mu_\ell}\rangle$) is an arbitrary symmetric traceless tensor and we abbreviated the kernel as
\be 
 G^{\mu_1 \cdots \mu_\ell}(x,y;\xi) \equiv \left( \frac{|y-\xi||\xi-x| }{|y-x|} \right)^{\Delta_{\cal O}-d}
\frac{s^{\mu_1}\ldots s^{\mu_{\ell}}}{ (|y-\xi||\xi-x||y-x|)^{\ell} } \,.
\ee

\subsection{Two-dimensional case}
\label{sec:2dConv}

Let us briefly make the statements of the previous subsection more explicit in the case of two-dimensional CFTs (and AdS$_3$, respectively). In this case, we can work in right- and left-moving coordinates 
\begin{equation}
ds_{\text{\tiny CFT}_2}^2 = -dt^2 + dx^2 = d\xi \, d\bar\xi \,.
\end{equation}
In these coordinates, the non-zero generators \eqref{eq:rotationGen} can be written as:
\begin{equation}\label{eq:2dGen}
\begin{split}
M_{01}  |{\cal O}\rangle= - M_{10}  |{\cal O}\rangle&= i(\xi \partial_\xi - \bar \xi \partial_{\bar\xi})  |{\cal O}\rangle\equiv i (\bar L_0 - L_0 ) |{\cal O}\rangle \,,\\
P_0 |{\cal O}\rangle &= i (\partial_{\bar\xi} - \partial_\xi )  |{\cal O}\rangle\equiv i(\bar L_{-1}  - L_{-1} ) |{\cal O}\rangle \,, \\
P_1  |{\cal O}\rangle&= i (\partial_{\bar\xi} + \partial_\xi )  |{\cal O}\rangle\equiv i(\bar L_{-1}  + L_{-1} )  |{\cal O}\rangle\,, \\
Q_0  |{\cal O}\rangle&= i ( \bar\xi^2 \partial_{\bar\xi} - \xi^2 \partial_\xi + (\bar \xi - \xi) \Delta_{\cal O} ) |{\cal O}\rangle\equiv i(\bar L_1 - L_1) |{\cal O}\rangle \,,\\
Q_1  |{\cal O}\rangle&= i  (- \bar\xi^2 \partial_{\bar\xi} - \xi^2 \partial_\xi - (\bar \xi + \xi) \Delta_{\cal O}) |{\cal O}\rangle \equiv -i(\bar L_1 + L_1)  |{\cal O}\rangle\,,\\
D  |{\cal O}\rangle&= i (\bar\xi \partial_{\bar\xi} + \xi \partial_\xi + \Delta_{\cal O} )  |{\cal O}\rangle\equiv -i ( \bar L_0 + L_0 ) |{\cal O}\rangle \,.
\end{split}
\end{equation} 
This defines conformal generators $L_n$ satisfying the usual de Witt algebra 
\begin{equation}\label{eq:deWitt}
[L_{n},L_m] = (n-m) L_{n+m} \,,\qquad [\bar L_n,\bar L_m] = (n-m) \bar L_{m+n} \,, \qquad [L_m,\bar L_n] = 0\,,  
\end{equation}
for $n,m=-1,0,1$. The conformal Casimir defined in \eqref{eq:CasimirDef} reads as follows in terms of $L_n$:
\begin{equation}
\begin{split}
 {\cal C}_2  |{\cal O}\rangle &= 2 \left({\cal C}_2^{(d=2)} + \bar {\cal C}_2^{(d=2)}  \right)|{\cal O}\rangle  \,,
\end{split}
\end{equation}
where we make the factorization into natural left- and right-moving Casimir operators explicit by defining
\begin{equation} \label{eq:LRcasimir}
{\cal C}_2^{(d=2)}  \equiv  -L_0^2 +\frac{1}{2} ( L_1 L_{-1} + L_{-1} L_1 ) \quad \text{and} \quad  \bar {\cal C}_2^{(d=2)} \equiv -\bar L_0^2 +\frac{1}{2} (  \bar L_1  \bar L_{-1} +  \bar L_{-1}  \bar L_1 )  \,.
\end{equation}

The above discussion concerned the action of conformal generators on CFT states. There is an analogous set of identities for AdS$_3$ isometry generators. We work in Poincar\'{e} coordinates
\begin{equation}
ds_{\text{\tiny AdS}_3}^2 = \frac{\RAdS^2}{z^2} \left( dz^2 + d\xi \, d\bar\xi \right) \,.
\end{equation}
Using the general definitions of section \ref{sec:ConvGeneral}, we then find the following isometry generators in AdS$_3$:
\be
L_{-1} = \partial_{\xi}, \quad L_{0} = - \frac{1}{2}\,z \, \partial_{z} - \xi \, \partial_{\xi} \quad \mathrm{and} \quad L_{1}= z \, \xi \, \partial_{z} + \xi^2 \, \partial_{\xi} -  z^{2} \, \partial_{\bar{\xi}}\, ,
\ee
and similarly for $\bar L_n$ with $\xi$ and $\bar\xi$ interchanged. We then have that the combinations appearing in the Casimir ${\cal C}_2$, and its left- and right-moving parts defined in \eqref{eq:LRcasimir}, all correspond to the scalar Laplacian on AdS$_3$:
\begin{equation}
 {\cal C}_2 \,\phi(u) = -\RAdS^2 \,\nabla^2_{\text{\tiny AdS}_3} \phi(u) \,,\qquad
 {\cal C}_2^{(d=2)}\,\phi(u) =  \bar {\cal C}_2^{(d=2)} \,\phi(u) = -\frac{1}{4}\RAdS^2 \,\nabla^2_{\text{\tiny AdS}_3} \phi(u)  \,.
\end{equation}

\section{Relative normalization of CFT and bulk quantities}
\label{app:normalization}

In this appendix we demonstrate how to fix the relative normalization between $\dSO$ as defined in Eq.~\eqref{gen:firstlaw} and its holographic couterpart $\dSOh$ in Eq.~\eqref{gen:bulk}. Our strategy will be to exploit the fact that the normalization can be determined in the limit of very small diamonds, or equivalently with $\langle \mathcal{O} \rangle = \text{constant}$. For simplicity, we assume the centre of the diamond is located at $\frac{1}{2}(x^\mu+y^\mu)= 0$, and we work on a time slice such that $\frac{1}{2} (y^\mu-x^\mu) = R \, \delta^\mu_0$.

Consider first the field theory observable $\dS x y $ in the limit $x\rightarrow y$, \ie for a constant expectation value $\langle {\cal O}\rangle$ throughout the causal diamond:
\be \label{eq:deltaSlimit}
\begin{split}
\dS x y  &= \Cbdy\, \langle {\cal O}\rangle \int_{D(x,y)} d^d \xi\ \left( \frac{(y-\xi)^2(\xi-x)^2 }{-(y-x)^2} \right)^{\half(\Delta_{\cal O}-d)} \,.
\end{split}
\ee
To evaluate the integral, it is useful to parameterize the causal diamond as follows:
\be
 \xi^\mu=\left(\frac{\zeta-\bar\zeta}{2}\,R,\,\frac{\zeta+\bar\zeta}{2}\,R\, \vec{\omega} \right) \,,\quad
 y^\mu = (R,\vec{0}\,) \,,\quad x^\mu = (-R,\vec{0}\,) \,,
\ee
where $\vec{\omega}\in S^{d-2}$ is a unit vector that parameterizes the spacelike spherical slices. The full range $\zeta,\bar\zeta\in[-1,1]$ would cover the diamond twice. Considering the symmetries of the integrand in \eqref{eq:deltaSlimit}, we can effectively integrate over the range $\zeta\in[-1,1]$ and $\bar\zeta\in[0,1]$:
\be\label{eq:delSbdy}
\begin{split}
  &\dS x y =\\
  &\quad= \frac{\Cbdy\, \langle {\cal O}\rangle}{2^{\Delta_\mathcal{O}-1}}\,\Omega_{d-2}\,R^{\Delta_\mathcal{O}} \int_{-1}^1 d\zeta \, \int_{0}^1 d\bar\zeta  \,\left(1-\zeta^2\right)^{(\Delta_\mathcal{O}-d)/2}\, \left(1-\bar\zeta^2\right)^{(\Delta_\mathcal{O}-d)/2} \left(\zeta+\bar\zeta\right)^{d-2} 
 \\
 &\quad= \frac{\Cbdy\, \langle {\cal O}\rangle}{2^{\Delta_\mathcal{O}+1}} \,\Omega_{d-2}\, \Gamma\left(\frac{\Delta_\mathcal{O}+2-d}{2}\right)^2\,R^{\Delta_\mathcal{O}}\, \sum_{n=0}^{d-2} \binom{d-2}{n} \frac{\left(1+(-1)^n\right)\Gamma\left(\frac{d-n-1}{2}\right)\Gamma\left(\frac{n+1}{2}\right)}{\Gamma\left(\frac{\Delta_\mathcal{O}-n+1}{2}\right)\Gamma\left(\frac{\Delta_\mathcal{O}-d+n+3}{2}\right)} 
 \\
 &\quad= \frac{\Cbdy\, \langle {\cal O}\rangle}{4\, \pi^{1/2}} \,\Omega_{d-2} \,\frac{\Gamma\left(\frac{d-1}{2}\right) \Gamma\left(\frac{\Delta_\mathcal{O}+2-d}{2}\right)^2\Gamma\left(\frac{\Delta_\mathcal{O}}{2}\right)^2}{\Gamma\left(\Delta_\mathcal{O}+1-\frac{d}{2}\right)\Gamma\left(\Delta_\mathcal{O}\right)} \,(2R)^{\Delta_\mathcal{O}}\,,
 \end{split}
\ee
where we binomially expanded the measure factor $(\zeta+\bar\zeta)^{d-2}$ to perform a term-by-term integration. The final line can be simplified slightly by substituting $\Omega_{d-2}=2\pi^{\frac{d-1}{2}}/\Gamma\left(\frac{d-1}{2}\right)$ for the volume of a unit $(d-2)$-sphere, however, the present form is convenient for our comparison below.

Next, we compute $\dSOh$ as defined in Eq.~\eqref{gen:bulk} using standard holographic techniques. In particular, we will work in Poincar\'e coordinates 
\be
ds^2_\mt{AdS} = \frac{\Rads^2}{z^2} \left( dz^2 -dt^2 + dr^2 + r^2 \, d\Omega^2_{d-2} \right) \,.
\ee
If one considers the dual field $\phi(u)$ in a linearized approximation in this background, the asymptotic behaviour takes the following form:
\be
\phi(z\rightarrow 0, w^\mu) = \phi_0(w)\,z^{d-\Delta_\cO} + \phi_1(w)\, z^{\Delta_\cO}+\cdots\label{slap}
\ee
where
\be
\lambda=\phi_0\qquad{\rm and}\qquad
\langle \mathcal{O}\rangle = \frac{\Rads^{d-1}}{2\lp^{d-1}}\,
(2\Delta_\cO-d)\,\phi_1\,.\labell{dash}
\ee
Here $\lambda(\xi)$ is the coupling to the operator in the boundary CFT and we set it to zero in the following.\footnote{Eqs.~\reef{slap} and \reef{dash} present a standard set of holographic conventions, \eg see \cite{teddy}, although perhaps not unique. Further we note that the choice $\lambda=0$ means that we are only studying excitations the CFT ground state here. It would be interesting to extend the discussion in this paper to holographic RG flows where the boundary theory is deformed away from a conformal fixed point.} In keeping with the previous calculation, we also assume that $\langle \mathcal{O}\rangle$ is constant, at least within the boundary region of interest.

The boundary sphere in the previous calculation was chosen to be: $t=0$ and $r=R$. The corresponding extremal surface in the bulk is the hemisphere:
$t=0$ and $z^2+r^2=R^2$. We can parameterize this bulk surface with
$z=R\,\sin\lambda$ and $r=R\,\cos\lambda$ where $0\le\lambda\le \frac\pi 2$. Then keeping on the leading term in the asymptotic expansion of the bulk scalar, the computation of the observable $\dSOh$ reads as follows:
\be \label{eq:deltaSlimitHolo}
\begin{split}
 \dSOh 
 &=  \Cblk \,\phi_1 \, \frac{\RAdS^{d-1}}{\lp^{d-1}}\,\int\! d\Omega_{d-2} \  \int_0^{\pi/2} d\lambda \ \frac{\cos^{d-2}\!\lambda}{\sin^{d-1}\!\lambda}\,  (R\, \sin\lambda)^{\Delta_\mathcal{O}} 
 \\ 
 &=  \Cblk  \, \frac{2\,\langle \mathcal{O} \rangle}{2\Delta_\cO-d} \,\Omega_{d-2}\, \frac{\Gamma\left( \frac{d-1}{2}\right) \Gamma\left(\frac{\Delta_\mathcal{O}+2-d}{2}\right)\Gamma\left(\frac{\Delta_\mathcal{O}}{2}\right)}{4\,\pi^{1/2} \Gamma\left(\Delta_\mathcal{O}\right)}  \, (2R)^{\Delta_\mathcal{O}} \,,
\end{split}
\ee
where we have substituted $\lp^{d-1}=8\pi G_\mt{N}$ and applied Eq.~\reef{dash} in the second line.

We can now equate the two results \eqref{eq:delSbdy} and \eqref{eq:deltaSlimitHolo} and thus fix the relative normalization:
\be \label{eq:CbdyCblk}
  \Cblk =  \Cbdy  \,\frac{ \Gamma\left(\frac{\Delta_\mathcal{O}+2-d}{2}\right)\Gamma\left(\frac{\Delta_\mathcal{O}}{2}\right)}{\Gamma\left(\Delta_\mathcal{O}-\frac{d}{2}\right)}  \,.
\ee

\subsection{Holographic computation for a free scalar in AdS$_{3}$}
\label{exam}

We expect that the generalized first law \eqref{genfirst} provides the leading order contribution to a set of novel physical quantities in CFTs in an analogous way in which the entanglement first law provides the leading order perturbation of the vacuum entanglement entropy for excited states. In the present section we want to corroborate this proposal by providing the holographic dual of $\delta S_{\mathcal{O}}$ in a class of CFTs which admit a semi-classical gravity description.

In section \ref{sec:bulk}, we argued that $\dSO=\dSOh$ with an appropriate choice of the bulk normalization constant $\Cblk$. The latter was fixed above by comparing the two expressions in a situation where $\langle\cO\rangle$ was a constant. In the following, we explicitly demonstrate that the equivalence of the boundary and bulk expressions for a more nontrivial field configuration. To do so, we focus on AdS$_{3}$ with a free probe scalar field $\phi$ dual to a  primary operator $\cal O$ with $h = \bar{h}=\Delta_\cO/2$ in a two-dimensional holographic CFT. In this case, the `sphere' of interest becomes an interval of length $2R$, which for simplicity, we assume is centred at the origin on the $t=0$ time slice.
Further Eq.~\eqref{genfirst} becomes 
\be \label{genfirst2}
\dSO = 
\frac{\Cbdy}{2} \int_{-R}^R d\xi \left( \frac{R^2-\xi^2}{2R} \right)^{\frac{\Delta_{\cal O}-2}2} 
 \int_{-R}^R d\bar{\xi} \left( \frac{R^2-\bar{\xi}^2}{2R} \right)^{\frac{\Delta_{\cal O}-2}2}
 \langle {\cal O}(\xi,\bar{\xi}) \rangle \,.
\ee
The holographic expression in Eq.~\reef{gen:bulk} reduces to an integral of the bulk scalar over the spatial geodesic $\gamma$ connecting the endpoints of the interval in the boundary theory:
\be  \label{eq:FOholo}
  \dSOh = \frac{\Cbdy}{\lp}
 \,\frac{ \Gamma\left(\Delta_\mathcal{O}/2\right)^2}{\Gamma\left(\Delta_\mathcal{O}-1\right)}\ \int_\gamma d\lambda\,\sqrt{h} \; \phi(\lambda) \,,
\ee
where we have used $8\pi G_\mt{N}=\lp$ for $d=2$ and substituted for the normalization constant $\Cblk$ using Eq.~\reef{eq:CbdyCblk}. 
 
For our explicit computation, we pick a simple linearized perturbation by putting a delta-function source at a point $(\xi_0,\bar \xi_0)$ on the boundary. The linearized solution is given by the usual bulk-boundary propagator
\be \label{linfield}
\phi(r,\xi,\bar{\xi}) = \alpha \left( \frac{z}{z^2 + (\xi-\xi_0)^2} \right)^{\Delta_{\cal O}} \,,
\ee
Here, $\alpha$ is an arbitrary constant measuring the strength of the source and we are using Poincar\'e coordinates on AdS$_3$
\be
ds^2 = \frac{1}{z^2} \left(dz^2+d\xi\, d\bar{\xi}\right) 
\ee
where the curvature radius is set to unity and $w,\bar w$ denote the null coordinates introduced in Eq.~\reef{eq.LCdef}, \ie $\xi=x-t$ and $\bar \xi=x+t$.
For simplicity, we will assume that the source is spacelike separated from the interval, \ie $(\xi-\xi_0)^2>0$ for any point $\xi=\bar \xi=x\in [-R,R]$ in the interval.

The bulk geodesic spanning the boundary interval above may be parametrized by 
\be x=R\, \cos\lambda \quad \mathrm{and} \quad z=R\, \sin\lambda\, .
\ee
The line element along the geodesic is $d\lambda/\sin\lambda$ and then Eq.~\reef{eq:FOholo} yields
\be
\begin{split}
\dSOh&=\frac{\Cbdy}{\lp}
 \,\frac{ \Gamma(\frac{\Delta_{\cal O}}{2})^2}{\Gamma\left(\Delta_\mathcal{O}-1\right)}\,\alpha \int_0^{\pi} \frac{d\lambda}{\sin\lambda} \left( \frac{R \sin\lambda}{R^2 \sin^2\lambda +
 (R\cos\lambda-\xi_0)(R\cos\lambda-\bar\xi_0)} \right)^{\Delta_{\cal O}} \\
 &=\frac{\Cbdy}{2\lp}
 \,\frac{ \Gamma(\frac{\Delta_{\cal O}}{2})^2}{\Gamma\left(\Delta_\mathcal{O}-1\right)}\,\alpha\,(2R)^{\Delta_\cO} \int_0^{\pi} d\lambda\;   \frac{ \sin^{\Delta_{\cal O}-1}\!\frac{\lambda}{2} \ 
\cos^{\Delta_{\cal O}-1}\!\frac{\lambda}{2}}{
\big(|R-\xi_0|^2\, \cos^2\! \frac{\lambda}{2} + |R+\xi_0|^2\, \sin^2\! \frac{\lambda}{2}\big)^{\Delta_{\cal O}} } \\
 & = \frac{\Cbdy}{2\lp}\,\alpha\,
 (\Delta_\mathcal{O}-1)\, \frac{\Gamma(\frac{\Delta_{\cal O}}{2})^4}{\Gamma(\Delta_{\cal O})^2} \left( \frac{4R^2}{(\xi_0{}^2-R^2)(\bar\xi_0{}^2-R^2)} \right)^{\Delta_{\cal O}/2}\,.
\end{split} 
\label{eq:FOfinal}
\ee
where in a slight abuse of notation, we have defined $|R\pm\xi_0|^2\equiv(R\pm\xi_0)(R\pm\bar\xi_0)$ in the second line. The integral there can be found, \eg in \cite{gradshteyn2007}.
Note that the final result can be split into right- and left-moving factors, which was not at all clear from the initial expression.\footnote{In the limit $\xi_0^2\gg R^2$, the expectation value is essentially constant across the interval --- see Eq.~\reef{eq.Otest}. Hence in this limit, the leading contribution above can be matched with that in Eq.~\reef{eq:deltaSlimitHolo} with $d=2$. Note that in this case, $\Omega_0=2$.}
 
Now let us now turn to the boundary computation. First we should extract the expectation value from the  $\langle {\cal O} \rangle$ from our linearized solution \eqref{linfield} for the bulk scalar. As we take $z \rightarrow 0$ in Eq.~\eqref{linfield}, we immediately recognize the behavior of a normalizable mode
\be
\phi(z\rightarrow 0,\xi,\bar{\xi}) = \alpha \left( \frac{z}{(\xi-\xi_0)^2} \right)^{\Delta_{\cal O}}+\cdots \,.
\ee
Now applying Eq.~\reef{dash} with $d=2$, we find
\be
\label{eq.Otest}
\langle {\cal O}(\xi,\bar{\xi})\rangle = \frac{\Delta_\cO-1}{\lp} \, \frac{\alpha}{(\xi-\xi_0)^{\Delta_{\cal O}}(\bar\xi-\bar\xi_0)^{\Delta_{\cal O}}} \,.
\ee
Since this profile factorizes into right- and left-moving contributions, upon substitution into Eq.~\eqref{genfirst2}, we also find a factorized answer:
\be
 \dSO = \frac{\Cbdy}{2\lp}\,\alpha\,
 (\Delta_\mathcal{O}-1)\,  \left| \int_{-R}^R d\xi \left( \frac{R^2-\xi^2}{2R} \right)^{\frac{\Delta_{\cal O}-2}2} 
\frac{1}{(\xi-\xi_0)^{\Delta_{\cal O}}}\right|^2\,,
\ee
where as above, we are using $|f(\xi)|^2=f(\xi) f(\bar\xi)$ in the notation of complex coordinates. 
This integral can also be performed, \eg see Eq.~(3.199) in\cite{gradshteyn2007} and one finds\footnote{Again in the limit $\xi_0^2\gg R^2$, the leading contribution in expression \eqref{eq:2dResult} agrees with Eq.~\reef{eq:delSbdy} for $d=2$.}
\be\label{eq:2dResult}
\dSO =  \frac{\Cbdy}{2\lp}\,\alpha\,
 (\Delta_\mathcal{O}-1)\, \frac{\Gamma(\frac{\Delta_{\cal O}}{2})^4}{\Gamma(\Delta_{\cal O})^2} \left( \frac{4R^2}{(\xi_0{}^2-R^2)(\bar\xi_0{}^2-R^2)} \right)^{\Delta_{\cal O}/2} \,,
\ee
which provides a perfect agreement with the holographic result in Eq.~(\ref{eq:FOfinal}). 

Since this is a linearized calculation, the agreement \reef{claim} readily extends to arbitrary field configurations that are generated by the insertion of sources that are spacelike separated from the interval of interest. Of course, Eqs.~\reef{eq:FOfinal} and \reef{eq:2dResult} show that there are singularities that appear when the sources cross the lightcones of the endpoints of the interval, \ie when the sources move into causal contact with the interval.
It would be interesting to investigate further here to understand if $\dSO=\dSOh$ still applies in the latter situation. Following the general arguments in section \ref{sec:bulk}, this is intimately related to the question of better understanding 
causal wedge reconstruction in the bulk.

\bibliography{enthol2}{}
\bibliographystyle{utphys}

\end{document}